\documentclass[12pt]{article}
\usepackage[utf8]{inputenc}
\usepackage[T1]{fontenc}

\usepackage[english,activeacute]{babel}
\usepackage{natbib}
\usepackage{comment}
\usepackage{float}
\usepackage[hidelinks]{hyperref}
\usepackage{mathrsfs}
\usepackage{enumitem}
\usepackage[font={small,it}]{caption}
\usepackage{amsmath,amsfonts,amsthm,amssymb}
\usepackage{bm,rotating,multirow,dsfont,graphicx}
\usepackage[usenames, dvipsnames]{color}
\usepackage{url}
\usepackage{multicol}
\usepackage{multirow}
\usepackage[T1]{fontenc}
\usepackage{flafter}
\usepackage{appendix}
\usepackage{subfigure}
\usepackage{xcolor}
\usepackage{soul}
\usepackage{setspace}
\usepackage{booktabs}
\usepackage{tabularx}
\usepackage{setspace}

\usepackage{tikz}
\usetikzlibrary{arrows.meta}
\usetikzlibrary{positioning}
\usetikzlibrary{calc}

\makeatletter
\def\hlinewd#1{%
	\noalign{\ifnum0=`}\fi\hrule \@height #1 %
	\futurelet\reserved@a\@xhline}
\makeatother
\def\spacingset#1{\renewcommand{\baselinestretch}{#1}\small\normalsize}
\spacingset{1}

\begin{document}

\title{\textbf{Peace Sells, But Whose Songs Connect? \\ Bayesian Multilayer Network Analysis of the Big 4 of Thrash Metal}}

\date{}

\author{
    Juan Sosa\footnote{Corresponding author: jcsosam@unal.edu.co.}\qquad\qquad Erika Martínez\qquad\qquad Danna L. Cruz-Reyes \\
    \hspace{1cm}\\
    Universidad Nacional de Colombia, Colombia 
}

\maketitle

\begin{abstract}
We propose a Bayesian framework for multilayer song similarity networks and apply it to the complete studio discographies of the ``Big 4'' of thrash metal (Metallica, Slayer, Megadeth, Anthrax). Starting from raw audio, we construct four feature–specific layers (loudness, brightness, tonality, rhythm), augment them with song exogenous information, and represent each layer as a $k$–nearest neighbor graph. We then fit a family of hierarchical probit models with global and layer–specific baselines, node– and layer–specific sociability effects, dyadic covariates, and alternative forms of latent structure (bilinear, distance–based, and stochastic block communities), comparing increasingly flexible specifications using posterior predictive checks, discrimination and calibration metrics (AUC, Brier score, log–loss), and information criteria (DIC, WAIC). Across all bands, the richest stochastic block specification attains the best predictive performance and posterior predictive fit, while revealing sparse but structured connectivity, interpretable covariate effects (notably album membership and temporal proximity), and latent communities and hubs that cut across albums and eras. Taken together, these results illustrate how Bayesian multilayer network models can help organize high–dimensional audio and text features into coherent, musically meaningful patterns.
\end{abstract}


\spacingset{1.1} 

\newpage

\section{Introduction}

Music is fundamentally characterized as a tool of cultural expression, functioning as a universal system of socialization and identity formation. Through combinations of pitch, timbre, rhythm, and dynamics, musical expressions condense aesthetic, emotional, and social dimensions that reflect collective identities, historical processes, and technological transformations. The scientific study of music has gained renewed momentum with the development of computational audio tools that enable the quantitative analysis of large sound corpora, giving rise to the field of \emph{Music Information Retrieval} (MIR), whose primary objective is the automated, reproducible, and scalable extraction, modeling, and comparison of musical patterns, \cite{Muller2015Fundamentals} and \cite{Lerch2012Audio}.

Within this landscape, \emph{thrash metal} emerges as a particularly compelling object of study. Originating in the early 1980s as a synthesis between the speed and attitude of punk and the instrumental complexity of traditional heavy metal, thrash metal transformed heavy music by introducing extreme tempos, highly distorted guitars, intense rhythmic patterns, and a compositional approach dominated by fast riffs and dynamic structures. This style represented a historical turning point that gave rise to multiple subsequent subgenres, such as death metal, groove metal, and metalcore, significantly expanding the musical lexicon of contemporary metal.

The consolidation of thrash metal was marked by the emergence of four bands that defined its global projection, Slayer, Megadeth, Metallica, and Anthrax, collectively known as the \emph{Big 4 of Thrash Metal}. Although united by a common aesthetic identity and instrumentation, these groups developed clearly differentiated stylistic trajectories. Slayer became characterized by extreme rhythmic aggressiveness and dark atmospheres, Megadeth by remarkable instrumental complexity and elaborate harmonic structures, Metallica by a compositional evolution toward broader and more accessible arrangements, and Anthrax by a distinctive hybridity incorporating influences from punk and hardcore. These contrasting developments render the Big 4 an ideal controlled laboratory for comparative musical analysis, combining genre homogeneity with long discographies, marked temporal evolution, and substantial within and across band stylistic heterogeneity.

\begin{figure}[!htb]
    \centering
    \subfigure[\textsf{METALLICA}.]  {\includegraphics[scale=0.15]{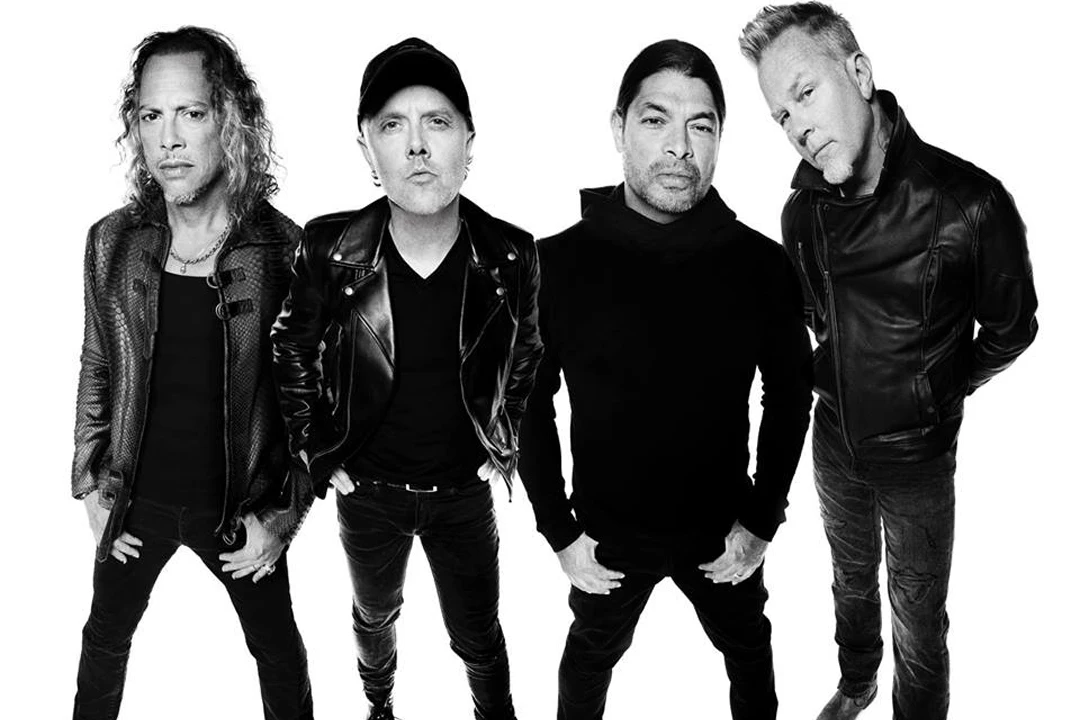}}
    \subfigure[\textsf{SLAYER}.   ]  {\includegraphics[scale=0.15]{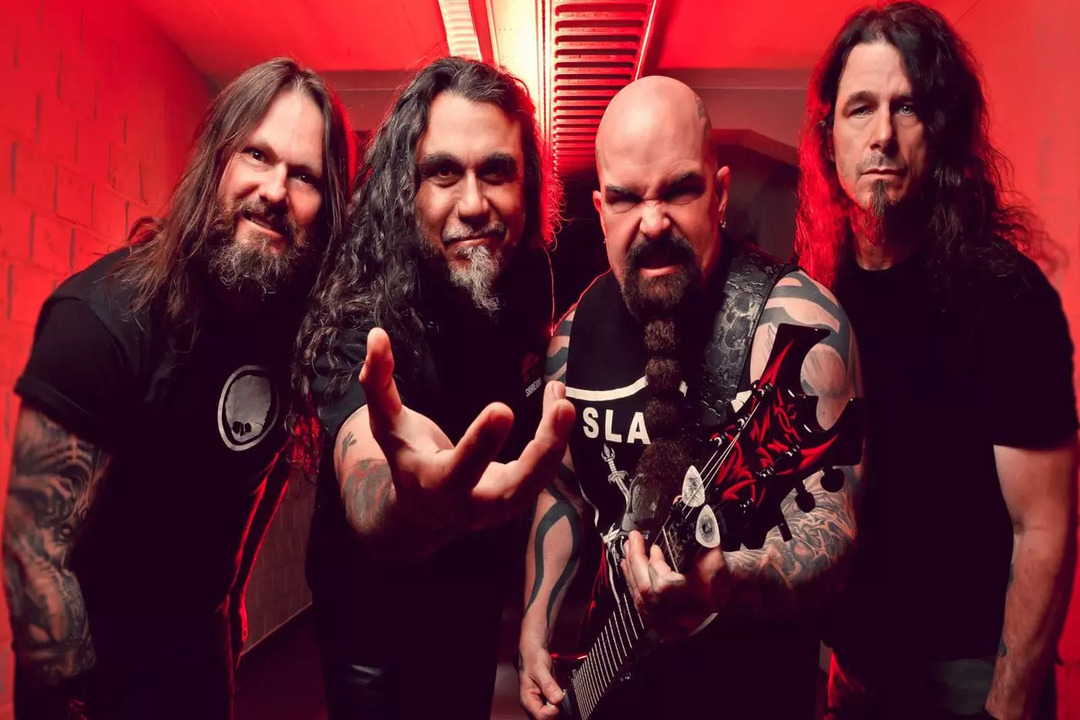}}
    \subfigure[\textsf{MEGADETH}. ]  {\includegraphics[scale=0.15]{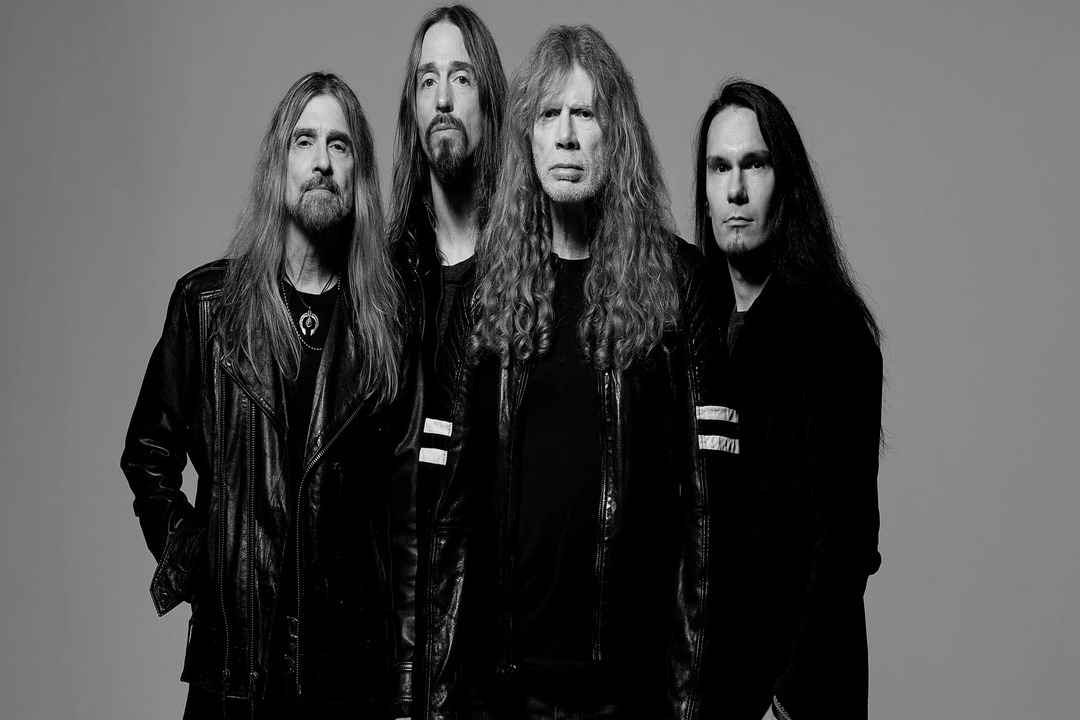}}
    \subfigure[\textsf{ANTHRAX}.  ]  {\includegraphics[scale=0.15]{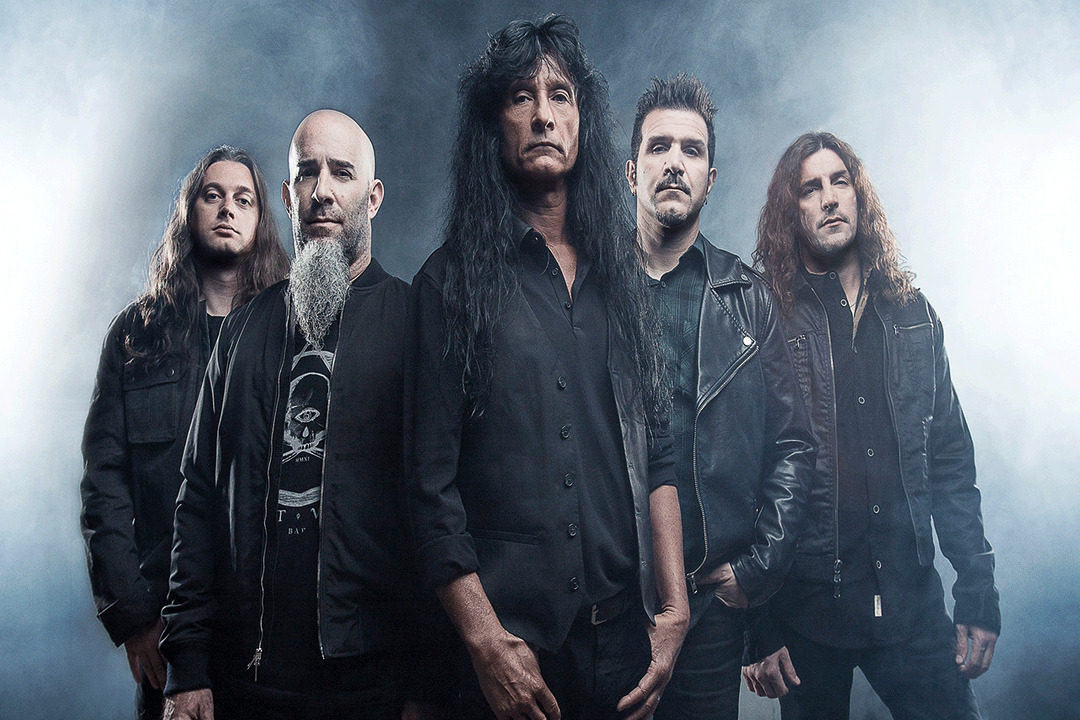}}
    \caption{Metallica, Slayer, Megadeth, and Anthrax.}
    \label{fig_bands}
\end{figure}

From a modern statistical standpoint, music may be addressed as a complex system of interactions among multiple acoustic components, including frequencies, notes, intervals, rhythms, timbres, and intensities, which can be formalized as networks. In this framework, nodes represent musical events (notes, chords, motifs, or spectral segments), while edges quantify relationships of co-occurrence, transition, or sonic similarity, \cite{Serra2015MusicNetworks}, \cite{Gallagher2014MusicalNetworks} and \cite{Park2019MusicGraphs}. Several studies in MIR have leveraged network representations to explore stylistic similarity, genre structure, and artist influence patterns \cite{Serra2012UnsupervisedMusicStructure} and \cite{McFee2014Metrics}. More recently, this perspective has been extended toward multilayer representations, in which distinct musical facets, such as dynamics (loudness), timbral brightness, tonality, and rhythmic patterns, are modeled as concurrent layers of the same system \cite{Pons2017MusicMultilayer}. These developments align closely with the broader theory of multilayer and multiplex networks \cite{Kivela2014MultilayerReview}, \cite{Battiston2014Multilayer}, \cite{Boccaletti2014MultilayerReview} and \cite{DeDomenico2013Multilayer}, which provides a unified mathematical framework for representing interconnected relational data arising from heterogeneous sources.

Despite the potential of these methods, most existing applications in music networks remain focused on descriptive or deterministic approaches, frequently restricted to single layer or aggregated similarity networks and relying on clustering or modularity based community detection techniques. Such methods typically do not provide a fully probabilistic treatment of uncertainty, nor do they permit explicit modeling of cross-layer dependence or the joint estimation of covariate effects and latent network structure.

In this context, Bayesian latent space models provide a flexible inferential framework for analyzing complex network data. Initially developed for social networks \cite{hoff2002latent}, these models were subsequently generalized to multirelational settings through joint and hierarchical formulations that enable information sharing across layers while preserving layer-specific latent structures \cite{Gollini2016JointNetworks} and \cite{Sosa2022Multilayer}. Additional Bayesian developments on community based modeling include mixed membership stochastic block models \cite{Airoldi2008MixedMembership} and degree corrected and hierarchical block specifications \cite{Peixoto2014HierarchicalSBM}. 

The multilayer Bayesian literature also intersects with a vast body of research on \emph{dynamic network models}, where network structure evolves over time and dependence across temporal layers is explicitly encoded. Representative examples include dynamic latent space models \cite{Sewell2015DynamicLatentSpace}, \cite{Durante2017DynamicNetworks} and dynamic stochastic block models \cite{Yang2011DynamicSBM}, \cite{Matias2017DynamicReview}. While these approaches are conceptually related, the present work does not consider dynamic network modeling, our layers correspond to different audio feature spaces rather than time indexed network realizations. Temporal information is incorporated exclusively through exogenous covariates, rather than via latent temporal evolution of the network itself.

Despite the growing body of work on Bayesian multilayer network modeling, applications to networks constructed directly from raw musical audio remain largely unexplored. In particular, no unified probabilistic framework has yet been developed that simultaneously integrates heterogeneous acoustic descriptors while modeling sociability effects, exogenous covariates, alternative forms of latent structure, and predictive uncertainty.

This work introduces several methodological, theoretical, and applied contributions. First, we propose a general methodology for constructing similarity networks directly from audio signals, integrating heterogeneous acoustic descriptors into multilayer graph representations. Second, we develop and evaluate five Bayesian models for multilayer network data aimed at studying musical sociability patterns, incorporating covariates, and exploring latent structures through visualization and clustering techniques. Third, we formulate a systematic framework for prior elicitation tailored specifically to musical network analysis. Fourth, we provide the complete development of the proposed models and their inference through \emph{Markov Chain Monte Carlo} (MCMC) methods, including both theoretical foundations and fully reproducible computational implementations. Fifth, we apply our framework to the empirical analysis of the complete studio discographies of the four emblematic thrash metal bands. Sixth, we conduct comparative performance studies on both this dataset and independent external datasets. Finally, we release a freely accessible public repository enabling full reproducibility of all results reported in this study.

This paper is organized as follows. Section \ref{sec_data} describes the construction of the multilayer audio networks, including the preprocessing pipeline, feature extraction, and exploratory network analysis for the Big 4 discographies. Section \ref{sec_models} presents the suite of Bayesian multilayer network models developed in this work, from baseline sociability specifications to increasingly rich latent structure formulations. Section \ref{sec_computation} details the Bayesian inferential strategy and MCMC algorithms used for posterior estimation. Section \ref{sec_illustration} reports model comparison results and provides a comprehensive empirical analysis of the Big 4 multilayer networks under the best performing specification. Section \ref{sec_community} focuses on posterior based community detection and the interpretation of latent structures relative to album organization. Section \ref{sec_more_dataset} extends the evaluation to additional real world multilayer network datasets to assess the robustness and generality of our findings. Finally, Section \ref{sec_discussion} concludes with a summary of results, methodological implications, and directions for future research.


\section{The Big 4 data}\label{sec_data}

This section outlines the end–to–end audio pipeline used in our study. First, we curate the “Big 4” studio discographies and harmonize the raw audio via a uniform decoding and normalization procedure. Next, we analyze each track in short, overlapping frames to extract four interpretable descriptors, loudness (RMS), brightness (SC), tonality (SFM), and rhythmic onset strength (Flux), that summarize time–varying timbral and rhythmic structure. Finally, we construct feature–specific song–song similarity graphs and conduct an exploratory data analysis that characterizes network structure and informs subsequent modeling.

\subsection{Source material and digitization}

We consider the discographies of the ``Big 4'' of thrash metal (Metallica, Slayer, Megadeth, and Anthrax) restricted to official studio albums listed on each band's website at the time of data collection. For each album, we compile the track list and obtain the corresponding audio files, along with complementary metadata for each song (year, band, album, song, duration in seconds, beats per minute, and lyrics). By band, the data comprises: Metallica: 13 studio albums (136 songs) spanning 1983--2023; Slayer: 11 studio albums (114 songs) spanning 1983--2015; Megadeth: 16 studio albums (173 songs) spanning 1985--2022; and Anthrax: 11 studio albums (123 songs) spanning 1984--2016. In Appendix~\ref{sec_albums} we list the album coverage for each band. All audio is decoded via FFmpeg (an open–source, cross–platform multimedia framework) to uncompressed WAV (Waveform Audio File Format) at the original sampling rate and native channels, then converted to mono by channel averaging. When the source encoding employs pulse–code modulation with integer samples, amplitudes are linearly rescaled to $[-1,1]$ prior to analysis to ensure comparability across sources. This decoding strategy enforces a consistent analysis back end irrespective of the container/codec.

\subsection{Feature extraction}

To capture how the timbral and rhythmic content of each song evolves over time, we analyze every track by dividing the audio signal into short, overlapping segments and computing spectral descriptors on each segment. Let $x[t]$ denote the mono waveform sampled at rate $f_s$. We use a Hann window
\[
w[n]=\tfrac12\!\left(1-\cos\!\frac{2\pi n}{N-1}\right),\qquad n=0,\ldots,N-1,
\]
with window length $N=\lfloor 0.046\,f_s\rfloor$ (approximately $46$\,ms) and hop size $H=\lfloor 0.023\,f_s\rfloor$ (approximately $23$\,ms). For frame index $m = 0,1,\ldots,M_f-1$, where $m$ denotes the sequential position of each analysis frame, the windowed signal is $x_m[n]=x[mH+n]\,w[n]$. Let $X_m(k)$ denote the length–$N$ discrete Fourier transform of $x_m$, and define the one–sided magnitude spectrum $M_m(k)=|X_m(k)|$ for $k=1,\ldots,K$, where $K=\lfloor N/2\rfloor$ (i.e., the direct–current bin at $k=0$ is excluded) and $f_k = k\,f_s/N$ is the center frequency of bin $k$. With a small constant $\epsilon>0$ to ensure numerical stability and $\log$ denoting the natural logarithm, we compute four standard per–frame metrics:
\begin{enumerate}
\item Loudness proxy: Log-root-mean-square energy,
\[
\mathrm{RMS}_m=\log\left(\sqrt{\tfrac{1}{N}\textstyle\sum_{n=0}^{N-1} x_m[n]^2} + \epsilon \right).
\]
Higher values indicate greater acoustic energy.

\item Spectral brightness: Log-frequency spectral centroid,
\[
\mathrm{SC}_m=\frac{\sum_{k} (\log f_k)\,M_m(k)}{\sum_{k} M_m(k)}.
\]
Larger values correspond to brighter (more high-frequency) spectra.

\item Spectral flatness: Logit ratio of geometric to arithmetic mean of magnitudes,
\[
\mathrm{SFM}_m=\text{logit}\,\frac{\exp\!\big(\tfrac{1}{K}\sum_{k}\log(M_m(k)+\epsilon)\big)}{\tfrac{1}{K}\sum_{k} M_m(k)+\epsilon}.
\]
High spectral flatness reflects noise-like or percussive content, whereas low spectral flatness reflects harmonic, peak-dominated spectra.

\item Spectral flux: Log-half-wave rectified inter-frame spectral change,
\[
\mathrm{Flux}_m=\log\left(\textstyle\sum_{k}\max\big(M_m(k)-M_{m-1}(k),\,0\big) + \epsilon\right).
\]
Larger values mark onsets/accents and rhythmic activity.
\end{enumerate}

Intuitively, $\mathrm{RMS}_m$ tracks instantaneous acoustic energy in each frame; $\mathrm{SC}_m$ indicates where that energy is concentrated on the frequency axis, with larger values meaning a brighter, high–frequency tilt; $\mathrm{SFM}_m$ differentiates tone–like frames (low, peak–dominated spectra) from noise–like or percussive frames (high, flat spectra); and $\mathrm{Flux}_m$ shows rapid spectral changes, peaking at onsets and accents while remaining low during sustained passages. Taken together, these measures disentangle loudness, brightness, harmonicity, and rhythmic activity, respectively.

\subsection{Curve construction}

For each track and metric we obtain a frame–indexed sequence \(\{y_m\}\). We map frames to normalized time \(t_m\in[0,1]\) and smooth \(\{y_m\}\) with cubic smoothing splines \citep{wang2011smoothing}, which penalize curvature and attenuate frame–level noise while preserving broad temporal structure. The resulting smooth function is evaluated on a common grid $\{u_\ell\}$ with $M$ points via cubic interpolation for every song–metric pair. To ensure comparability across tracks, each smoothed curve is then standardized to zero mean and unit variance, removing global gain/offset differences and emphasizing shape.

\subsection{Song–song similarity}

Let \(g,g_j\in\mathbb{R}^{M}\) denote standardized curves evaluated on a common grid \(\{u_\ell\}\) with $M$ points for songs $i$ and $j$ under a fixed metric. Write \(g_{i,\ell}=g_i(u_\ell)\) and \(g_{j,\ell}=g_j(u_\ell)\). Our primary dissimilarity between songs $i$ and $j$ is the Canberra distance given by
\[
\textsf{d}_{i,j}
=\sum_{\ell=1}^{M}\frac{\lvert g_{i,\ell}-g_{j,\ell}\rvert}{\lvert g_{i,\ell}\rvert+\lvert g_{j,\ell}\rvert},
\]
with the convention that the summand $\ell$ is zero when \(g_{i\ell}=g_{j\ell}=0\). This distance is robust to scale and gives greater weight to relative differences near zero. Alternative distances include the correlation distance, the cosine distance, and the Euclidean distance.

\subsection{From distances to graphs}

Within each band and for each metric separately, we construct an undirected, unweighted similarity network whose nodes are songs. Let \(\mathbf{D}=[\textsf{d}_{i,j}]\) be the pairwise distance matrix and define affinities \(\textsf{w}_{i,j} = \textsf{d}_{i,j}^{-1}\). For each node \(i\), let $\mathrm{NN}_k(i)$ be the index set of the \(k\) largest affinities from \(i\) (ties broken deterministically). This yields a directed \(k\)-nearest-neighbor graph, which we symmetrize with an OR rule to obtain the undirected adjacency \(\mathbf{A}=[a_{i,j}]\), with
\[
a_{i,j} = I\!\left\{\, j\in \mathrm{NN}_k(i)\ \text{ or }\ i\in \mathrm{NN}_k(j)\,\right\},\qquad a_{i,i}=0.
\]
This construction produces one layer per metric, hence \(K=4\) layers in total: Loudness (\textsf{RMS}), Brightness (\textsf{SC}), Tonality (\textsf{SFM}), and Rhythm (\textsf{Flux}). Thus, each band therefore yields a four-layer multilayer network over the same node set of songs, enabling cross-metric comparisons at the layer level. To construct the Big 4 data, we evaluate each curve on a grid of $M=1000$ points and form $k$-nearest–neighbor graphs with $k=3$.

\subsection{Dyadic covariates}

We also derive song-level textual covariates from lyrics when available. Lyrics are read line by line, tokenized to lowercase words, and filtered using standard stopword lists (accented characters are normalized to ASCII). From the \texttt{Bing} polarity and \texttt{NRC} emotion lexica we compute lexicon coverage and emotion shares (anger, anticipation, disgust, fear, joy, sadness, surprise, trust). Using the \texttt{NRC VAD} lexicon we extract mean valence, arousal, and dominance. These features are then mapped to dyadic covariate matrices: absolute differences in year/BPM/duration, a same–album indicator, cosine similarity of emotion profiles, and Euclidean distance in standardized VAD space. Nonbinary covariates are standardized for comparability (the binary indicator is left on its original scale). These matrices are used as exogenous regressors in the models.

\subsection{Exploratory data analysis}

\begin{figure}[!htb]
	\centering
	\setlength{\tabcolsep}{0pt}
	\begin{tabular}{ccccc}
		& \textsf{METALLICA} & \textsf{SLAYER} & \textsf{MEGADETH} & \textsf{ANTHRAX} \\
		\begin{sideways} \hspace{0.9cm} \textbf{Loudness} \end{sideways}     &
		\includegraphics[scale = 0.28]{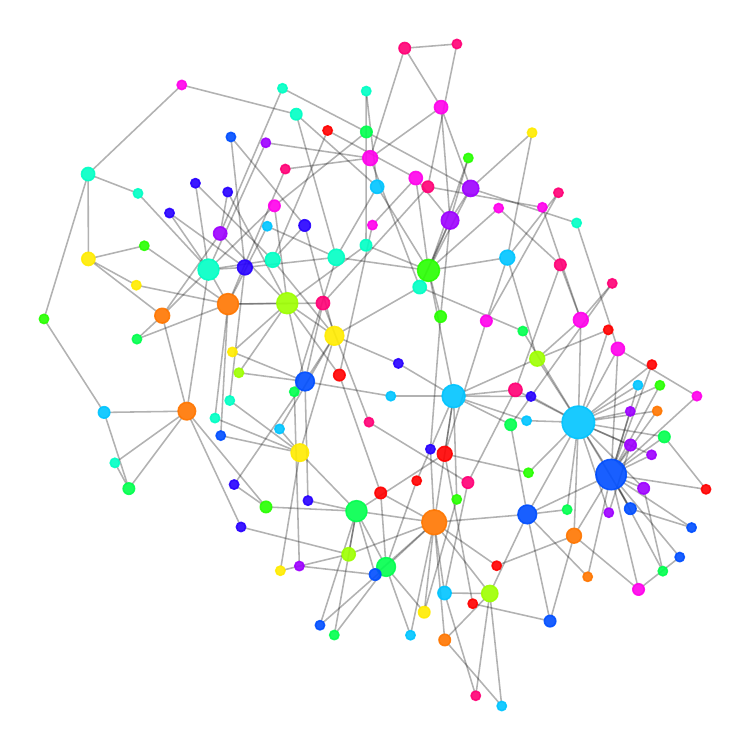} &
		\includegraphics[scale = 0.28]{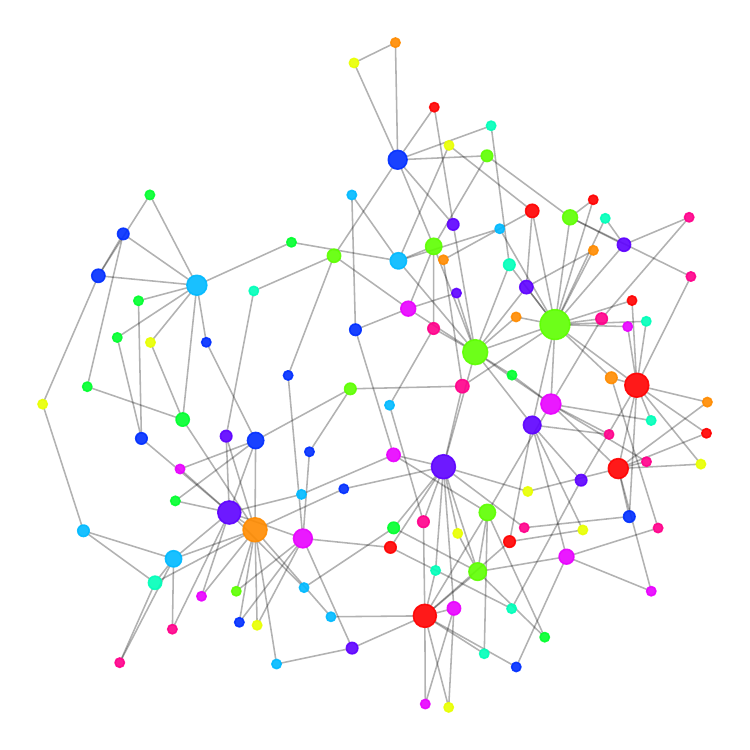}    &
		\includegraphics[scale = 0.28]{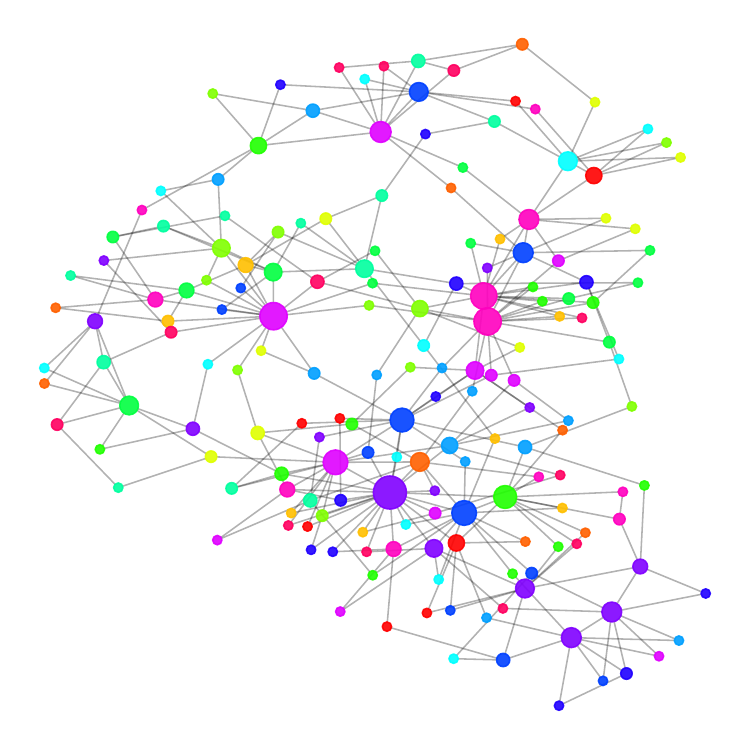}  &
		\includegraphics[scale = 0.28]{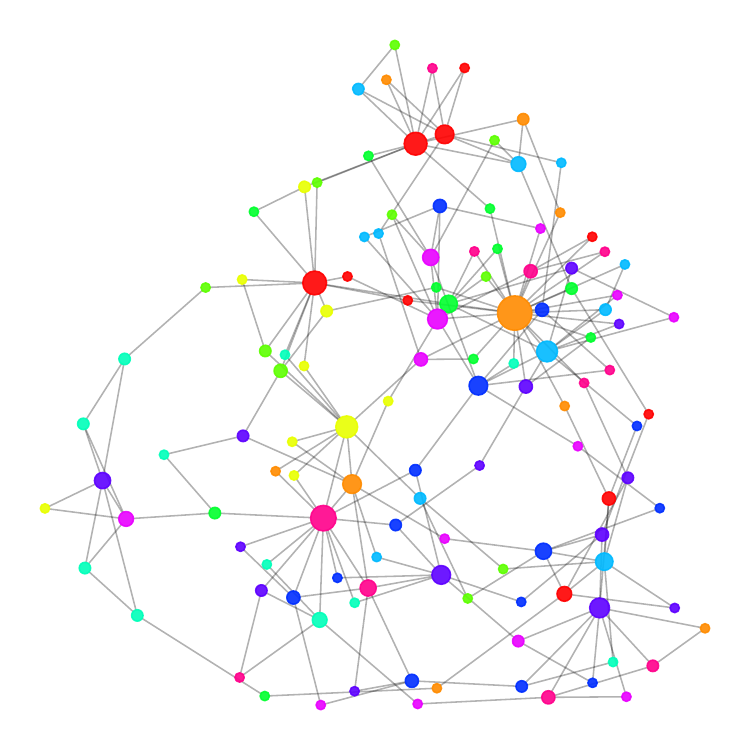}  \\
		\begin{sideways} \hspace{0.9cm} \textbf{Brightness} \end{sideways}     &
		\includegraphics[scale = 0.28]{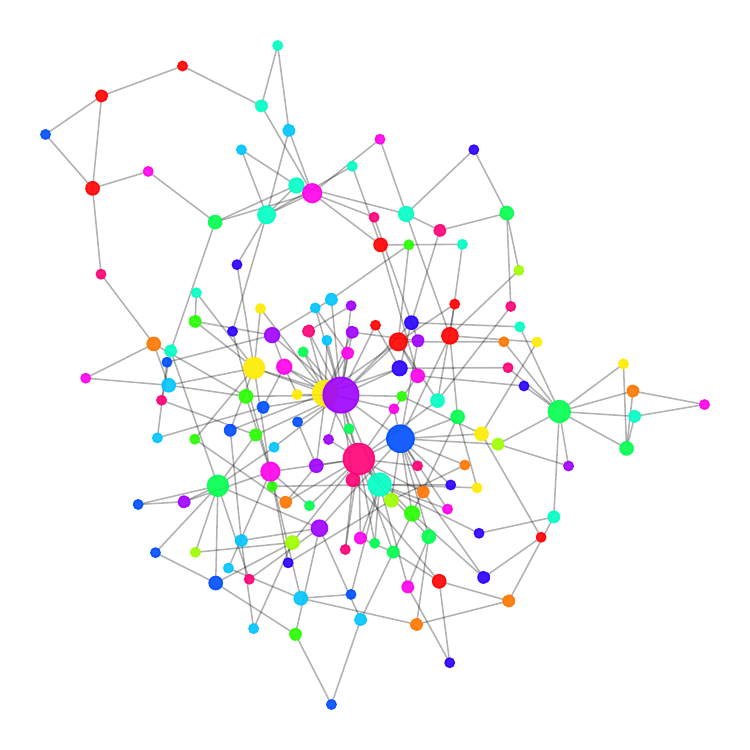} &
		\includegraphics[scale = 0.28]{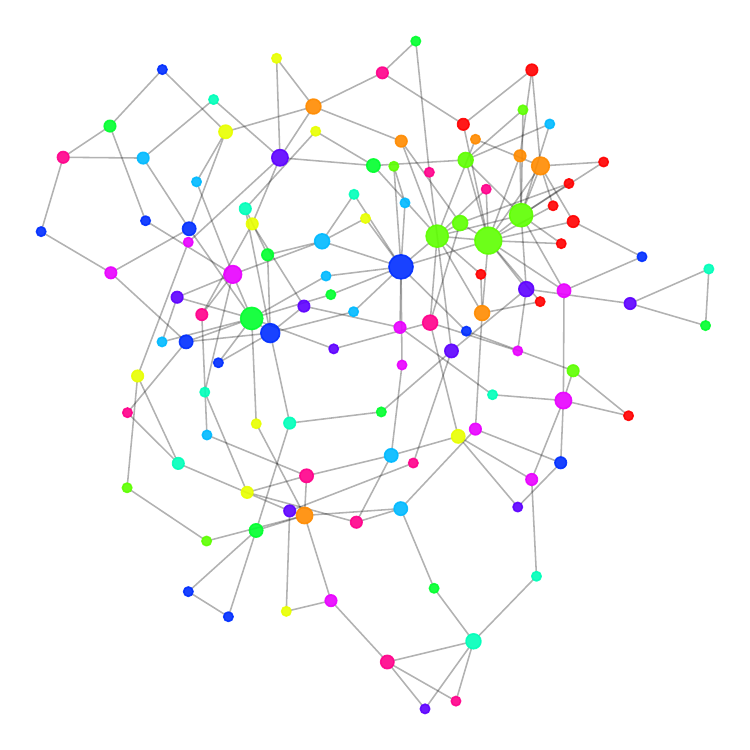}    &
		\includegraphics[scale = 0.28]{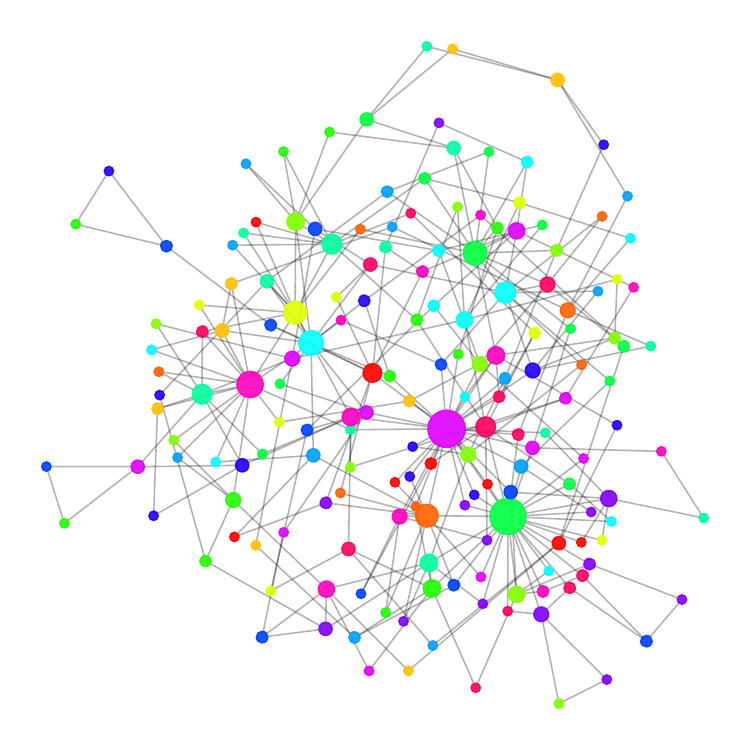}  &
		\includegraphics[scale = 0.29]{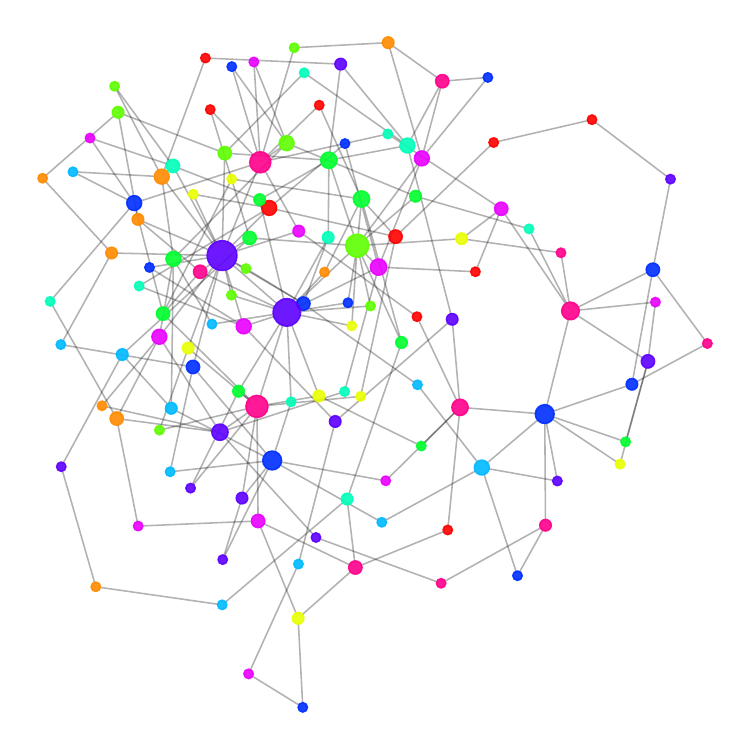}  \\
        \begin{sideways} \hspace{0.9cm} \textbf{Tonality} \end{sideways}     &
		\includegraphics[scale = 0.28]{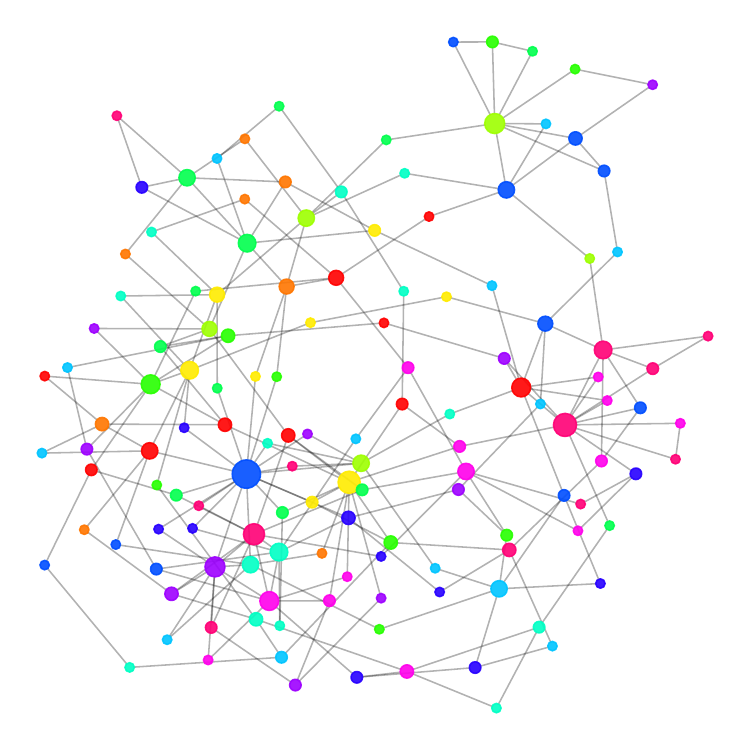} &
		\includegraphics[scale = 0.28]{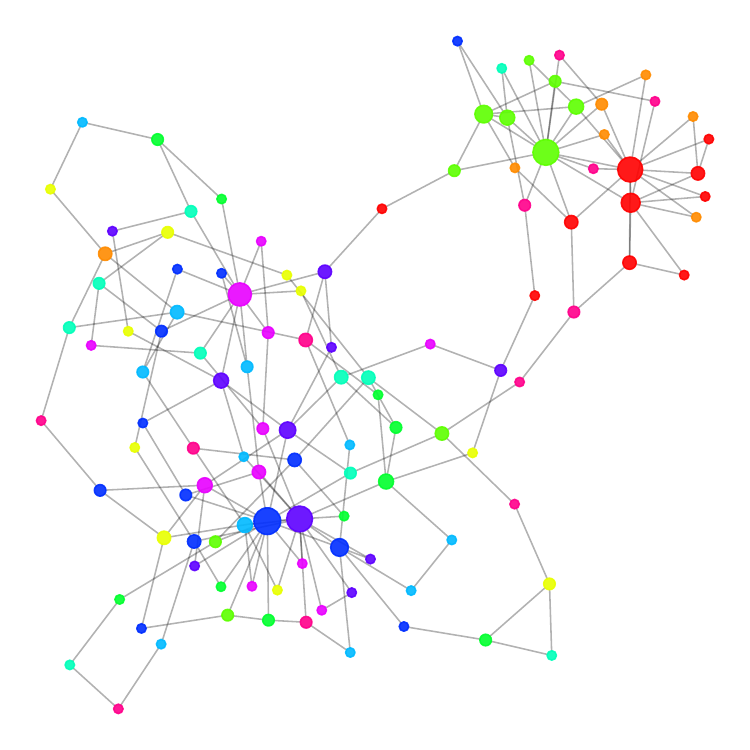}    &
		\includegraphics[scale = 0.28]{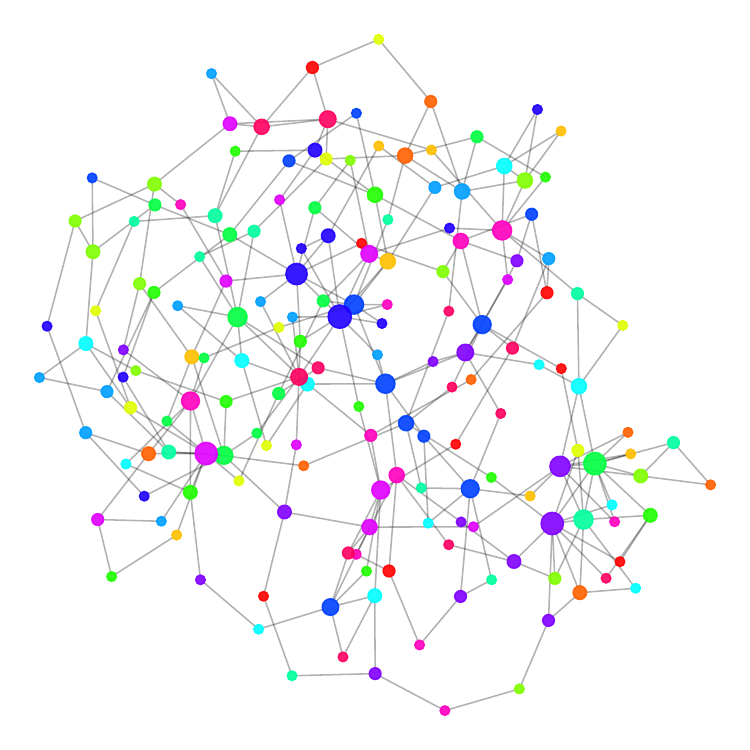}  &
		\includegraphics[scale = 0.29]{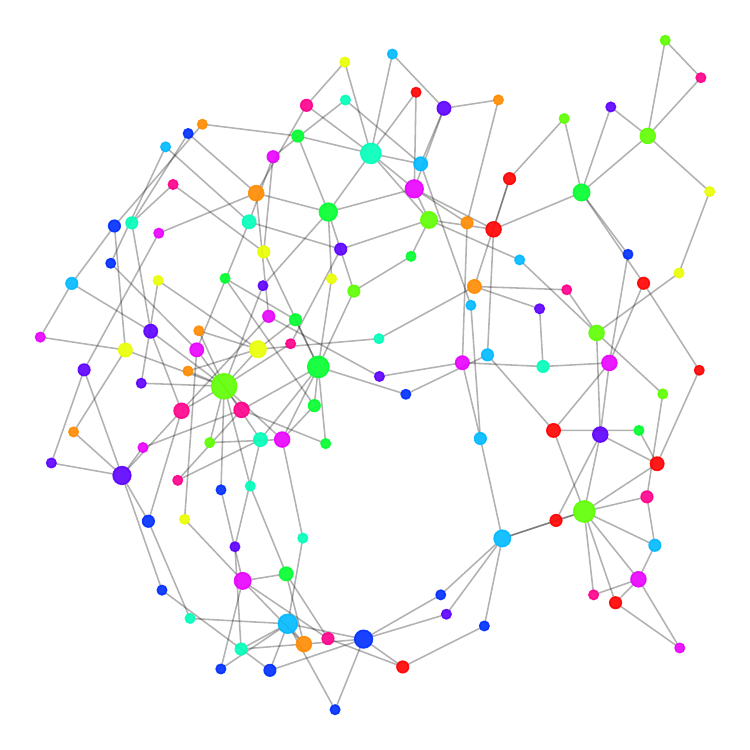}  \\
        \begin{sideways} \hspace{0.9cm} \textbf{Rhythm} \end{sideways}     &
		\includegraphics[scale = 0.28]{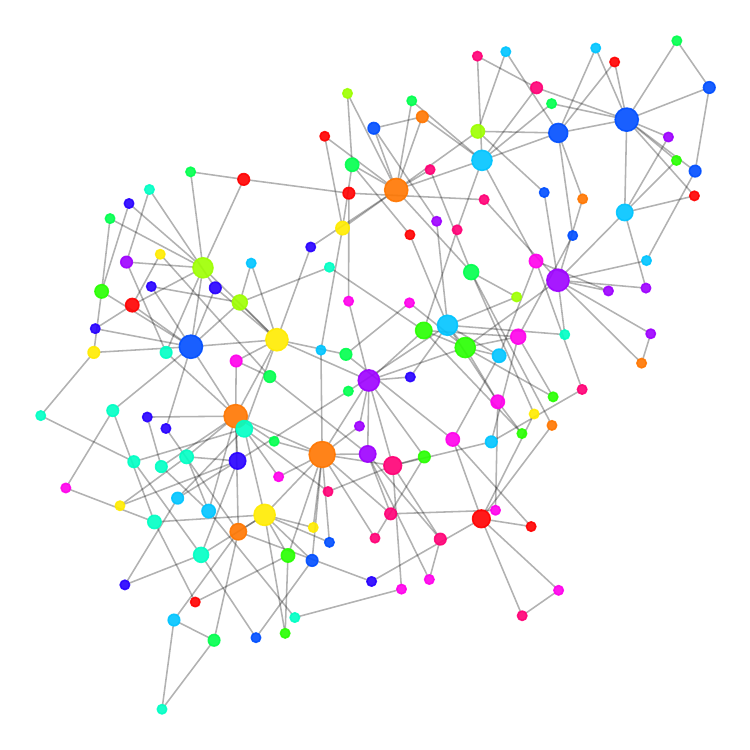} &
		\includegraphics[scale = 0.28]{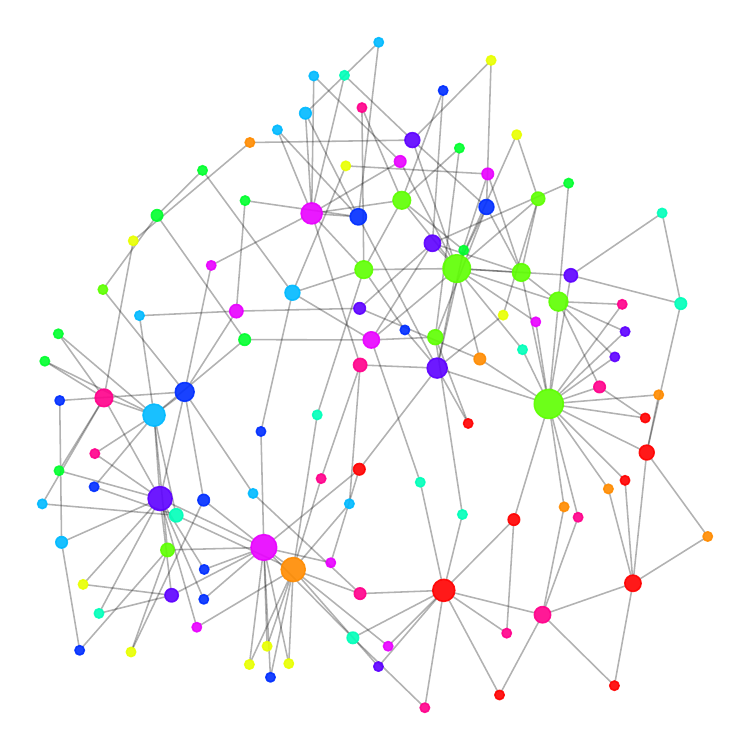}    &
		\includegraphics[scale = 0.28]{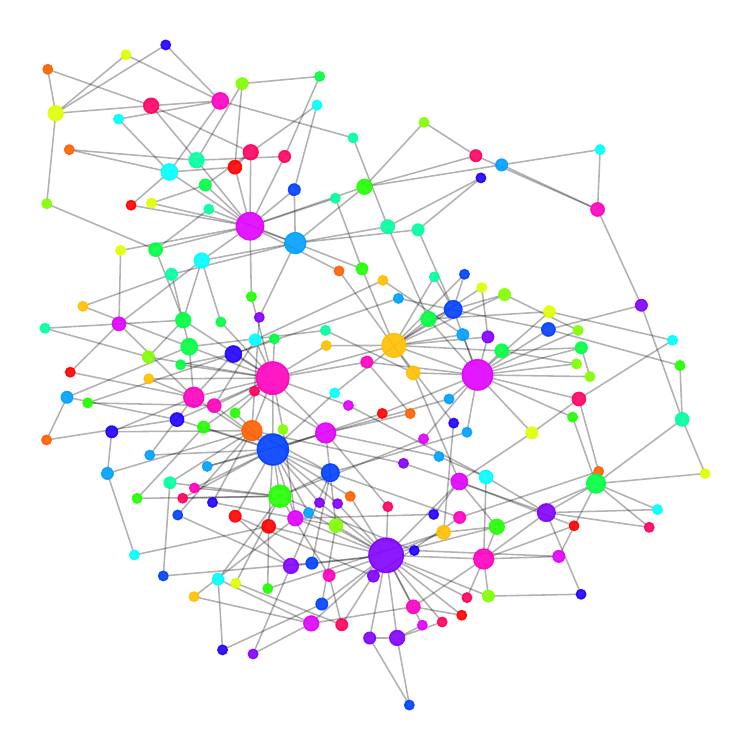}  &
		\includegraphics[scale = 0.29]{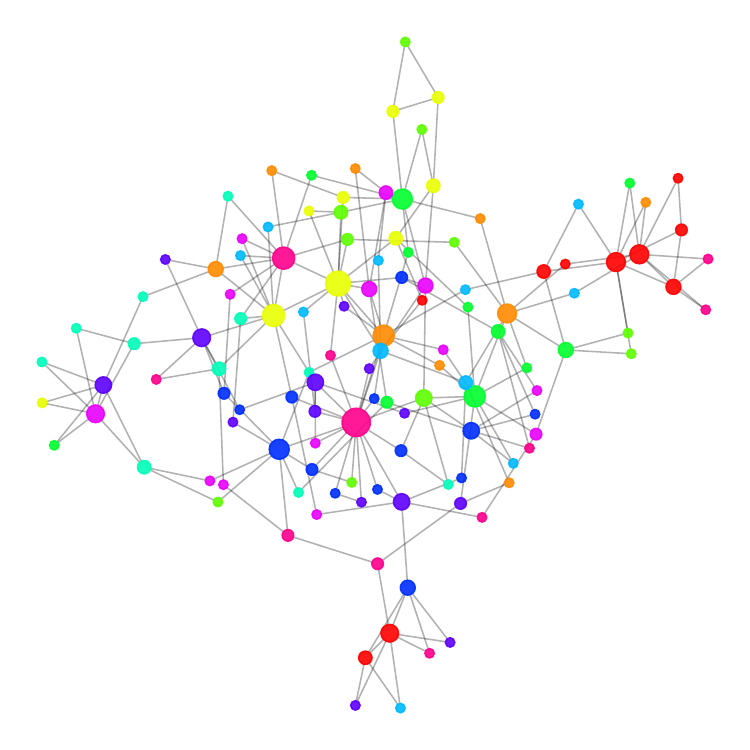}  \\
	\end{tabular}
	\caption{Song–similarity multilayer networks for the Big 4 data. Columns correspond to bands (Metallica, Slayer, Megadeth, Anthrax) and and rows to audio layers (Loudness, Brightness, Tonality, Rhythm). Nodes represent songs, with node color indicating album membership. Edges connect $k$-nearest neighbors ($k=3$) under the feature-specific curve distance (affinities from inverse Canberra distance), yielding undirected, unweighted graphs in each layer.}
	\label{fig_big_4_data}
\end{figure}

The Big 4 dataset constructed as described above is visualized in Figure~\ref{fig_big_4_data}, which displays the four layers, Loudness (RMS), Brightness (SC), Tonality (SFM), and Rhythm (Flux), for each band (Metallica, Slayer, Megadeth, Anthrax). Nodes correspond to songs, and nodes with the same color belong to the same album. Furthermore, Table~\ref{tab_big_4_data_eda} reports descriptive structural metrics, including 
edge density, global transitivity, degree assortativity, mean degree, degree 
variability, mean geodesic distance, diameter, and triangle counts. 
The definitions of these classical network measures follow standard treatments 
in network science \citep{newman2010networks, watts1998collective, albert2002statistical}.

At the network level, shorter mean geodesic distances and smaller diameters indicate a repertoire in which most songs find close neighbors under a given feature (greater self–similarity), whereas longer path lengths reflect broader dispersion (greater originality). Local redundancy is captured by global transitivity, as higher transitivity and tighter album–colored pockets in the layouts signal clusters of mutually similar songs (mini “families” or formulas), while lower transitivity suggests more idiosyncratic constructions. Degree heterogeneity and negative degree assortativity further reveal \emph{templates}: a few hub songs act as prototypes that connect otherwise distinct material. Stronger disassortativity implies that these prototypes bridge many lower–degree, more specialized tracks rather than forming hub–hub cliques.

Across bands, all layers are sparse yet well connected (densities $\approx 0.02$–$0.03$, mean geodesic distances $3.8$–$5.1$, diameters $8$–$11$). Tonality consistently exhibits the longest mean path lengths and among the highest clustering levels (e.g., Slayer \((4.896,\,0.147)\), Megadeth \((5.093,\,0.146)\), and Anthrax \((4.780,\,0.124)\); ordered pairs denote mean geodesic distance and global transitivity). This pattern indicates localized similarity (tight within-album families) but weaker cross-album connectivity, consistent with greater originality in spectral texture across a career. By contrast, Brightness often yields the shortest paths (e.g., Megadeth mean geodesic distance $3.799$, diameter $8$) and relatively high degree variability (e.g., Metallica SD degree $3.616$ vs. mean degree $3.691$), suggesting a stable production palette together with a few bright–sounding prototypes that many songs resemble. Rhythm layers exhibit moderate clustering and short–to–moderate path lengths—most notably for Slayer and Metallica, consistent with recurrent rhythmic blueprints that promote self–similarity while avoiding graph collapse. Loudness lies between these extremes, with mid–range path lengths and clustering, capturing energy–level signatures that are shared yet less homogenizing than spectral balance.

\begin{table}[!htb]
\centering
\begin{tabular}{cccccccc}
\hline
Layer & Dens. & Trans. & Assor. & M. Deg. & SD Deg. & M. Geo. & Diam. \\
\hline
\multicolumn{8}{c}{\textsf{METALLICA}} \\
\hline
1 & 0.027 & 0.121 & -0.254 & 3.706 & 3.280 & 4.070 &  9 \\
2 & 0.027 & 0.110 & -0.137 & 3.691 & 3.616 & 3.807 & 10 \\
3 & 0.025 & 0.113 & -0.249 & 3.441 & 2.394 & 4.428 &  9 \\
4 & 0.027 & 0.150 & -0.224 & 3.603 & 2.722 & 4.246 & 10 \\
\hline
\multicolumn{8}{c}{\textsf{SLAYER}} \\
\hline
1 & 0.033 & 0.129 & -0.297 & 3.702 & 3.170 & 3.907 &  8 \\
2 & 0.030 & 0.130 & -0.070 & 3.386 & 2.385 & 4.325 &  9 \\
3 & 0.030 & 0.147 & -0.158 & 3.368 & 2.668 & 4.896 & 11 \\
4 & 0.033 & 0.117 & -0.341 & 3.719 & 3.263 & 3.797 &  8 \\
\hline
\multicolumn{8}{c}{\textsf{MEGADETH}} \\
\hline
1 & 0.022 & 0.134 & -0.247 & 3.723 & 3.373 & 4.642 & 11 \\
2 & 0.022 & 0.102 & -0.166 & 3.711 & 3.797 & 3.799 &  8 \\
3 & 0.020 & 0.146 & -0.179 & 3.399 & 2.059 & 5.093 & 11 \\
4 & 0.022 & 0.099 & -0.250 & 3.769 & 3.756 & 4.073 &  9 \\
\hline
\multicolumn{8}{c}{\textsf{ANTHRAX}} \\
\hline
1 & 0.029 & 0.132 & -0.286 & 3.593 & 3.271 & 4.078 &  8 \\
2 & 0.028 & 0.096 & -0.116 & 3.463 & 2.687 & 3.966 &  8 \\
3 & 0.027 & 0.124 & -0.212 & 3.350 & 2.000 & 4.780 & 10 \\
4 & 0.029 & 0.115 & -0.267 & 3.528 & 2.693 & 4.312 & 10 \\
\hline
\end{tabular}
\caption{Network summary statistics by layer for the Big 4 (Metallica, Slayer, Megadeth, Anthrax). Layers correspond to audio–feature networks: 1 = Loudness (RMS), 2 = Brightness (SC), 3 = Tonality (SFM), 4 = Rhythm (Flux). Columns report density (Dens.), global transitivity (Trans.), degree assortativity (Assor.), mean degree (M. Deg.), standard deviation of degree (SD Deg.), mean geodesic distance (M. Geo.), and diameter (Diam.). All graphs are undirected, unweighted $k$-nearest–neighbor networks with $k=3$.}
\label{tab_big_4_data_eda}
\end{table}

Negative degree assortativity is pervasive (approximately \( -0.070 \) to \( -0.341 \)), strongest in Slayer’s Rhythm layer (\( -0.341 \)) and salient in Loudness for Metallica (\( -0.254 \)) and Anthrax (\( -0.286 \)). Musically, disassortativity indicates that a few archetypal tracks act as hubs linking many lower–degree, more specialized songs, thereby stitching together disparate regions of the repertoire and mitigating fragmentation (less self–copying within cliques and more \emph{templates bridging variety}). Album–colored clusters are most pronounced when global transitivity is higher (e.g., Tonality for Slayer and Megadeth), consistent with era–specific textural families, whereas Brightness and Rhythm exhibit more cross–color edges, pointing to stable timbral brightness and rhythmic drive that cut across albums. Taken together, the layers partition “musicality” into complementary facets. Energy (RMS) and spectral brightness (SC) foster band–wide coherence; spectral texture (SFM) sustains originality through localized families; and onset strength (Flux) provides rhythmic common ground while still admitting band–specific connectors. These empirical regularities motivate multilayer models with layer–specific mechanisms and partial pooling across layers to capture both recurrent templates and innovation within each discography.

\section{Models}\label{sec_models}

In this section, we develop a sequence of fully Bayesian models for multilayer network data that build on one another. We begin with random sociability effects (e.g., \citealt{krivitsky2009representing}), then incorporate dyadic covariates, and finally introduce several multilayer latent structures (see \citealt{sosa2021review} for a review). We consider multilayer network data with $K$ layers, $\mathcal{Y}=\{\mathbf{Y}_1,\ldots,\mathbf{Y}_K\}$, where each layer $\mathbf{Y}_k=[y_{i,j,k}]$ is an $n\times n$ symmetric binary adjacency matrix on the common node set $V=\{1,\ldots,n\}$. Entries satisfy $y_{i,i,k}=0$ (no self-loops), $y_{i,j,k}=y_{j,i,k}$ (undirected), and $y_{i,j,k}\in\{0,1\}$ (binary ties). Here, $y_{i,j,k}=1$ indicates a link between nodes $i$ and $j$ in layer $k$, and $y_{i,j,k}=0$ otherwise.

\subsection{Sociability Multilayer Network (SMN) model}

For $1 \le i<j \le n$ and $k = 1,\ldots,K$, the likelihood is $y_{i,j,k}\overset{\text{ind}}{\sim}\textsf{Ber}(\theta_{i,j,k})$, with $\theta_{i,j,k}=\Phi(\eta_{i,j,k})$, where $\eta_{i,j,k}$ is a linear predictor given by 
$$
\eta_{i,j,k}=\zeta+\mu_k+\delta_{i,k}+\delta_{j,k}.
$$ 
In this parametrization, $\zeta\in\mathbb{R}$ is a global connectivity effect shared by all layers, $\mu_k\in\mathbb{R}$ is a layer–specific intercept capturing the baseline tie propensity in layer $k$, and $\delta_{i,k}\in\mathbb{R}$ is a node–specific sociability effect in layer $k$, which makes $\delta_{i,k}+\delta_{j,k}$ to account for within–layer dyadic heterogeneity.

To complete the Bayesian specification, we assign hierarchical Gaussian priors to the additive effects,
$$
\zeta \mid \omega^2 \sim \textsf{N}(0,\omega^2),\qquad
\mu_k \mid \sigma^2 \overset{\text{iid}}{\sim}\textsf{N}(0,\sigma^2),\qquad
\delta_{i,k} \mid \vartheta_i,\tau^2 \overset{\text{ind}}{\sim}\textsf{N}(\vartheta_i,\tau^2),
$$
where $\vartheta_i$ is a node–specific baseline sociability shared across layers (the “center” about which the layer–level effects $\delta_{i,k}$ for node $i$ fluctuate), inducing partial pooling of sociability for node $i$ over $k=1,\ldots,K$. 
In this way, the model captures across-layer heterogeneity through $\mu_k$ and within-layer degree variability through $\delta_{i,k}$, while partially pooling $\delta_{i,k}$ toward $\vartheta_i$ across layers. 
We further set
$\vartheta_i \overset{\text{iid}}{\sim}\textsf{N}(0,\kappa^2)$, and place inverse–gamma priors on the variance components,
$$
\omega^2 \sim \textsf{IG}(a_\omega,b_\omega),\qquad
\sigma^2 \sim \textsf{IG}(a_\sigma,b_\sigma),\qquad
\tau^2   \sim \textsf{IG}(a_\tau,b_\tau),\qquad
\kappa^2 \sim \textsf{IG}(a_\kappa,b_\kappa),
$$
with fixed hyperparameters $a_\omega,b_\omega,a_\sigma,b_\sigma,a_\tau,b_\tau,a_\kappa,b_\kappa$. Unless otherwise noted, all priors are mutually independent.

\subsection{Sociability Multilayer Network with Covariates (\textsf{SMN-C}) model}

Building on the \textsf{SMN} specification, we extend the linear predictor to incorporate exogenous dyadic information. Specifically, we retain the probit likelihood $y_{i,j,k}\overset{\text{ind}}{\sim}\textsf{Ber}\!\big(\Phi(\eta_{i,j,k})\big)$ and set
$$
\eta_{i,j,k}= \mu_k + \zeta + \delta_{i,k} + \delta_{j,k} + \boldsymbol{x}_{i,j}^{\top}\boldsymbol{\beta}_k,
$$
where $\boldsymbol{x}_{i,j} = (x_{i,j,1},\ldots,x_{x_{i,j,p}})\in\mathbb{R}^{p}$ is a dyadic covariate vector symmetric in $(i,j)$ (i.e., $\boldsymbol{x}_{i,j}=\boldsymbol{x}_{j,i}$), and $\boldsymbol{\beta}_k = (\beta_{i,j,1},\ldots,\beta_{x_{i,j,p}})\in\mathbb{R}^{p}$ is a layer-specific coefficient vector quantifying how covariates are associated with tie propensity in layer $k$. The covariate linear term $\boldsymbol{x}_{i,j}^{\top}\boldsymbol{\beta}_k = \sum_{\ell=1}^p \beta_{k,\ell}\,x_{i,j,\ell}$ accounts for systematic dyadic variation not captured by latent sociability effects. Allowing $\boldsymbol{\beta}_k$ to vary across layers isolates layer–specific mechanisms.

We set $\boldsymbol{\beta}_k \mid \varsigma^2 \overset{\text{iid}}{\sim} \textsf{N}_p(\boldsymbol{0},\,\varsigma^{2} \mathbf{I})$, with $\varsigma^{2} \sim \textsf{IG}(a_\varsigma,b_\varsigma)$, inducing shrinkage of the layer-specific coefficients $\boldsymbol{\beta}_k$ toward zero. 
The global shrinkage variance $\varsigma^{2}$ controls the prior dispersion of the coefficients, stabilizing estimation and mitigating overfitting when $p$ is moderate or when covariates are correlated. We recommend standardizing each quantitative column of $\boldsymbol{x}_{i,j}$ to zero mean and unit variance so that coefficients are comparable and the prior on $\boldsymbol{\beta}_k$ has a consistent interpretation across covariates.
The remainder of the model follows the \textsf{SMN} hierarchical structure.

\subsection{Sociability Multilayer Network with Covariates and Bilinear Geometry (\textsf{SMN-C-BG}) model}

Building on the \textsf{SMN-C} specification, we enrich the linear predictor with a shared bilinear latent-space term to capture higher-order affinity beyond degree effects and covariates. Specifically, we retain the probit likelihood $y_{i,j,k}\overset{\text{ind}}{\sim}\textsf{Ber}\!\big(\Phi(\eta_{i,j,k})\big)$ and set
$$
\eta_{i,j,k}
= \zeta + \mu_k + \delta_{i,k} + \delta_{j,k}
+ \boldsymbol{x}_{i,j}^{\top}\boldsymbol{\beta}_k
+ \lambda_k\,\boldsymbol{u}_i^{\top}\boldsymbol{u}_j,
$$
where $\boldsymbol{u}_i=(u_{i,1},\ldots,u_{i,d})\in\mathbb{R}^d$ are node-specific latent positions shared across layers, with fixed latent dimension $d$, and $\lambda_k\in\mathbb{R}$ is a layer-specific geometry scale. 
The bilinear inner product $\boldsymbol{u}_i^{\top}\boldsymbol{u}_j=\sum_{\ell=1}^{d} u_{i,\ell}\,u_{j,\ell}$ captures similarity-based attraction or repulsion that is shared across layers and scaled by $\lambda_k$. See \cite{hoff2005bilinear} for more details about bilinear models.

We set $\boldsymbol{u}_i \overset{\text{iid}}{\sim} \textsf{N}_d(\boldsymbol{0},\mathbf{I})$ to fix the overall scale of the latent space and prevent confounding between the latent positions $\boldsymbol{u}_i$ and the layer scales $\lambda_k$. Indeed, for any $c>0$,
\[ 
\lambda_k\,\boldsymbol{u}_i^{\top}\boldsymbol{u}_j
=\frac{\lambda_k}{c^{2}}\,(c \ \boldsymbol{u}_i)^{\top}(c \ \boldsymbol{u}_j),
\]
so fixing the latent-space scale assigns layer strength to $\lambda_k$ while enforcing a shared geometry to the $\boldsymbol{u}_i$.
We also set $\lambda_k \overset{\text{iid}}{\sim} \textsf{N}(0,\upsilon^2)$, with $\upsilon^2 \sim \textsf{IG}(a_{\upsilon},b_{\upsilon})$, which centers the bilinear term $\lambda_k\,\boldsymbol{u}_i^{\top}\boldsymbol{u}_j$ at zero (allowing both assortative and disassortative patterns) and induces shrinkage of the layer–specific geometry scales $\lambda_k$ toward zero. 
The remainder of the model follows the \textsf{SMN-C} hierarchical specification.

\subsection{Sociability Multilayer Network with Covariates and Latent Distance (\textsf{SMN-C-LD}) model}

Building on the \textsf{SMN-C} specification, we replace the bilinear term with a shared latent distance–decay effect to model geometric attenuation of ties. Specifically, we retain the probit likelihood $y_{i,j,k}\overset{\text{ind}}{\sim}\textsf{Ber}\!\big(\Phi(\eta_{i,j,k})\big)$ and set
\[
\eta_{i,j,k}
=\zeta+\mu_k+\delta_{i,k}+\delta_{j,k}
+\boldsymbol{x}_{i,j}^{\top}\boldsymbol{\beta}_k
- e^{\lambda_k}\,\lVert \boldsymbol{u}_i-\boldsymbol{u}_j\rVert,
\]
where $\boldsymbol{u}_i=(u_{i,1},\ldots,u_{i,d})\in\mathbb{R}^d$ are node-specific latent positions shared across layers, with fixed latent dimension $d$, $\lambda_k\in\mathbb{R}$ is a layer-specific log–distance scale, and $\lVert\cdot\rVert$ is the Euclidean norm on $\mathbb{R}^d$. 
The nonnegative distance term $e^{\lambda_k}\,\lVert \boldsymbol{u}_i-\boldsymbol{u}_j\rVert$ enforces a monotone decay of tie propensity with latent distance: more positive $\lambda_k$ implies stronger decay (more local ties), whereas more negative $\lambda_k$ implies weaker decay (more long-range ties). See \cite{hoff2002latent} for more details about distance models.

Similar to the SMN-CBG model, we set $\boldsymbol{u}_i \overset{\text{iid}}{\sim} \textsf{N}_d(\boldsymbol{0},\mathbf{I}_d)$ and, in addition, center the latent configuration \emph{post hoc} so that $\sum_{i=1}^n \boldsymbol{u}_i=\boldsymbol{0}$. This addresses overall scale and location indeterminacies in the latent space and avoids confounding between the latent positions $\boldsymbol{u}_i$ and the log–distance scales $\lambda_k$, since, as in the SMN-CBG model, for any $c>0$,
\[
e^{\lambda_k}\,\lVert \boldsymbol{u}_i-\boldsymbol{u}_j\rVert
= e^{\lambda_k+\log c}\,\big\lVert c^{-1}\boldsymbol{u}_i - c^{-1}\boldsymbol{u}_j\big\rVert.
\]
We further set $\lambda_k \overset{\text{iid}}{\sim} \textsf{N}(0,\upsilon^2)$, with $\upsilon^2 \sim \textsf{IG}(a_{\upsilon},b_{\upsilon})$, which centers the log–distance scale $\lambda_k$ at zero (neutral prior decay) and induces shrinkage of the $\lambda_k$ toward zero. The remainder of the model follows the \textsf{SMN-C} hierarchical specification.

\subsection{Sociability Multilayer Network with Covariates and Stochastic Blocks (\textsf{SMN-C-SB}) model}

Building on the \textsf{SMN-C} specification, we introduce a layer–specific stochastic block component to capture community-level structure beyond degree effects and covariates. Specifically, we retain the probit likelihood $y_{i,j,k}\overset{\text{ind}}{\sim}\textsf{Ber}\big(\Phi(\eta_{i,j,k})\big)$ and set
\[
\eta_{i,j,k}
= \zeta + \mu_k + \delta_{i,k} + \delta_{j,k}
+ \boldsymbol{x}_{i,j}^{\top}\boldsymbol{\beta}_k
+ \gamma_{\phi(\xi_{i,k},\xi_{j,k}),k},
\]
where $\xi_{i,k}\in\{1,\ldots,C\}$ is the block label of node $i$ in layer $k$, with fixed number of blocks $C$, and $\phi(x,y)=(\min(x,y),\max(x,y))$ because $\mathbf{\Gamma}_k = [\gamma_{a,b,k}]$ is a symmetric $C\times C$ matrix of within/between–block affinities for layer $k$.
Unlike the bilinear and latent–distance variants, which encode geometry through shared latent positions, the \textsf{SMN-C-SB} model represents modular structure using discrete communities that may change across layers.

We place a layer–specific prior on the block structure following Dirichlet–Multinomial formulation. Given the mixing weights $\boldsymbol{\omega}_k=(\omega_{k,1},\ldots,\omega_{k,C})\in\Delta^{C}$, where $\Delta^{C}$ is the $C$-probability simplex, cluster assignments are are assigned $\xi_{i,k}\mid\boldsymbol{\omega}_k \overset{\text{ind}}{\sim} \textsf{Cat}(\boldsymbol{\omega}_k)$, allowing community proportions to vary by layer. 
In addition, we set $\boldsymbol{\omega}_k \mid \alpha \overset{\text{ind}}{\sim} \textsf{Dir}(\alpha/C,\ldots,\alpha/C)$, with $\alpha\sim\textsf{G}(a_\alpha,b_\alpha)$, yielding an exchangeable prior whose concentration controls dispersion and the expected number of active blocks. Finally, block affinities are modeled as $\gamma_{a,b,k}\mid\rho^2 \overset{\text{ind}}{\sim}\textsf{N}(0,\rho^2)$, for $1\le a\le b\le C$, with $\rho^2\sim\textsf{IG}(a_\rho,b_\rho)$, imposing no prior sign preference and shrinking minor effects toward zero. The remainder of the model follows the \textsf{SMN-C} hierarchical specification.

\subsection{Model summary}

For all models, let $n$ denote the number of nodes, $K$ the number of layers, $p$ the number of dyadic covariates, $d$ the latent dimension, and $C$ the number of communities. For each model, we summarize: the additional latent structure introduced, the number of model parameters, and the number of hyperparameters. The baseline model is \textsf{SMN}. This model adds no latent structure beyond additive effects and uses $8$ hyperparameters. The model parameters are $\mathbf{\Theta} = \big(\zeta, \{\mu_k\}, \{\delta_{i,k}\}, \{\vartheta_i\}, \omega^2, \sigma^2, \tau^2, \kappa^2\big)$, which corresponds to a total of $K(n+1) + n + 5$ unknowns.
Next, the \textsf{SMN-C} model augments the \textsf{SMN} model with layer-specific regression vectors $\boldsymbol{\beta}_k\in\mathbb{R}^p$, adding the covariate coefficients $\{\boldsymbol{\beta}_k\}$ and the variance component $\varsigma^2$ to $\mathbf{\Theta}$ and bringing the total number of hyperparameters to $10$. This extention raises the total number of model parameters to $K(n+p+1)+n+6$.

The \textsf{SMN-C-BG} model extends the \textsf{SMN-C} model by adding a shared latent structure $\mathbf{U}=[\boldsymbol{u}_1^{\top},\ldots,\boldsymbol{u}_n^{\top}]^{\top}\in\mathbb{R}^{n\times d}$ and layer-specific geometry scales $\lambda_k \in \mathbb{R}$ (with variance component $\upsilon^2$) to $\mathbf{\Theta}$. This extension brings the total number of hyperparameters to $12$ and increases the total number of model parameters to $K(n+p+2)+n(d+1)+7$. Similarly, the \textsf{SMN-C-LD} model replaces the bilinear term with a shared latent distance effect while retaining the same parameter dimensionality as the \textsf{SMN-C-BG} model.

Finally, the \textsf{SMN-C-SB} model generalizes \textsf{SMN-C} by incorporating a layer-specific community (block) structure. Specifically, it augments $\mathbf{\Theta}$ with cluster indicators $\xi_{i,k}\in\{1,\ldots,C\}$, layer-specific symmetric block-affinity matrices $\mathbf{\Gamma}_k\in\mathbb{R}^{C\times C}$, layer-specific mixing weights $\boldsymbol{\omega}_k \in \Delta^{C}$, and a concentration parameter $\alpha>0$. This extension yields a total of $14$ hyperparameters and makes the total number of model parameters $K\big(2n+p+C+\binom{C+1}{2}+1\big)+n+8$.
\subsection{Identifiability}

Across the latent–geometry variants, parameters are identifiable only up to 
natural group actions that leave the likelihood invariant. In the 
\textsf{SMN\text{-}C\text{-}BG} model, the latent structure 
$\mathbf{U}=[\boldsymbol{u}_1^\top,\ldots,\boldsymbol{u}_n^\top]^\top\in\mathbb{R}^{n\times d}$ 
is identifiable only up to orthogonal rotations and reflections, a well-known 
invariance in bilinear latent–space models \citep{hoff2005bilinear}. Indeed, 
for any orthogonal matrix $\mathbf{Q}$ with $\mathbf{Q}^\top\mathbf{Q}=\mathbf{I}_d$, 
the reparameterization $\tilde{\mathbf{U}}=\mathbf{U}\mathbf{Q}$ preserves all 
inner products, $\tilde{\boldsymbol{u}}_i^\top \tilde{\boldsymbol{u}}_j=
\boldsymbol{u}_i^\top \boldsymbol{u}_j$, and hence the likelihood. Thus, while 
$\mathbf{U}$ is not identifiable, the identifiable functionals are the pairwise 
inner products $\boldsymbol{u}_i^\top \boldsymbol{u}_j$.

Similarly, in the \textsf{SMN\text{-}C\text{-}LD} model, the latent configuration 
is identifiable only up to rigid motions (translations, rotations, and 
reflections), mirroring the invariances of classical latent–distance network 
models \citep{hoff2002latent}. Because the likelihood depends on $\mathbf{U}$ 
solely through pairwise distances, for any orthogonal matrix $\mathbf{Q}$ and 
vector $\boldsymbol{c}\in\mathbb{R}^d$ we have 
$\|(\boldsymbol{u}_i+\boldsymbol{c})\mathbf{Q}-(\boldsymbol{u}_j+\boldsymbol{c})\mathbf{Q}\|=
\|\boldsymbol{u}_i-\boldsymbol{u}_j\|$, leaving the likelihood unchanged. Hence, 
the identifiable functionals are the distances $\|\boldsymbol{u}_i-\boldsymbol{u}_j\|$. 
In both latent–geometry cases we mitigate these indeterminacies by imposing 
spherical priors and by post–hoc centering such that 
$\sum_i \boldsymbol{u}_i=\boldsymbol{0}$ when necessary.

For the \textsf{SMN\text{-}C\text{-}SB} model, cluster labels are identifiable only up to within–layer permutations (label switching). Inference and reporting therefore rely on permutation–invariant summaries such as posterior co–clustering probabilities or relabeled draws obtained by post–processing. Finally, in all additive models (\textsf{SMN} and extensions), location is only defined up to shifts among $\zeta$, $\{\mu_k\}$, and $\{\delta_{i,k}\}$. The hierarchical centering adopted in our priors (and soft constraints such as $\sum_i \delta_{i,k}=0$ when necessary) resolves this confounding without altering the implied likelihood.

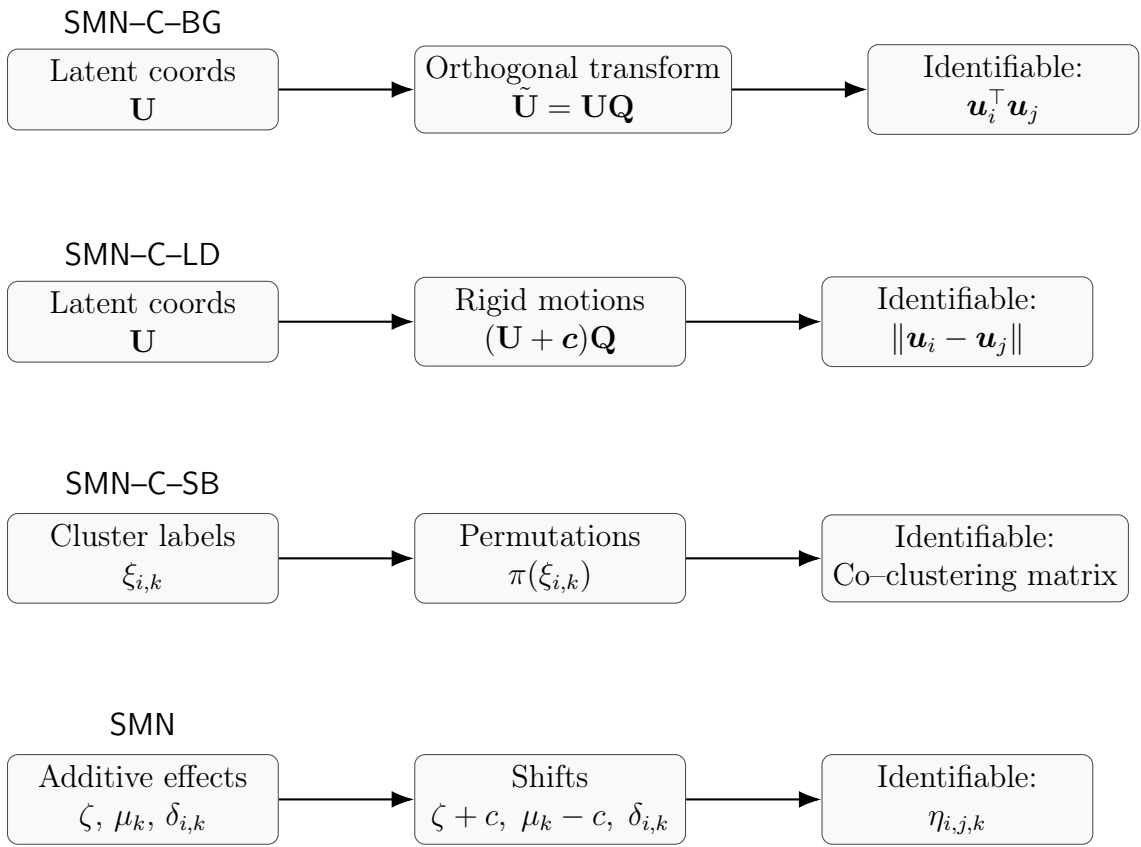
\begin{figure}[!htb]
\centering
\begin{tikzpicture}[
    node distance=1.8cm,
    box/.style={rectangle, rounded corners, draw=black!70, fill=gray!5, 
                minimum width=3.6cm, minimum height=1cm, align=center},
    arrow/.style={-{Latex[length=3mm]}, thick}
]

\node[box] (bgU) {Latent coords \\ $\mathbf{U}$};
\node[box, right=of bgU] (bgQ) {Orthogonal transform \\ $\tilde{\mathbf{U}}=\mathbf{UQ}$};
\node[box, right=of bgQ] (bgid) {Identifiable: \\ $\boldsymbol{u}_i^\top\boldsymbol{u}_j$};

\draw[arrow] (bgU) -- (bgQ);
\draw[arrow] (bgQ) -- (bgid);

\node[above of=bgU, yshift=-0.9cm] {\textsf{SMN--C--BG}};

\node[box, below=2cm of bgU] (ldU) {Latent coords \\ $\mathbf{U}$};
\node[box, right=of ldU] (ldT) {Rigid motions \\ $(\mathbf{U}+\boldsymbol{c})\mathbf{Q}$};
\node[box, right=of ldT] (ldid) {Identifiable: \\ $\|\boldsymbol{u}_i-\boldsymbol{u}_j\|$};

\draw[arrow] (ldU) -- (ldT);
\draw[arrow] (ldT) -- (ldid);

\node[above of=ldU, yshift=-0.9cm] {\textsf{SMN--C--LD}};

\node[box, below=2cm of ldU] (sbL) {Cluster labels \\ $\xi_{i,k}$};
\node[box, right=of sbL] (sbP) {Permutations \\ $\pi(\xi_{i,k})$};
\node[box, right=of sbP] (sbid) {Identifiable: \\ Co--clustering matrix};

\draw[arrow] (sbL) -- (sbP);
\draw[arrow] (sbP) -- (sbid);

\node[above of=sbL, yshift=-0.8cm] {\textsf{SMN--C--SB}};

\node[box, below=2cm of sbL] (add) {Additive effects \\ $\zeta,\,\mu_k,\,\delta_{i,k}$};
\node[box, right=of add] (addshift) {Shifts \\ $\zeta+c,\; \mu_k-c,\; \delta_{i,k}$};
\node[box, right=of addshift] (addid) {Identifiable: \\ $\eta_{i,j,k}$};

\draw[arrow] (add) -- (addshift);
\draw[arrow] (addshift) -- (addid);

\node[above of=add, yshift=-0.8cm] {\textsf{SMN}};

\end{tikzpicture}
\caption{
Summary of non-identifiability patterns across the proposed SMN formulations. Each row corresponds to one of the latent specifications and lists the transformations that leave the likelihood invariant together with the identifiable likelihood-invariant functionals (inner products, distances, co-clustering, or linear predictors).}
\label{fig:identifiability}
\end{figure}

Figure~\ref{fig:identifiability} summarizes these identifiability properties by 
grouping the latent specifications according to the transformations that leave 
their likelihoods unchanged. 

\subsection{Prior elicitation}

Our prior specification adheres to four principles: (i) exchangeability and centering, by assigning zero–mean priors to all additive effects; (ii) weakly informative global shrinkage on variance components, to stabilize estimation while preserving signal; (iii) approximately uniform prior interaction probabilities, so that no dyad is favored before hand; and (iv) modularity across model variants, ensuring priors remain comparable as additional structure is introduced.

For the \textsf{SMN} model, we set \((a_\omega,b_\omega)=(a_\sigma,b_\sigma)=(a_\tau,b_\tau)=(a_\kappa,b_\kappa)=(3,3)\). Under this choice, each variance component has mean \(3/2\), variance \(9/4\), and coefficient of variation 100\%. This centers the variance components at \(1.5\) with substantial dispersion, yielding a weakly informative prior that encourages, but does not force, shrinkage of the additive effects. Further more, for the \textsf{SMN\text{-}C} model, we set \((a_{\varsigma},b_{\varsigma})=(3,200)\). Under this choice, \(\textsf{E}(\varsigma^2)=100\), \(\textsf{Var}(\varsigma^2)=10{,}000\), and \(\textsf{CV}(\varsigma^2)=100\%\). Thus, the regression coefficients receive a zero–centered, highly diffuse prior (especially after covariate standardization), allowing a broad range of plausible magnitudes without favoring any particular scale a priori.

For both latent–geometry variants, we place spherical Gaussian priors on the shared latent positions, i.e., \(\boldsymbol{u}_i \overset{\text{iid}}{\sim} \textsf{N}_d(\boldsymbol{0},\mathbf{I})\), which helps resolve scale identifiability. We also work with post hoc Procrustes–aligned (e.g., \citealt{hoff2002latent}) and centered configurations satisfying \(\sum_{i=1}^n \boldsymbol{u}_i=\boldsymbol{0}\) to mitigate rotation and translation indeterminacies when needed (e.g., latent visualization).
For \textsf{SMN\text{-}C\text{-}BG} model, we use \((a_{\upsilon},b_{\upsilon})=(3,100)\), yielding a weakly informative prior on \(\upsilon^2\) and hence a diffuse, zero–centered prior on the geometry scales \(\lambda_k\) in the bilinear term. In contrast, for \textsf{SMN\text{-}C\text{-}LD} model, we use \((a_{\upsilon},b_{\upsilon})=(3,1)\) to concentrate \(\upsilon^2\) and keep \(\lambda_k\) tightly centered near zero, which prevents overly flat decay and avoids U–shaped prior densities for edge probabilities under the distance–decay specification. Finally, we set the latent dimension to \(d=3\), which strikes a practical balance between expressive power and computational burden: three dimensions suffice to capture simultaneous assortative and disassortative tendencies, transitivity, and layer–specific deformations of the shared geometry, while keeping the per–iteration cost \(\mathcal{O}(K\,n^{2} d)\) tractable.

For the \textsf{SMN\text{-}C\text{-}SB} model, we adopt a Dirichlet–Multinomial prior for the layer–specific community proportions, with $(a_\alpha,b_\alpha) = (1,1)$, which preserves exchangeability across blocks while allowing variability in community proportions by layer. We set \((a_{\rho},b_{\rho})=(3,200)\) for the variance component for the symmetric within/between–block affinities; this specification imposes no prior sign preference and regularizes small block effects toward zero. In applications, we choose the candidate number of communities \(C\) as the maximum number of communities obtained by clustering each observed layer with the Louvain algorithm \citep{blondel2008fast}, and we initialize \(\{\xi_{i,k}\}\) with the resulting labels to improve mixing and reduce label–switching transients.

\begin{figure}[!htb]
    \centering
    \subfigure[\textsf{SMN}]      {\includegraphics[scale=0.22]{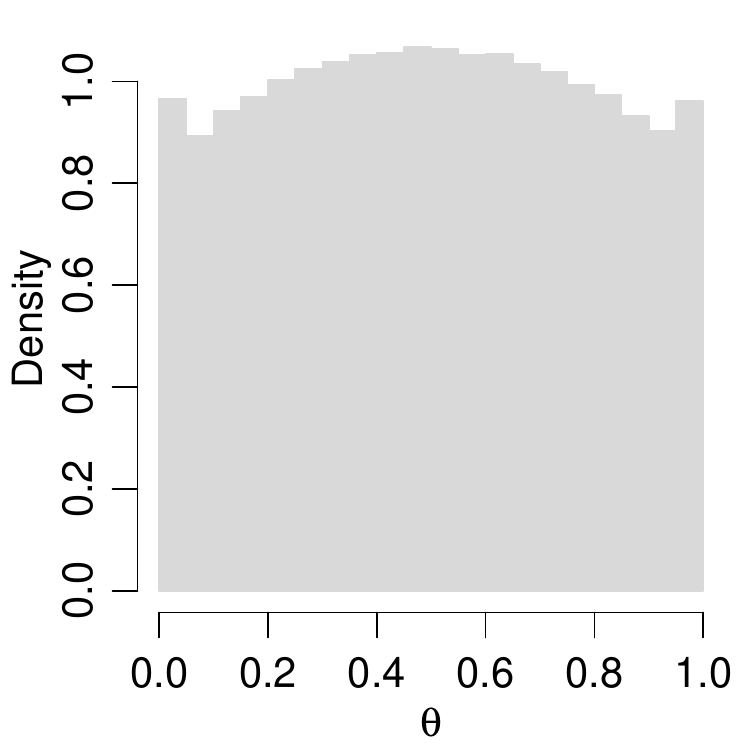}}
    \subfigure[\textsf{SMN-C}]    {\includegraphics[scale=0.22]{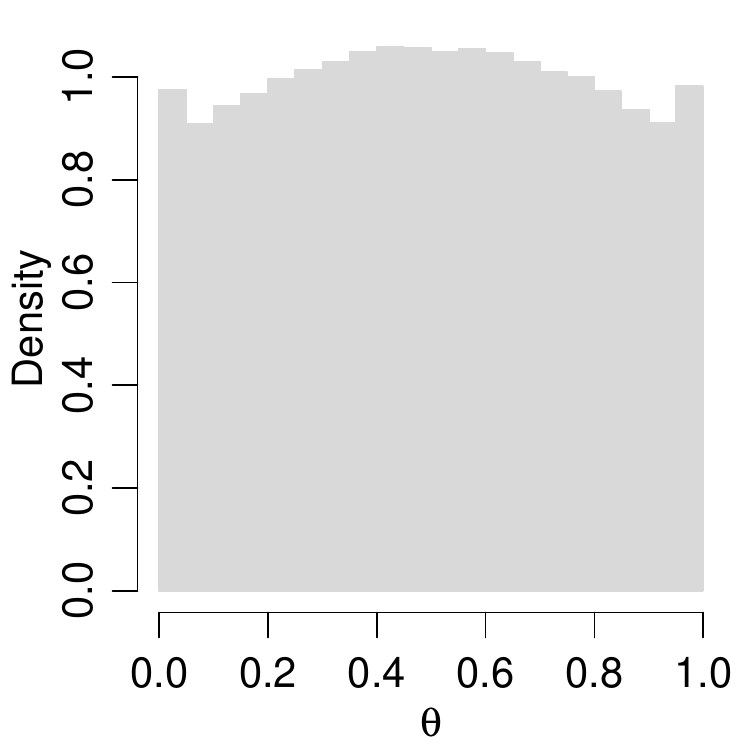}}
    \subfigure[\textsf{SMN-C-BG}] {\includegraphics[scale=0.22]{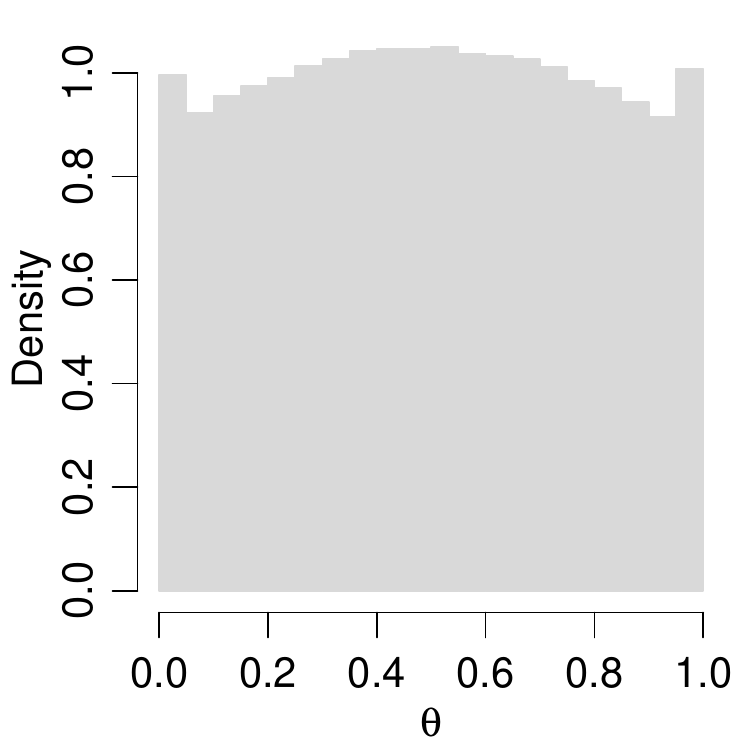}}
    \subfigure[\textsf{SMN-C-LD}] {\includegraphics[scale=0.22]{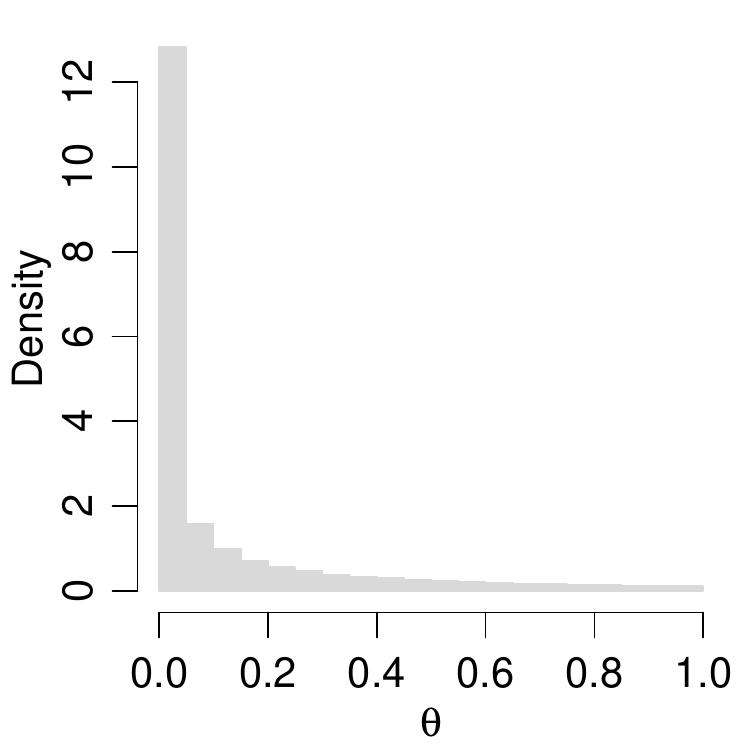}}
    \subfigure[\textsf{SMN-C-SB}] {\includegraphics[scale=0.22]{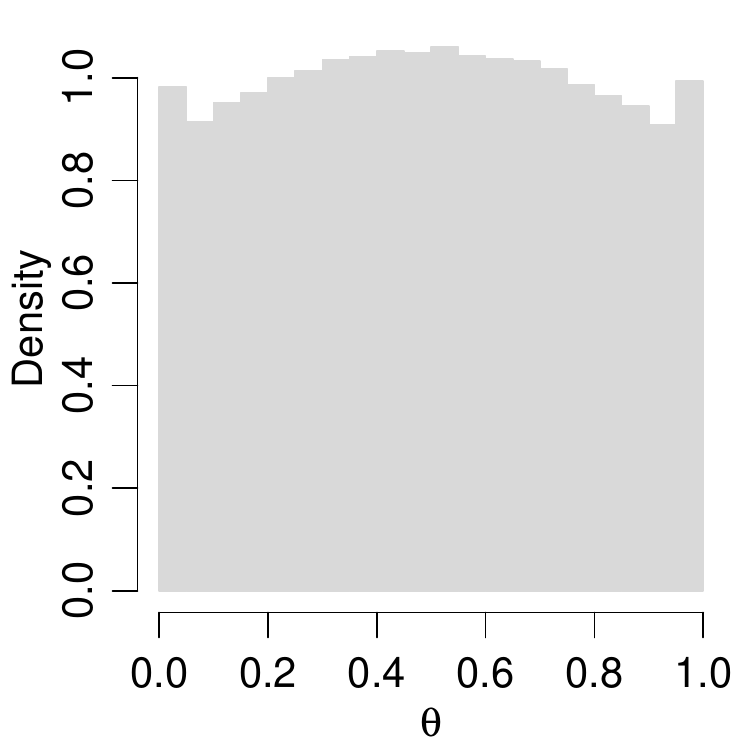}}
    \caption{Prior predictive distributions of interaction probabilities under the proposed hyperparameter elicitation for each model.}
    \label{fig_prior_sim}
\end{figure}

With the above hyperparameters, the prior on edge probabilities \(\theta_{i,j,k}=\Phi(\eta_{i,j,k})\) is approximately uniform on \((0,1)\) for all models except \textsf{SMN\text{-}C\text{-}LD}, reflecting zero–centered additive effects and moderate prior dispersion in the layer–specific and node–specific terms. By contrast, for \textsf{SMN\text{-}C\text{-}LD} the prior predictive distribution is right–skewed: it places substantial mass near \([0,0.05]\), decreases thereafter, and becomes approximately flat for \(\theta\gtrsim 0.4\). This behavior arises because the distance penalty enters the linear predictor with a negative sign and, under diffuse log–distance scales, more readily drives \(\eta_{i,j,k}\) downward at typical latent separations. These patterns are shown in Figure \ref{fig_prior_sim} using prior–predictive simulations.

\section{Computation}\label{sec_computation}

The posterior distribution synthesizes information from the observed data $\mathcal{Y}$ and prior beliefs about the model parameters $\mathbf{\Theta}$, balancing goodness of fit with the regularization induced by the priors. By Bayes' theorem, $p(\Theta \mid \mathcal{Y}) \propto p(\mathcal{Y} \mid \Theta)\, p(\Theta)$, where $p(\mathbf{Y} \mid \Theta)$ is the likelihood and $p(\Theta)$ is the prior. We explore the posterior distribution $p(\Theta \mid \mathcal{Y})$ using Markov chain Monte Carlo (MCMC; e.g., \citealt{gamerman2006markov}). MCMC generates a dependent sequence $\Theta^{(1)},\ldots,\Theta^{(S)}$ whose stationary distribution is the posterior. After burn-in, these draws approximate $p(\Theta \mid \mathcal{Y})$, so point estimates and uncertainty summaries follow from their empirical distribution. The computational algorithm employs Gibbs sampling, with Metropolis–Hastings steps (e.g., \citealt{gelman2014bayesian}) where conjugacy is not available.

To facilitate posterior computation, we adopt the probit augmentation of \citet{albert1993bayesian} by introducing latent Gaussian variables $z_{i,j,k}$ such that $z_{i,j,k}\mid \eta_{i,j,k}\ \overset{\text{ind}}{\sim}\ \textsf{N}(\eta_{i,j,k},1)$, with $y_{i,j,k}=I\{z_{i,j,k}>0\}$, so that, conditional on $y_{i,j,k}$, 
\[
z_{i,j,k}\mid \eta_{i,j,k},\, y_{i,j,k}\ \overset{\text{ind}}{\sim}\
\begin{cases}
\textsf{TN}_{(0,\infty)}\!\big(\eta_{i,j,k},\,1\big), & y_{i,j,k}=1,\\[2pt]
\textsf{TN}_{(-\infty,0]}\!\big(\eta_{i,j,k},\,1\big), & y_{i,j,k}=0,
\end{cases}
\]
where $\textsf{TN}_{A}(\mu,\sigma^{2})$ denotes a $\textsf{N}(\mu,\sigma^{2})$ distribution truncated to the set $A$. Integrating out $z_{i,j,k}$ recovers the original Bernoulli model, while the augmentation typically simplifies the computation of full conditional distributions, ensuring they take standard forms, which facilitates the implementation of a Gibbs sampler. If a logit link is used instead, an analogous augmentation is available via P\'olya--Gamma auxiliary variables \citep{polson2013bayesian}. Complete details of the MCMC algorithms used to fit all models are provided below in subsection \ref{sec_MCMC_algorithms}.

In all applications we fitted each of the five models to each dataset using long MCMC runs. For every chain, we discarded the first 200{,}000 iterations as burn-in and then collected an additional 1{,}000{,}000 iterations, which were thinned by keeping every 20th draw, yielding a total of 50{,}000 posterior samples for inference. For each fitted model and dataset, we carried out an exhaustive convergence assessment by computing Monte Carlo standard errors for all parameters and by inspecting trace plots of the log-likelihood. These diagnostics are not included in the manuscript for space reasons, but they can be fully reproduced from the code repository, which is is publicly available at \url{https://github.com/jstats1702/the-big-4}. In all cases, the diagnostics provided no evidence of lack of convergence.

\subsection{Posterior computation}\label{sec_MCMC_algorithms}

Here we provide full details of the MCMC algorithms for all models in the paper, including the posterior distribution, the full conditional distributions, and the iterative procedure used to fit each model. Unless otherwise stated, we initialize the chains by sampling all parameters from their priors. The only exception is the block labels $\{\xi_{i,k}\}$ in the \textsf{SMN-C-SB} model, which we initialize via the Louvain algorithm \citep{blondel2008fast} on the observed network to mitigate label switching and accelerate mixing.

\subsubsection{\textsf{SMN} model}

Let $\boldsymbol{\Theta}=\big(\zeta, \{\mu_k\}, \{\delta_{i,k}\}, \{\vartheta_i\}, \omega^2, \sigma^2, \tau^2, \kappa^2\big)$ be the set of model parameters (cardinality $K(n+1)+n+5$) and let $\mathbf{Z}=[z_{i,j,k}]$ be the array of Gaussian auxiliary variables with
$z_{i,j,k}\mid \eta_{i,j,k}\ \overset{\text{ind}}{\sim}\ \textsf{N}(\eta_{i,j,k},1)$, where $\eta_{i,j,k}=\zeta+\mu_k+\delta_{i,k}+\delta_{j,k}$. Up to a normalizing constant, the corresponding augmented posterior is:
\begin{align*}
p(\mathbf{\Theta},\mathbf{Z}\mid \mathcal{Y})
&\propto \prod_k \prod_{i<j} p(y_{i,j,k}\mid z_{i,j,k}) \times \prod_k \prod_{i<j} \textsf{TN}(z_{i,j,k} \mid \eta_{i,j,k},y_{i,j,k}) \\
&\quad \times \textsf{N}(\zeta \mid 0,\omega^2) \times \prod_k \textsf{N}(\mu_k \mid 0,\sigma^2) \times \prod_i \prod_k \textsf{N}(\delta_{i,k} \mid \vartheta_i,\tau^2) \prod_i \textsf{N}(\vartheta_i \mid 0,\kappa^2) \\
&\quad\quad \times \textsf{IG}(\omega^2 \mid a_\omega,b_\omega) \times \textsf{IG}(\sigma^2 \mid a_\sigma,b_\sigma) \times \textsf{IG}(\tau^2 \mid a_\tau,b_\tau) \times \textsf{IG}(\kappa^2 \mid a_\kappa,b_\kappa).
\end{align*}

Following parameter initialization, the Gibbs sampler for drawing from the posterior distribution $p(\boldsymbol{\Theta}, \mathbf{Z} \mid \mathcal{Y})$ consists of sequentially sampling each parameter from its corresponding full conditional distribution (FCD), conditioning on the most recently updated values of the remaining paramters. The FCDs are derived from the augmented posterior by isolating the terms involving the target block while treating all other quantities as fixed. Consequently, the FCDs are given by:
\begin{itemize}
    \item $z_{i,j,k}\mid \cdot$ follows a truncated Normal distribution:
    \[
    z_{i,j,k}\mid \cdot \sim
    \begin{cases}
    \textsf{TN}_{(0,\infty)}\!\big(\eta_{i,j,k},\,1\big), & y_{i,j,k}=1,\\[2pt]
    \textsf{TN}_{(-\infty,0]}\!\big(\eta_{i,j,k},\,1\big), & y_{i,j,k}=0,
    \end{cases}
    \qquad
    \eta_{i,j,k}=\zeta+\mu_k+\delta_{i,k}+\delta_{j,k}.
    \]

    \item $\zeta \mid \cdot  \sim \textsf{N}(M,V^2)$, with
    \[
    M = V^2 \sum_k\sum_{i<j}(z_{i,j,k}-\mu_k-\delta_{i,k}-\delta_{j,k}),\qquad
    V^2 = \left(\frac{1}{\omega^2} + K\,\frac{n(n-1)}{2} \right)^{-1}.
    \]
    
    \item $\mu_k \mid \cdot \sim \textsf{N}(M_k,V^2_k)$, with
    \[
    M_k = V^2_k \sum_{i<j}(z_{i,j,k}-\zeta-\delta_{i,k}-\delta_{j,k}),\qquad
    V^2_k = \left(\frac{1}{\sigma^2} + \frac{n(n-1)}{2} \right)^{-1}.
    \]

    \item $\delta_{i,k}\mid \cdot \sim \textsf{N}(M_{i,k},V^2_{i,k})$, with
    \[
    M_{i,k} = V^2_{i,k}\left(\frac{\vartheta_i}{\tau^{2}} + \sum_{j\ne i}^{n}(z_{i,j,k}-\mu_k-\zeta-\delta_{j,k})\right),\qquad
    V^2_{i,k} = \left(\frac{1}{\tau^2} + n-1 \right)^{-1}.
    \]

    \item $\vartheta_i\mid \cdot \sim \textsf{N}(M_i, V^2_i)$, with
    \[
    M_i = V^2_i\,\frac{1}{\tau^2}\sum_k \delta_{i,k},\qquad
    V^2_i = \left(\frac{1}{\kappa^{2}} + \frac{1}{\tau^{2}}K\right)^{-1}.
    \]

    \item $\omega^2\mid\cdot\sim\textsf{IG}(A,B)$, with
    \[
    A = a_\omega + \frac{1}{2}, \qquad
    B = b_\omega + \frac{1}{2}\zeta^2.
    \]
    
    \item $\sigma^2\mid\cdot\sim\textsf{IG}(A,B)$, with
    \[
    A = a_\sigma + \frac{K}{2},\qquad
    B = b_\sigma + \frac{1}{2}\sum_k \mu_k^2.
    \]

    \item $\tau^2\mid\cdot\sim\textsf{IG}(A,B)$, with
    \[
    A = a_\tau + \frac{nK}{2},\qquad
    B = b_\tau + \frac{1}{2}\sum_i\sum_k(\delta_{i,k} - \vartheta_i)^2.
    \]
    
    \item $\kappa^2\mid\cdot\sim\textsf{IG}(A,B)$, with
    \[
    A = a_\kappa + \frac{n}{2},\qquad 
    B = b_\kappa + \frac{1}{2}\sum_i \vartheta_i^2.
    \]
\end{itemize}

\subsubsection{\textsf{SMN-C} model}

Let $\boldsymbol{\Theta}=\big(\zeta,\{\mu_k\},\{\delta_{i,k}\},\{\vartheta_i\},\{\boldsymbol{\beta}_k\},\upsilon^2,\sigma^2,\tau^2,\kappa^2,\varsigma^2\big)$ be the set of model parameters (cardinality $K(n+p+1)+n+6$), and let $\mathbf{Z}=[z_{i,j,k}]$ be the array of Gaussian auxiliary variables with $z_{i,j,k}\mid \eta_{i,j,k}\ \overset{\text{ind}}{\sim}\ \textsf{N}(\eta_{i,j,k},1)$, where $\eta_{i,j,k}=\zeta+\mu_k+\delta_{i,k}+\delta_{j,k}+\mathbf{x}_{i,j}^\top\boldsymbol{\beta}_k$, with
$\mathbf{x}_{i,j}\in\mathbb{R}^p$ and $\boldsymbol{\beta}_k\in\mathbb{R}^p$. Up to a normalizing constant, the augmented posterior is:
\begin{align*}
p(\boldsymbol{\Theta},\mathbf{Z}\mid \mathcal{Y})
&\propto \prod_{k}\prod_{i<j} p(y_{i,j,k}\mid z_{i,j,k})
\times \prod_{k}\prod_{i<j} \textsf{TN}\big(z_{i,j,k}\mid \eta_{i,j,k},y_{i,j,k}\big)\\
&\quad \times\ \textsf{N}(\zeta\mid 0,\upsilon^2)\times\prod_{k}\textsf{N}(\mu_k\mid 0,\sigma^2)\times\prod_{i}\prod_{k}\textsf{N}(\delta_{i,k}\mid \vartheta_i,\tau^2)\times\prod_{i}\textsf{N}(\vartheta_i\mid 0,\kappa^2)\\
&\quad\quad \times\ \prod_{k}\textsf{N}_p\big(\boldsymbol{\beta}_k\mid \boldsymbol{0},\varsigma^2\mathbf{I}\big)\times\textsf{IG}(\omega^2\mid a_\omega,b_\omega)\times \textsf{IG}(\sigma^2\mid a_\sigma,b_\sigma)\\
&\quad\quad\quad \times\ \textsf{IG}(\tau^2\mid a_\tau,b_\tau)\times \textsf{IG}(\kappa^2\mid a_\kappa,b_\kappa)\times \textsf{IG}(\varsigma^2\mid a_{\varsigma},b_{\varsigma}).
\end{align*}

The FCDs are give by:
\begin{itemize}
    \item $z_{i,j,k}\mid \cdot$ is identical to that in the \textsf{SMN} model, except that $\eta_{i,j,k}=\zeta+\mu_k+\delta_{i,k}+\delta_{j,k}+\boldsymbol{x}_{i,j}^\top\boldsymbol{\beta}_k$. 

    \item $\zeta \mid \cdot$ is identical to that in the \textsf{SMN} model, except that
    \[
    M = V^2 \sum_k\sum_{i<j}(z_{i,j,k}-\mu_k-\delta_{i,k}-\delta_{j,k}-\mathbf{x}_{i,j}^\top\boldsymbol{\beta}_k).
    \]

    \item $\mu_k \mid \cdot)$ is identical to that in the \textsf{SMN} model, except that
    \[
    M_k = V^2_k \sum_{i<j}(z_{i,j,k}-\zeta-\delta_{i,k}-\delta_{j,k}-\mathbf{x}_{i,j}^\top\boldsymbol{\beta}_k).
    \]

    \item $\delta_{i,k}\mid \cdot$ is identical to that in the \textsf{SMN} model, except that
    \[
    M_{i,k} = V^2_{i,k}\left(\frac{\vartheta_i}{\tau^{2}} + \sum_{j\ne i}^{n}(z_{i,j,k}-\mu_k-\zeta-\delta_{j,k}-\mathbf{x}_{i,j}^\top\boldsymbol{\beta}_k)\right).
    \]

    \item $\vartheta_i\mid \cdot$ is identical to that in the \textsf{SMN} model.

    \item $\boldsymbol{\beta}_k\mid \cdot \sim \textsf{N}_p(\mathbf{m}_k,\ \mathbf{V}_k)$, with
    \[
    \mathbf{m}_{k} = \mathbf{V}_{k}\sum_{i<j}(z_{i,j,k}-\zeta-\mu_k-\delta_{i,k}-\delta_{j,k}) \ \boldsymbol{x}_{i,j},\qquad
    \mathbf{V}_{k} = \left(\frac{1}{\varsigma^2}\mathbf{I} + \sum_{i<j}\boldsymbol{x}_{i,j} \ \boldsymbol{x}_{i,j}^\top\right)^{-1}.
    \]

    \item $\omega^2\mid\cdot$ is identical to that in the \textsf{SMN} model.

    \item $\sigma^2\mid\cdot$ is identical to that in the \textsf{SMN} model.

    \item $\tau^2\mid\cdot$ is identical to that in the \textsf{SMN} model.

    \item $\kappa^2\mid\cdot$ is identical to that in the \textsf{SMN} model.

    \item $\varsigma^2\mid\cdot\sim\textsf{IG}(A, B)$, with
    \[
    A = a_{\varsigma} + \frac{Kp}{2},\qquad
    B = b_{\varsigma} + \frac{1}{2}\sum_{k}\|\boldsymbol{\beta}_k\|^2.
    \]   
\end{itemize}

\subsubsection{\textsf{SMN-C-BG} model}

Let $\boldsymbol{\Theta}=\big(\zeta,\{\mu_k\},\{\delta_{i,k}\},\{\vartheta_i\},\{\boldsymbol{\beta}_k\},\{\boldsymbol{u}_i\},\{\lambda_k\},\omega^2,\sigma^2,\tau^2,\kappa^2,\varsigma^2,\upsilon^2\big)$ be the set of model parameters (cardinality $K(n+p+2)+n(d+1)+7$), and let $\mathbf{Z}=[z_{i,j,k}]$ be the array of Gaussian auxiliary variables with
$z_{i,j,k}\mid \eta_{i,j,k}\ \overset{\text{ind}}{\sim}\ \textsf{N}(\eta_{i,j,k},1)$, where
$\eta_{i,j,k}=\zeta+\mu_k+\delta_{i,k}+\delta_{j,k}+\boldsymbol{x}_{i,j}^\top\boldsymbol{\beta}_k+\lambda_k\,\boldsymbol{u}_i^\top\boldsymbol{u}_j$, with $\lambda_k\in\mathbb{R}$ and $\boldsymbol{u}_i\in\mathbb{R}^d$. Up to a normalizing constant, the augmented posterior is:
\begin{align*}
p(\boldsymbol{\Theta},\mathbf{Z}\mid \mathcal{Y})
&\propto \prod_{k}\prod_{i<j} p(y_{i,j,k}\mid z_{i,j,k})
\times \prod_{k}\prod_{i<j} \textsf{TN}\big(z_{i,j,k}\mid \eta_{i,j,k},y_{i,j,k}\big)\\
&\quad \times \textsf{N}(\zeta\mid 0,\omega^2) \times \prod_{k}\textsf{N}(\mu_k\mid 0,\sigma^2) \times \prod_{i}\prod_{k}\textsf{N}(\delta_{i,k}\mid \vartheta_i,\tau^2) \times \prod_{i=1}^{n}\textsf{N}(\vartheta_i\mid 0,\kappa^2)\\
&\quad\quad \times \prod_{k}\textsf{N}_p\big(\boldsymbol{\beta}_k\mid \boldsymbol{0},\varsigma^2\mathbf{I}\big) \times \prod_{k}\textsf{N}(\lambda_k\mid 0,\upsilon^2) \times \prod_{i}\textsf{N}_d\big(\boldsymbol{u}_i\mid \boldsymbol{0},\mathbf{I}\big)\\
&\quad\quad\quad \times \textsf{IG}(\omega^2\mid a_\omega,b_\omega) \times \textsf{IG}(\sigma^2\mid a_\sigma,b_\sigma) \times \textsf{IG}(\tau^2\mid a_\tau,b_\tau)\\
&\quad\quad\quad\quad \times \textsf{IG}(\kappa^2\mid a_\kappa,b_\kappa) \times \textsf{IG}(\varsigma^2\mid a_{\varsigma},b_{\varsigma}) \times \textsf{IG}(\upsilon^2\mid a_{\upsilon},b_{\upsilon}).
\end{align*}

The FCDs are given by:
\begin{itemize}
    \item $z_{i,j,k}\mid \cdot$ is identical to that in the \textsf{SMN} model, except that $\eta_{i,j,k}=\zeta+\mu_k+\delta_{i,k}+\delta_{j,k}+\boldsymbol{x}_{i,j}^\top\boldsymbol{\beta}_k+\lambda_k\,\boldsymbol{u}_i^\top\boldsymbol{u}_j$. 

    \item $\zeta\mid \cdot$ is identical to that in the \textsf{SMN} model, except that
    \[
    M = V^2 \sum_{k}\sum_{i<j}(z_{i,j,k}-\mu_k-\delta_{i,k}-\delta_{j,k}-\boldsymbol{x}_{i,j}^\top\boldsymbol{\beta}_k-\lambda_k\,\boldsymbol{u}_i^\top\boldsymbol{u}_j).
    \]
    
    \item $\mu_k\mid \cdot$ is identical to that in the \textsf{SMN} model, except that 
    \[
    M_k=V_k^2\sum_{i<j}(z_{i,j,k}-\zeta-\delta_{i,k}-\delta_{j,k}-\boldsymbol{x}_{i,j}^\top\boldsymbol{\beta}_k-\lambda_k\,\boldsymbol{u}_i^\top\boldsymbol{u}_j).
    \]
    
    \item $\delta_{i,k}\mid \cdot$ is identical to that in the \textsf{SMN} model, except that
    \[
    M_{i,k}=V_{i,k}^2\left(\frac{\vartheta_i}{\tau^{2}} + \sum_{j\ne i}(z_{i,j,k}-\zeta-\mu_k-\delta_{j,k}-\boldsymbol{x}_{i,j}^\top\boldsymbol{\beta}_k-\lambda_k\,\boldsymbol{u}_i^\top\boldsymbol{u}_j)\right).
    \]
    
    \item $\vartheta_i\mid \cdot$ is identical to that in the \textsf{SMN} model.

    \item $\boldsymbol{\beta}_k\mid \cdot$ is identical to that in the \textsf{SMN-C} model, except that
    \[
    \mathbf{m}_{k}=\mathbf{V}_{k} \sum_{i<j} (z_{i,j,k}-\zeta-\mu_k-\delta_{i,k}-\delta_{j,k}-\lambda_k\,\boldsymbol{u}_i^\top\boldsymbol{u}_j) \ \boldsymbol{x}_{i,j}.
    \]

    \item $\boldsymbol{u}_i\mid \cdot \sim \textsf{N}_d(\mathbf{m}_i, \mathbf{V}_i)$, with
    \[
    \mathbf{m}_i=\mathbf{V}_i\sum_{k}\lambda_k\sum_{j\ne i}^{n} (z_{i,j,k}-\zeta-\mu_k-\delta_{i,k}-\delta_{j,k})\,\boldsymbol{u}_j,\,
    \mathbf{V}_i=\left(\mathbf{I}+\sum_{k}\lambda_k^2\sum_{j\ne i}\boldsymbol{u}_j\boldsymbol{u}_j^\top\right)^{-1}.
    \]
    
    \item $\lambda_k\mid \cdot \sim \textsf{N}(M_{k},V^2_{k})$, with
    \[
    M_{k}=V^2_{k}\sum_{i<j} (z_{i,j,k}-\zeta-\mu_k-\delta_{i,k}-\delta_{j,k}) \ (\boldsymbol{u}_i^\top\boldsymbol{u}_j),\,
    V^2_{k}=\left(\frac{1}{\upsilon^2}+\sum_{i<j}(\boldsymbol{u}_i^\top\boldsymbol{u}_j)^2\right)^{-1}.
    \]

    \item $\omega^2\mid\cdot$ is identical to that in the \textsf{SMN} model.

    \item $\sigma^2\mid\cdot$ is identical to that in the \textsf{SMN} model.

    \item $\tau^2\mid\cdot$ is identical to that in the \textsf{SMN} model.

    \item $\kappa^2\mid\cdot$ is identical to that in the \textsf{SMN} model.

    \item $\varsigma^2\mid\cdot$ is identical to that in the \textsf{SMN-C} model.

    \item $\upsilon^2\mid \cdot \sim \textsf{IG}(A,B)$, with
    \[
    A = a_{\upsilon}+\frac{K}{2},\qquad 
    B = b_{\upsilon}+\frac{1}{2}\sum_{k}\lambda_k^2.
    \]
\end{itemize}

\subsubsection{\textsf{SMN-C-LD} model}

Let $\boldsymbol{\Theta}=\big(\zeta,\{\mu_k\},\{\delta_{i,k}\},\{\vartheta_i\},\{\boldsymbol{\beta}_k\},\{\boldsymbol{u}_i\},\{\lambda_k\},\omega^2,\sigma^2,\tau^2,\kappa^2,\varsigma^2,\upsilon^2\big)$ be the set of model parameters (cardinality $K(n+p+2)+n(d+1)+7$), and let $\mathbf{Z}=[z_{i,j,k}]$ be the array of Gaussian auxiliary variables with $z_{i,j,k}\mid \eta_{i,j,k}\ \overset{\text{ind}}{\sim}\ \textsf{N}(\eta_{i,j,k},1)$, where $\eta_{i,j,k}=\zeta+\mu_k+\delta_{i,k}+\delta_{j,k}+\boldsymbol{x}_{i,j}^\top\boldsymbol{\beta}_k
- \exp(\lambda_k)\,\|\boldsymbol{u}_i-\boldsymbol{u}_j\|$, with $\lambda_k\in\mathbb{R}$ and $\boldsymbol{u}_i\in\mathbb{R}^d$. 
Up to a normalizing constant, the augmented posterior is identical to that in the \textsf{SMN-C-BG} model.

The FCDs are given by:
\begin{itemize}
    \item $z_{i,j,k}\mid \cdot$ is identical to that in the \textsf{SMN} model, except that $\eta_{i,j,k}=\zeta+\mu_k+\delta_{i,k}+\delta_{j,k}+\boldsymbol{x}_{i,j}^\top\boldsymbol{\beta}_k-\exp(\lambda_k)\,d_{i,j}$.

    \item $\zeta\mid \cdot$ is identical to that in the \textsf{SMN} model, except that
    \[
    M = V^2 \sum_{k}\sum_{i<j}(z_{i,j,k}-\mu_k-\delta_{i,k}-\delta_{j,k}-\boldsymbol{x}_{i,j}^\top\boldsymbol{\beta}_k+\exp(\lambda_k)\,d_{i,j}),
    \]
    where $d_{i,j}=\|\boldsymbol{u}_i-\boldsymbol{u}_j\|$.

    \item $\mu_k\mid \cdot$ is identical to that in the \textsf{SMN} model, except that 
    \[
    M_k=V_k^2\sum_{i<j}(z_{i,j,k}-\zeta-\delta_{i,k}-\delta_{j,k}-\boldsymbol{x}_{i,j}^\top\boldsymbol{\beta}_k+\exp(\lambda_k)\,d_{i,j}).
    \]

    \item $\delta_{i,k}\mid \cdot$ is identical to that in the \textsf{SMN} model, except that
    \[
    M_{i,k}=V_{i,k}^2\left(\frac{\vartheta_i}{\tau^{2}} + \sum_{j\ne i}(z_{i,j,k}-\zeta-\mu_k-\delta_{j,k}-\boldsymbol{x}_{i,j}^\top\boldsymbol{\beta}_k+\exp(\lambda_k)\,d_{i,j})\right).
    \]

    \item $\vartheta_i\mid \cdot$ is identical to that in the \textsf{SMN} model.

    \item $\boldsymbol{\beta}_k\mid \cdot$ is identical to that in the \textsf{SMN-C} model, except that
    \[
    \mathbf{m}_{k}=\mathbf{V}_{k}\sum_{i<j}(z_{i,j,k}-\zeta-\mu_k-\delta_{i,k}-\delta_{j,k}+\exp(\lambda_k)\,d_{i,j})\,\boldsymbol{x}_{i,j}.
    \]

    \item $\boldsymbol{u}_i\mid \cdot$ is updated using a random-walk Metropolis step \citep{andrieu2008tutorial}. The log-FCD is:
    \[
    \log p(\boldsymbol{u}_i\mid\cdot) = -\sum_{k}\exp(\lambda_k)\sum_{j\ne i} r_{i,j,k}\,d_{i,j} - \frac{1}{2}\sum_{k}\exp(2\lambda_k)\sum_{j\ne i} d_{i,j}^2 - \frac{1}{2}\,\boldsymbol{u}_i^\top\boldsymbol{u}_i,
    \]
    where $r_{i,j,k}=z_{i,j,k}-\zeta-\mu_k-\delta_{i,k}-\delta_{j,k}-\boldsymbol{x}_{i,j}^\top\boldsymbol{\beta}_k$. 

    At MCMC iteration $t$, given the current latent positions $\boldsymbol{u}_1,\dots,\boldsymbol{u}_n \in \mathbb{R}^d$ and the corresponding log–proposal scales $\ell_1,\dots,\ell_n$, we update each $\boldsymbol{u}_i$ with an adaptive random–walk Metropolis step as follows:
    \begin{enumerate}
      \item Proposal scale:
      Compute the current proposal standard deviation $s_i = \exp(\ell_i)$.
      
      \item Random–walk proposal:
      Draw a multivariate standard normal perturbation $\boldsymbol{\epsilon}_i \sim \textsf{N}_d(\boldsymbol{0},\mathbf{I})$ 
      and propose
      \[
        \boldsymbol{u}_i' = \boldsymbol{u}_i + \frac{s_i}{\sqrt{d}}\,\boldsymbol{\epsilon}_i.
      \]
      
      \item Acceptance probability:
      Evaluate the log–FCD of $\boldsymbol{u}_i$ at the current and proposed values,
      $\log p(\boldsymbol{u}_i \mid \cdot)$ and $\log p(\boldsymbol{u}_i' \mid \cdot)$, and compute
      \[
        \alpha_i 
        = \min\left\{1,\,
          \exp\big(\log p(\boldsymbol{u}_i' \mid \cdot) 
                  - \log p(\boldsymbol{u}_i \mid \cdot)\big)
          \right\}.
      \]
      
      \item Accept/reject:
      Draw $u_i \sim \textsf{U}(0,1)$. 
      If $u_i \leq \alpha_i$, set $\boldsymbol{u}_i \leftarrow \boldsymbol{u}_i'$ and define the acceptance 
      indicator $\text{I}_i^{(t)} = 1$, otherwise keep $\boldsymbol{u}_i$ unchanged and set 
      $\text{I}_i^{(t)} = 0$.
      
      \item Adaptive update of the proposal scale (burn–in period only): 
      During the burn–in phase, adapt the log–scale $\ell_i$ using a 
      Robbins–Monro update targeting a prescribed acceptance rate $\alpha^\star$
      (e.g., $\alpha^\star = 0.234$) as follows:
      \[
        \ell_i \leftarrow \ell_i + \gamma_t\big(\text{I}_i^{(t)} - \alpha^\star\big),
        \qquad
        \gamma_t = \frac{\eta_0}{\sqrt{1 + t}},
      \]
      where $\eta_0 > 0$ is a tuning constant (e.g., $\eta_0 = 0.05$). The updated $\ell_i$ is then constrained to lie in a 
      fixed interval, for example
      \[
        \ell_i \in (\log s_{\min},\, \log s_{\max}),
      \]
      to avoid excessively small or large proposal variances. After the burn–in period, the values $\ell_i$ 
      (and thus $s_i$) are kept fixed and the Metropolis updates proceed with 
      non–adaptive, node–specific step sizes.
    \end{enumerate}
        
    \item $\lambda_k\mid \cdot$ is updated using an adaptive random-walk Metropolis step \citep{andrieu2008tutorial}. The log-FCD is:
    \[
    \log p(\lambda_k\mid\cdot) = -\exp(\lambda_k)\,S_{1,k} - \frac{1}{2}\exp(2\lambda_k)\,S_{2,k} - \frac{1}{2}\frac{\lambda_k^2}{\upsilon^2},
    \]
    where $S_{1,k}=\sum_{i<j} r_{i,j,k}\,d_{i,j}$ and $S_{2,k}=\sum_{i<j} d_{i,j}^2$. 
    
    At MCMC iteration $t$, given the current values  $\lambda_1,\dots,\lambda_K \in \mathbb{R}$ and the corresponding log–proposal scales $\ell_1,\dots,\ell_K$, we update each $\lambda_k$ with an adaptive random–walk Metropolis step as follows:

    \begin{enumerate}
      \item Proposal scale:
      Compute the current proposal standard deviation $s_k = \exp(\ell_k)$.
    
      \item Random–walk proposal: 
      Draw a standard normal perturbation $\epsilon_k \sim \textsf{N}(0,1)$ 
      and propose $\lambda_k' = \lambda_k + s_k\,\epsilon_k$.
      
      \item Acceptance probability:
      Evaluate the log–FCD $\lambda_k$ at the current and proposed values,
      $\log p(\lambda_k \mid \cdot)$ and $\log p(\lambda_k' \mid \cdot)$, and compute
      \[
        \alpha_k 
        = \min\left\{1,\,
          \exp\big(\log p(\lambda_k' \mid \cdot) 
                  - \log p(\lambda_k \mid \cdot)\big)
          \right\}.
      \]
    
      \item Accept/reject:
      Draw $u_k \sim \textsf{U}(0,1)$. 
      If $u_k \leq \alpha_k$, set $\lambda_k \leftarrow \lambda_k'$ and define the acceptance 
      indicator $\text{I}_k^{(t)} = 1$, otherwise keep $\lambda_k$ unchanged and set 
      $\text{I}_k^{(t)} = 0$.
    
      \item Adaptive update of the proposal scale (burn–in period only): 
      During the burn–in phase, adapt the log–scale $\ell_k$ using a 
      Robbins–Monro update targeting a prescribed acceptance rate $\alpha^\star$
      (e.g., $\alpha^\star = 0.44$) as follows:
      \[
        \ell_k \leftarrow \ell_k + \gamma_t\big(\text{I}_k^{(t)} - \alpha^\star\big),
        \qquad
        \gamma_t = \frac{\eta_0}{\sqrt{1 + t}},
      \]
      where $\eta_0 > 0$ is a tuning constant (e.g., $\eta_0 = 0.05$). The updated $\ell_k$ is then constrained to lie in a 
      fixed interval, for example
      \[
        \ell_k \in (\log s_{\min},\, \log s_{\max}),
      \]
      to avoid excessively small or large proposal variances. After burn–in period, the values $\ell_k$ 
      (and thus $s_k$) are kept fixed and the Metropolis updates proceed with 
      non–adaptive, layer–specific step sizes.
    \end{enumerate}

    \item $\omega^2\mid\cdot$ is identical to that in the \textsf{SMN} model.

    \item $\sigma^2\mid\cdot$ is identical to that in the \textsf{SMN} model.

    \item $\tau^2\mid\cdot$ is identical to that in the \textsf{SMN} model.

    \item $\kappa^2\mid\cdot$ is identical to that in the \textsf{SMN} model.

    \item $\varsigma^2\mid\cdot$ is identical to that in the \textsf{SMN-C} model.

    \item $\upsilon^2\mid\cdot$ is identical to that in the \textsf{SMN-C-BG} model.
\end{itemize}

\subsubsection{\textsf{SMN-C-SB} model}

Let $\boldsymbol{\Theta}=\big(\zeta,\{\mu_k\},\{\delta_{i,k}\},\{\vartheta_i\},\{\boldsymbol{\beta}_k\},\{\gamma_{a,b,k}\},\{\boldsymbol{\omega}_k\},\{\xi_{i,k}\},\omega^2,\sigma^2,\tau^2,\kappa^2,\varsigma^2,\rho^2,\alpha\big)$ be the set of model parameters (cardinality $K\big(2n+p+C+\binom{C+1}{2}+1\big)+n+8$), and let $\mathbf{Z}=[z_{i,j,k}]$ be the array of Gaussian auxiliary variables with $z_{i,j,k}\mid \eta_{i,j,k}\ \overset{\text{ind}}{\sim}\ \textsf{N}(\eta_{i,j,k},1)$, where $\eta_{i,j,k}=\zeta+\mu_k+\delta_{i,k}+\delta_{j,k}+\boldsymbol{x}_{i,j}^\top\boldsymbol{\beta}_k+\gamma_{\phi(\xi_{i,k},\xi_{j,k}),k}$, with $\xi_{i,k}\in\{1,\ldots,C\}$ and $\gamma_{a,b,k}=\gamma_{b,a,k}$. Up to a normalizing constant, the augmented posterior is:
\begin{align*}
p(\boldsymbol{\Theta},\mathbf{Z}\mid \mathcal{Y})
&\propto \prod_{k}\prod_{i<j} p(y_{i,j,k}\mid z_{i,j,k})
\times \prod_{k}\prod_{i<j} \textsf{TN}\big(z_{i,j,k}\mid \eta_{i,j,k},y_{i,j,k}\big)\\
&\quad \times \textsf{N}(\zeta\mid 0,\omega^2)\times \prod_{k}\textsf{N}(\mu_k\mid 0,\sigma^2)\times \prod_{i}\prod_{k}\textsf{N}(\delta_{i,k}\mid \vartheta_i,\tau^2)\times \prod_{i}\textsf{N}(\vartheta_i\mid 0,\kappa^2)\\
&\quad \quad\times \prod_{k}\textsf{N}_p\big(\boldsymbol{\beta}_k\mid \boldsymbol{0},\varsigma^2\mathbf{I}\big)
\times \prod_{k}\prod_{a\le b}\textsf{N}(\theta_{a,b,k}\mid 0,\rho^2) \\
&\quad\quad\quad \times \prod_{k}\prod_{i}\textsf{Cat}\big(\xi_{i,k}\mid \boldsymbol{\omega}_k\big) \times  \prod_{k}\textsf{Dir}\big(\boldsymbol{\omega}_k\mid \tfrac{\alpha}{C}\mathbf{1}\big) \times \textsf{G}(\alpha\mid a_\alpha,b_\alpha)\\
&\quad\quad\quad\quad \times \textsf{IG}(\omega^2\mid a_\omega,b_\omega)\times \textsf{IG}(\sigma^2\mid a_\sigma,b_\sigma)\times \textsf{IG}(\tau^2\mid a_\tau,b_\tau)\\
&\quad\quad\quad\quad\quad \times \textsf{IG}(\kappa^2\mid a_\kappa,b_\kappa) \times \textsf{IG}(\varsigma^2\mid a_{\varsigma},b_{\varsigma})\times \textsf{IG}(\rho^2\mid a_\rho,b_\rho).
\end{align*}

The FCDs are given by:
\begin{itemize}
    \item $z_{i,j,k}\mid \cdot$ is identical to that in the \textsf{SMN} model, except that
    \[
    \eta_{i,j,k}=\zeta+\mu_k+\delta_{i,k}+\delta_{j,k}+\boldsymbol{x}_{i,j}^\top\boldsymbol{\beta}_k+\gamma_{\phi(\xi_{i,k},\xi_{j,k}),k}.
    \]
    
    \item $\zeta\mid$ is identical to that in the \textsf{SMN} model, except that
    \[
    M=V^2\sum_{k}\sum_{i<j}\!\big(z_{i,j,k}-\mu_k-\delta_{i,k}-\delta_{j,k}-\boldsymbol{x}_{i,j}^\top\boldsymbol{\beta}_k-\gamma_{\phi(\xi_{i,k},\xi_{j,k}),k}\big).
    \]
    
    \item $\mu_k\mid \cdot$ is identical to that in the \textsf{SMN} model, except that
    \[
    M_k=V_k^2\sum_{i<j}\!\big(z_{i,j,k}-\zeta-\delta_{i,k}-\delta_{j,k}-\boldsymbol{x}_{i,j}^\top\boldsymbol{\beta}_k-\gamma_{\phi(\xi_{i,k},\xi_{j,k}),k}\big).
    \]
    
    \item $\delta_{i,k}\mid \cdot$ is identical to that in the \textsf{SMN} model, except that
    \[
    M_{i,k}=V_{i,k}^2\!\left(\frac{\vartheta_i}{\tau^2}+\sum_{j\ne i}\!\big(z_{i,j,k}-\zeta-\mu_k-\delta_{j,k}-\boldsymbol{x}_{i,j}^\top\boldsymbol{\beta}_k-\gamma_{\phi(\xi_{i,k},\xi_{j,k}),k}\big)\right).
    \]
    
    \item $\vartheta_i\mid \cdot$ is identical to that in the \textsf{SMN} model.

    \item $\boldsymbol{\beta}_k\mid \cdot$ is identical to that in the \textsf{SMN} model, except that
    \[
    \mathbf{m}_k=\mathbf{V}_k\sum_{i<j}\big(z_{i,j,k}-\zeta-\mu_k-\delta_{i,k}-\delta_{j,k}-\gamma_{\phi(\xi_{i,k},\xi_{j,k}),k}\big)\,\boldsymbol{x}_{i,j}.
    \]
    
    \item $\theta_{a,b,k}\mid \cdot \sim \textsf{N}(M_{a,b,k},V_{a,b,k}^2)$, with
    \[
    M_{a,b,k}=V_{a,b,k}^2\sum_{D_{a,b,k}}\big(z_{i,j,k}-\zeta-\mu_k-\delta_{i,k}-\delta_{j,k}-\boldsymbol{x}_{i,j}^\top\boldsymbol{\beta}_k\big),\quad
    V_{a,b,k}^2=\left(\frac{1}{\rho^2}+N_{a,b,k}\right)^{-1},
    \]
    where $D_{a,b,k} = \{(i,j):i<j,\xi_{i,k}=a,\xi_{j,k}=b\}$ and $N_{a,b,k} =|D_{a,b,k}|$.
    
    \item $\xi_{i,k}\mid \cdot \sim \textsf{Cat}\big(\pi_{i,k,c}\big)$, with
    \[
    \log \pi_{c,i,k}=\log \omega_{c,k}
    -\frac{1}{2}\sum_{j\ne i}\!\Big(z_{i,j,k}-\zeta-\mu_k-\delta_{i,k}-\delta_{j,k}-\boldsymbol{x}_{i,j}^\top\boldsymbol{\beta}_k-\gamma_{\phi(c,\xi_{j,k}),k}\Big)^2,
    \]
    for each $c\in\{1,\ldots,C\}$.
    
    \item $\boldsymbol{\omega}_k\mid \cdot \sim \textsf{Dir}\big(\tfrac{\alpha}{C}+n_{k,1},\ldots,\tfrac{\alpha}{C}+n_{k,C}\big)$, where $n_{k,c}=\sum_{i}I\{\xi_{i,k}=c\}$.
    
    \item $\alpha\mid \cdot$ using an auxiliary variable $\eta$ as in \cite{escobar1995bayesian}, which consists in sampling $\eta\sim \textsf{Beta}(\alpha+1,n_\bullet)$, and then, sampling $\alpha$ from
    \[
    \alpha\mid \cdot \sim
    \begin{cases}
    \textsf{G}(a_\alpha+m_\bullet,\ b_\alpha-\log\eta) & \text{w.p. }\pi,\\[2pt]
    \textsf{G}(a_\alpha+m_\bullet-1,\ b_\alpha-\log\eta) & \text{w.p. }1-\pi,
    \end{cases}
    \]
    where $n_\bullet=nK$, $m_\bullet=\sum_{k} \sum_{c} I\{n_{k,c}>0\}$, and $\pi = (a_\alpha+m_\bullet-1)/(a_\alpha+m_\bullet-1 + n_\bullet(b_\alpha-\log\eta))$.

    \item $\omega^2\mid\cdot$ is identical to that in the \textsf{SMN} model.

    \item $\sigma^2\mid\cdot$ is identical to that in the \textsf{SMN} model.

    \item $\tau^2\mid\cdot$ is identical to that in the \textsf{SMN} model.

    \item $\kappa^2\mid\cdot$ is identical to that in the \textsf{SMN} model.

    \item $\varsigma^2\mid\cdot$ is identical to that in the \textsf{SMN-C} model.

    \item $\rho^2\mid \cdot \sim \textsf{IG}(A,B)$, with
    \[
    A=a_\rho+\frac{K\,C(C+1)}{4},\qquad
    B=b_\rho+\frac{1}{2}\sum_{k}\sum_{a\le b}\gamma_{a,b,k}^2.
    \]
\end{itemize}

\section{The Big 4 data revisited}\label{sec_illustration}

In this section, we revisit the Big~4 multilayer network and evaluate the 
empirical performance of the models introduced in Section~\ref{sec_models}. 
We first compare all specifications in terms of goodness-of-fit and predictive 
accuracy, and then present a detailed analysis of the data under the 
best-performing model.

\subsection{Model comparison}

To compare the competing models in terms of their ability to reproduce key structural features of the observed networks, we use posterior predictive checks (e.g., \citealt{gelman2014bayesian}). For each dataset–model combination, we generate synthetic multilayer networks at every iteration of the MCMC algorithm by sampling from the posterior predictive distribution, using the parameter values drawn at that iteration. At each iteration, we then compute a set of network summary statistics on each layer of the simulated multilayer networks, specifically density, global transitivity, degree assortativity, mean degree, standard deviation of degree, mean geodesic distance, and diameter. This procedure yields, for each statistic, model, and layer, an empirical approximation to its posterior predictive distribution.

We summarize the posterior predictive distribution of each statistic by its posterior mean and compare this mean to the corresponding value computed on the observed layer of the multilayer network. For each layer, we compute the root mean squared error (RMSE) between the posterior mean of the statistic and its observed value, and then average these layer-specific RMSEs to obtain a single measure of discrepancy for each statistic–model combination. Table~\ref{tab:ppc_rmse_metal_bands} reports these average RMSE values for all models and bands. Smaller values indicate that the model is better able to reproduce the corresponding network feature, thereby providing a basis for model comparison that is directly tied to the structural properties of the networks under study.

\begin{table}[!htb]
\centering
\scriptsize
\begin{tabularx}{0.9\textwidth}{l *{7}{>{\centering\arraybackslash}X}}
\toprule
Model & Dens. & Trans. & Assor. & M. Deg. & SD Deg. & M. Geo. & Diam. \\
\midrule
\multicolumn{8}{c}{\textsf{METALLICA}} \\
\midrule
\textsf{SMN}      & 0.001 & 0.069 & 0.131 & 0.093          & 0.412          & 0.772          & 2.083          \\
\textsf{SMN-C}    & 0.001 & 0.067 & 0.132 & \textbf{0.088} & 0.413          & 0.769          & 2.070          \\
\textsf{SMN-C-BG} & 0.001 & 0.037 & \textbf{0.096} & 0.089          & 0.433          & 0.674          & 1.914          \\
\textsf{SMN-C-LD} & 0.001 & 0.036 & 0.124 & 0.094          & 0.373          & 0.620          & 1.678          \\
\textsf{SMN-C-SB} & 0.001 & \textbf{0.015} & 0.110 & 0.102          & \textbf{0.289} & \textbf{0.355} & \textbf{1.017} \\
\midrule
\multicolumn{8}{c}{\textsf{SLAYER}} \\
\midrule
\textsf{SMN}      & 0.001 & 0.067 & 0.143 & 0.100          & 0.403          & 0.997          & 2.008          \\
\textsf{SMN-C}    & 0.001 & 0.066 & 0.146 & 0.100          & 0.393          & 0.990          & 1.962          \\
\textsf{SMN-C-BG} & 0.001 & 0.048 & \textbf{0.110} & \textbf{0.094} & 0.344          & 0.892          & 1.821          \\
\textsf{SMN-C-LD} & 0.001 & 0.047 & 0.137 & 0.100          & 0.360          & 0.898          & 1.825          \\
\textsf{SMN-C-SB} & 0.001 & \textbf{0.023} & 0.116 & 0.097          & \textbf{0.315} & \textbf{0.555} & \textbf{1.025} \\
\midrule
\multicolumn{8}{c}{\textsf{MEGADETH}} \\
\midrule
\textsf{SMN}      & 0.000 & 0.077 & 0.118 & 0.086          & 0.434          & 0.989          & 2.241          \\
\textsf{SMN-C}    & 0.000 & 0.076 & 0.118 & \textbf{0.082} & 0.442          & 0.995          & 2.238          \\
\textsf{SMN-C-BG} & 0.000 & 0.079 & 0.112 & 0.085          & 0.431          & 0.991          & 2.232          \\
\textsf{SMN-C-LD} & 0.001 & 0.057 & 0.110 & 0.099          & 0.492          & 0.867          & 1.801          \\
\textsf{SMN-C-SB} & 0.000 & \textbf{0.008} & \textbf{0.092} & 0.085          & \textbf{0.323} & \textbf{0.475} & \textbf{0.940} \\
\midrule
\multicolumn{8}{c}{\textsf{ANTHRAX}} \\
\midrule
\textsf{SMN}      & 0.001 & 0.062 & 0.144 & \textbf{0.090} & 0.449          & 0.902          & 1.676          \\
\textsf{SMN-C}    & 0.001 & 0.061 & 0.143 & 0.101          & 0.475          & 0.913          & 1.705          \\
\textsf{SMN-C-BG} & 0.001 & 0.041 & \textbf{0.119} & 0.097          & 0.495          & 0.827          & 1.506          \\
\textsf{SMN-C-LD} & 0.001 & 0.039 & 0.159 & 0.107          & 0.515          & 0.788          & 1.375          \\
\textsf{SMN-C-SB} & 0.001 & \textbf{0.017} & 0.138 & 0.093          & \textbf{0.374} & \textbf{0.504} & \textbf{0.726} \\
\bottomrule
\end{tabularx}
\caption{Mean RMSE for posterior predictive network statistics across models and bands. Columns report density (Dens.), global transitivity (Trans.), degree assortativity (Assor.), mean degree (M. Deg.), standard deviation of degree (SD Deg.), mean geodesic distance (M. Geo.), and diameter (Diam.).}
\label{tab:ppc_rmse_metal_bands}
\end{table}

Table~\ref{tab:ppc_rmse_metal_bands} shows that the baseline \textsf{SMN} and its covariate extension \textsf{SMN-C} exhibit the largest discrepancies between posterior predictive summaries and observed network statistics, particularly for global transitivity, degree assortativity, and path-based measures such as mean geodesic distance and diameter. Introducing additional latent structure systematically reduces these RMSEs across all four bands, indicating an improved ability to reproduce higher-order network features beyond overall density and mean degree, for which all models perform similarly. Among the five specifications, \textsf{SMN-C-SB} consistently attains the smallest RMSEs for most statistics and bands, especially for measures related to degree variability and geodesic distances, suggesting that this model provides the closest match to the observed multilayer network structure and is therefore the most adequate in terms of posterior predictive fit.

Beyond assessing the models' ability to reproduce fundamental structural features of the multilayer networks, we now compare them in terms of predictive performance using several standard metrics, including the area under the ROC curve (AUC), the Brier score (BS), and the log-loss (LL); see, for example, \cite{fawcett2006introduction} and \citealt{gneiting2007strictly}. In addition, we consider the deviance information criterion (DIC) and the Watanabe--Akaike information criterion (WAIC); see, for example, \cite{spiegelhalter2002bayesian}, \cite{watanabe2010asymptotic}, \cite{spiegelhalter2014deviance}, and \cite{gelman2014understanding}. The AUC measures the ability of the model to discriminate between edges and non-edges, with values closer to one indicating better discrimination. The Brier score is the mean squared error between predicted edge probabilities and observed outcomes, with smaller values corresponding to better-calibrated predictions. The log-loss is the average negative log-likelihood of the observed edges given the predicted probabilities, and thus penalizes overconfident wrong predictions, with smaller values indicating better predictive accuracy.
The information criteria aim to provide scalar summaries of out-of-sample predictive performance that trade off goodness of fit and model complexity, with smaller values indicating models that are expected to generalize better to new data.

To compute AUC, BS, and LL, we first obtain the interaction probabilities at each layer and each MCMC iteration from the posterior draws, then compute the corresponding metric at every iteration and layer, and thereby obtain the posterior distribution of each metric for each layer. We then summarize these layer-specific distributions by their posterior means and average the resulting values across layers to obtain the values reported in Table~\ref{tab:model_metrics_metal_bands}. The DIC and WAIC are computed following their standard definitions, with DIC based on the difference between the posterior mean deviance and the deviance evaluated at a point estimate (typically the posterior mean), and WAIC obtained from the sum of pointwise log posterior predictive densities combined with a variance-based penalty that defines an effective number of parameters.

\begin{table}[!ht]
\centering
\scriptsize
\begin{tabularx}{0.9\textwidth}{l *{5}{>{\centering\arraybackslash}X}}
\toprule
Model & AUC & BS & LL & DIC & WAIC \\
\midrule
\multicolumn{6}{c}{\textsf{METALLICA}} \\
\midrule
\textsf{SMN}      & 0.715 & 0.025 & 0.115 & 8757.926 & 8786.659 \\
\textsf{SMN-C}    & 0.733 & 0.025 & 0.113 & 8667.141 & 8699.057 \\
\textsf{SMN-C-BG} & 0.837 & 0.023 & 0.096 & 7734.479 & 7750.032 \\
\textsf{SMN-C-LD} & 0.847 & \textbf{0.022} & 0.093 & 7460.323 & 7626.480 \\
\textsf{SMN-C-SB} & \textbf{0.919} & \textbf{0.022} & \textbf{0.082} & \textbf{6494.318} & \textbf{6669.791} \\
\midrule
\multicolumn{6}{c}{\textsf{SLAYER}} \\
\midrule
\textsf{SMN}      & 0.714 & 0.030 & 0.130 & 6969.393 & 6995.004 \\
\textsf{SMN-C}    & 0.721 & 0.029 & 0.129 & 6942.133 & 6973.644 \\
\textsf{SMN-C-BG} & 0.838 & \textbf{0.026} & 0.108 & 6153.818 & 6184.103 \\
\textsf{SMN-C-LD} & 0.847 & \textbf{0.026} & 0.106 & 5995.067 & 6134.855 \\
\textsf{SMN-C-SB} & \textbf{0.897} & \textbf{0.026} & \textbf{0.098} & \textbf{5486.245} & \textbf{5645.787} \\
\midrule
\multicolumn{6}{c}{\textsf{MEGADETH}} \\
\midrule
\textsf{SMN}      & 0.722 & 0.020 & 0.095 & 11722.663 & 11754.791 \\
\textsf{SMN-C}    & 0.724 & 0.020 & 0.095 & 11736.792 & 11774.724 \\
\textsf{SMN-C-BG} & 0.761 & 0.020 & 0.092 & 11672.784 & 11723.345 \\
\textsf{SMN-C-LD} & 0.845 & \textbf{0.018} & 0.078 & 10082.736 & 10286.486 \\
\textsf{SMN-C-SB} & \textbf{0.924} & \textbf{0.018} & \textbf{0.069} & \textbf{8716.262} & \textbf{8902.372} \\
\midrule
\multicolumn{6}{c}{\textsf{ANTHRAX}} \\
\midrule
\textsf{SMN}      & 0.695 & 0.027 & 0.123 & 7667.040 & 7695.918 \\
\textsf{SMN-C}    & 0.705 & 0.027 & 0.122 & 7638.994 & 7671.187 \\
\textsf{SMN-C-BG} & 0.805 & 0.025 & 0.105 & 6886.220 & 6917.169 \\
\textsf{SMN-C-LD} & 0.816 & \textbf{0.024} & 0.101 & 6646.713 & 6787.791 \\
\textsf{SMN-C-SB} & \textbf{0.895} & \textbf{0.024} & \textbf{0.093} & \textbf{6086.639} & \textbf{6293.337} \\
\bottomrule
\end{tabularx}
\caption{Predictive performance and information criteria across models and bands. 
Columns report mean area under the ROC curve (AUC), Brier score (BS), log-loss (LL), 
deviance information criterion (DIC), and Watanabe--Akaike information criterion (WAIC).}
\label{tab:model_metrics_metal_bands}
\end{table}

Table~\ref{tab:model_metrics_metal_bands} shows a clear and consistent pattern across the four bands. For all datasets, the latent-structure extensions improve upon the baseline \textsf{SMN} and the covariate-only \textsf{SMN-C} in terms of higher AUC and lower BS and LL, indicating more accurate and better-calibrated edge probability predictions. The gains are especially marked when moving to \textsf{SMN-C-LD} and \textsf{SMN-C-SB}, which attain the largest AUC values (often above 0.90) and the smallest BS and LL across bands. A similar ranking is reflected in the DIC and WAIC values, where \textsf{SMN-C-SB} consistently achieves the lowest scores, with substantial reductions relative to \textsf{SMN} and \textsf{SMN-C}, particularly for the Megadeth and Anthrax multilayer networks. Overall, the predictive criteria uniformly favor \textsf{SMN-C-SB}, reinforcing the conclusions drawn from the posterior predictive RMSE analysis and suggesting that this specification offers the best compromise between goodness of fit and model complexity among the candidates considered.

\subsection{Data analysis}

Here, we provide an exhaustive analysis of the Big 4 dataset using the \textsf{SMN-C-SB} model, which is consistently the most appealing specification in terms of goodness of fit and out-of-sample predictive performance.

\subsubsection{Baseline effects}

\begin{figure}[!htb]
    \centering
    \subfigure[Posterior $\zeta+\mu_1$]  {\includegraphics[scale=0.28]{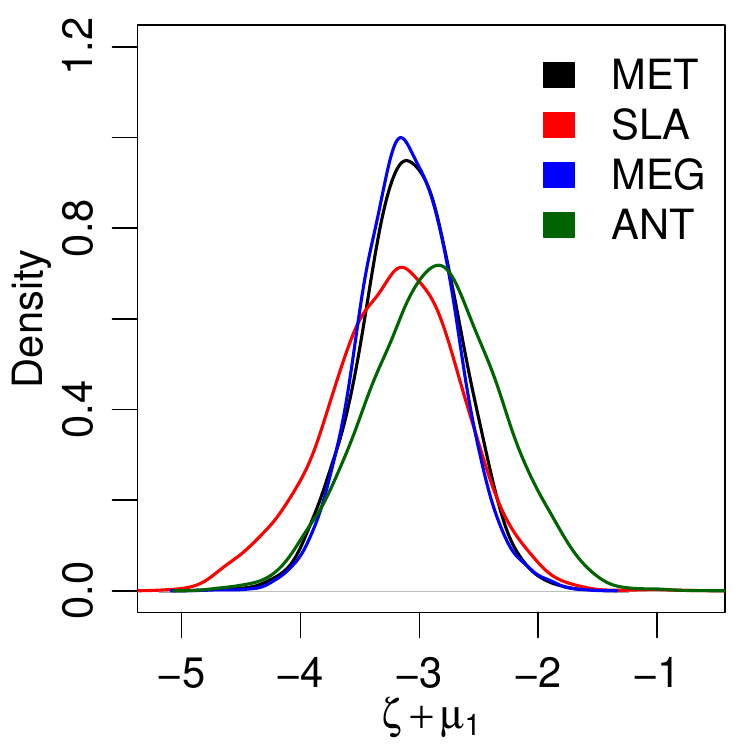}}
    \subfigure[Posterior $\zeta+\mu_2$]{\includegraphics[scale=0.28]{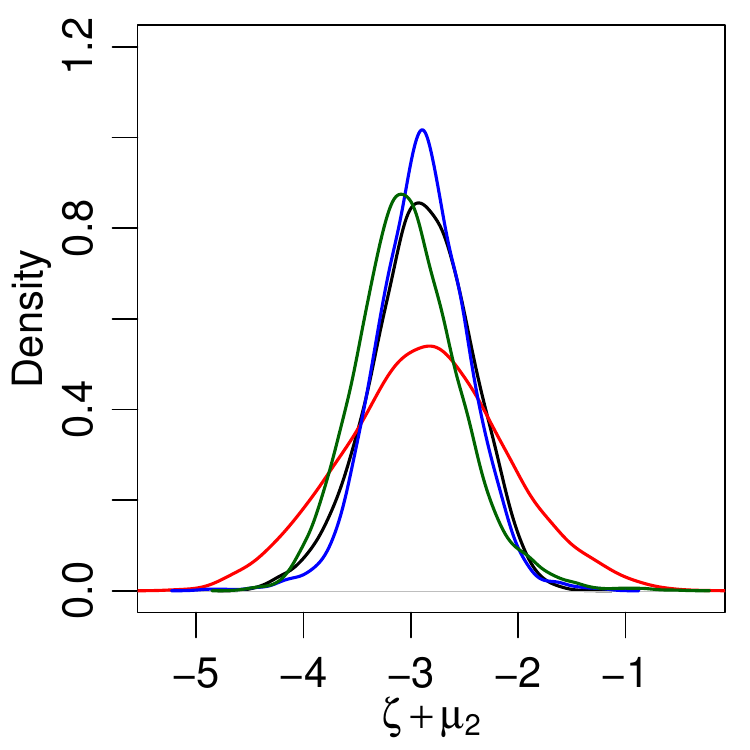}}
    \subfigure[Posterior $\zeta+\mu_3$]  {\includegraphics[scale=0.28]{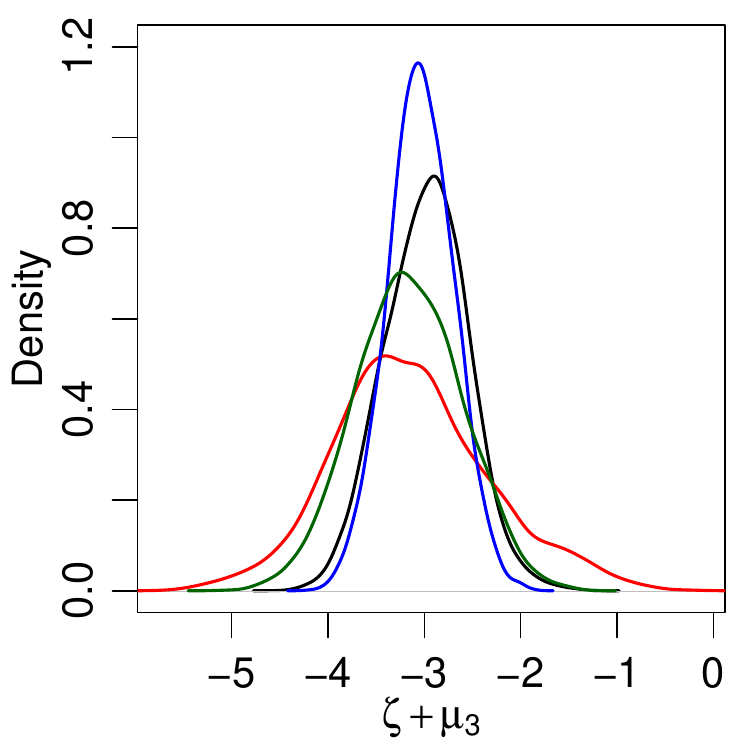}}
    \subfigure[Posterior $\zeta+\mu_4$]    {\includegraphics[scale=0.28]{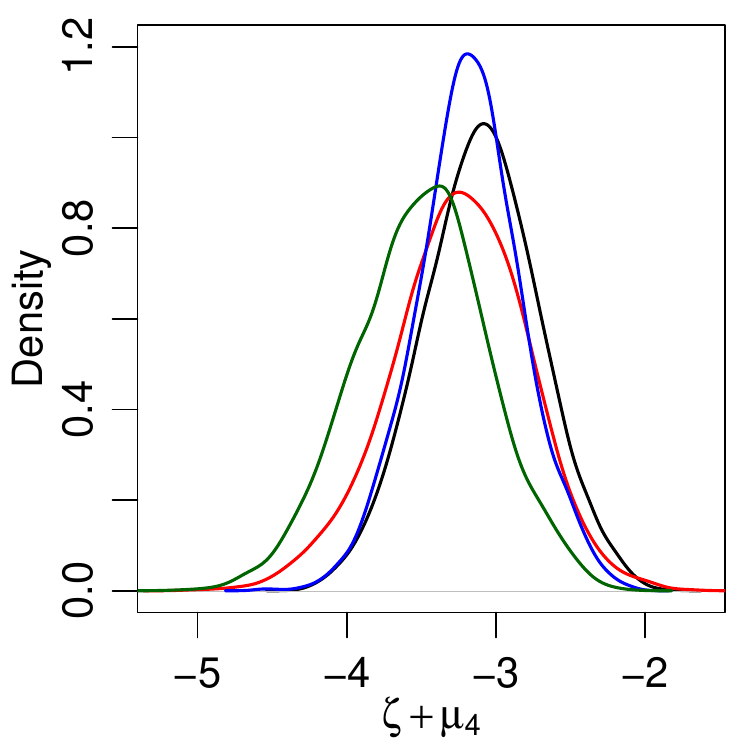}}
    \caption{Posterior inference on the baseline effects $\zeta + \mu_k$ for each layer and band under the \textsf{SMN-C-SB} model. Colors represent bands: Metallica (MET) in black, Slayer (SLA) in red, Megadeth (MEG) in blue, and Anthrax (ANT) in green.}
    \label{fig_mu}
\end{figure}

A joint examination of the posterior summaries for the band- and layer-specific baseline effects $\zeta + \mu_k$, displayed in Figure~\ref{fig_mu}, indicates that all four layers exhibit strongly negative values, with posterior means concentrated around $-3$ and 95\% credible intervals that substantially overlap across both bands and layers. On the probit scale, such values correspond to very low baseline probabilities of connection between two songs when sociability effects, covariates, and community structure are set to their reference levels, indicating that song-to-song ties are rare unless additional structured effects increase the linear predictor (and hence the edge probability). This pattern is consistent with the typical sparsity of song similarity networks, in which only a small fraction of song pairs are sufficiently alike to form an observed connection. The high degree of overlap between credible intervals across Metallica, Slayer, Megadeth, and Anthrax suggests that overall sparsity is broadly comparable across bands, so that differences in connectivity patterns are driven primarily by the structured components of the model rather than by systematic shifts in the baseline propensity to connect. The mild deviations observed for specific cases (for instance, slightly less negative values for Anthrax in one layer and more negative values in another) point to modest layer-specific variation in baseline density, but do not alter the overall conclusion that the four bands share a similar low-connectivity regime at the baseline level.

\begin{figure}[!htb]
	\centering
	\setlength{\tabcolsep}{0pt}
	\begin{tabular}{ccccc}
		& \hspace{0.6cm}\textsf{METALLICA} & \hspace{0.6cm}\textsf{SLAYER} & \hspace{0.6cm}\textsf{MEGADETH} & \hspace{0.6cm}\textsf{ANTHRAX} \\
		\begin{sideways} \hspace{0.9cm} \textbf{Loudness} \end{sideways}               &
		\includegraphics[scale = 0.28]{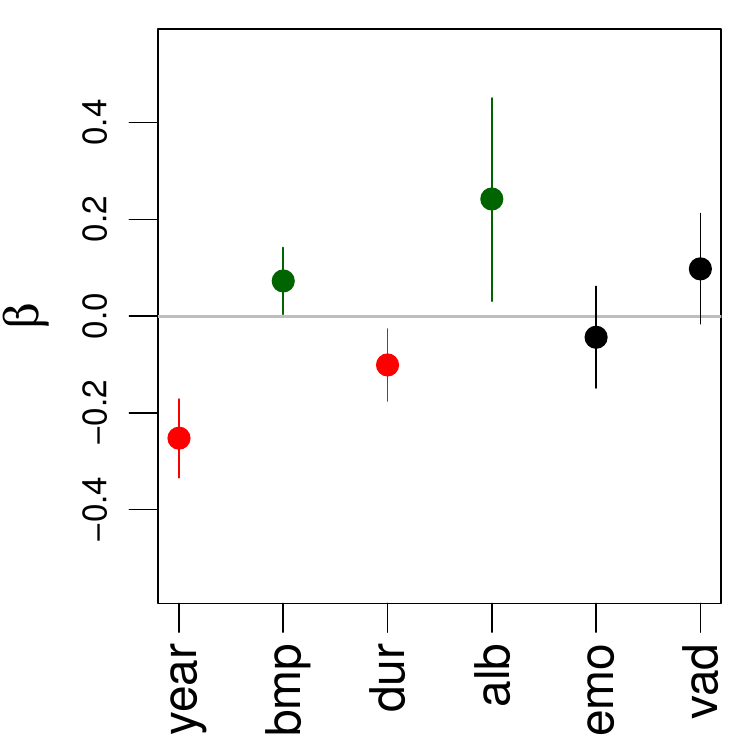}  &
		\includegraphics[scale = 0.28]{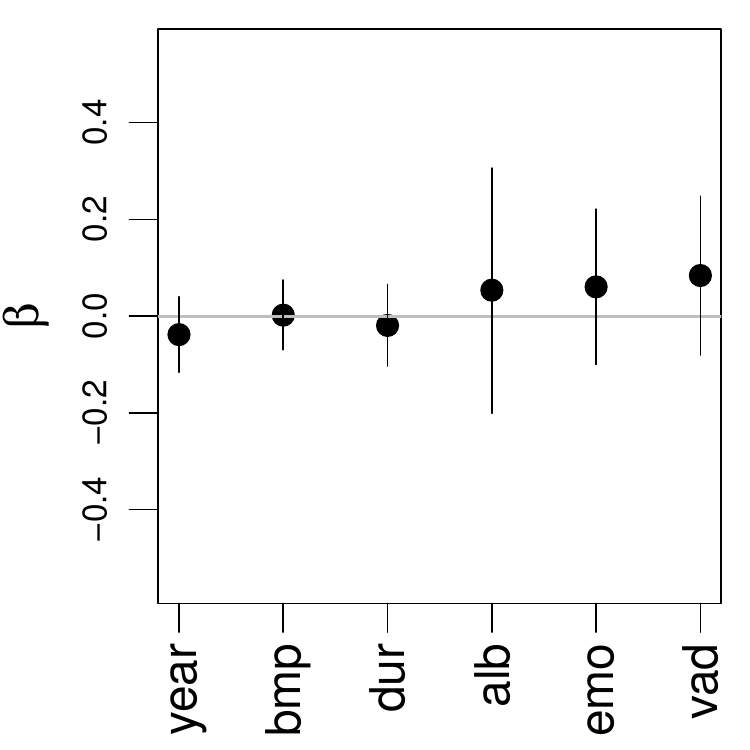}     &
		\includegraphics[scale = 0.28]{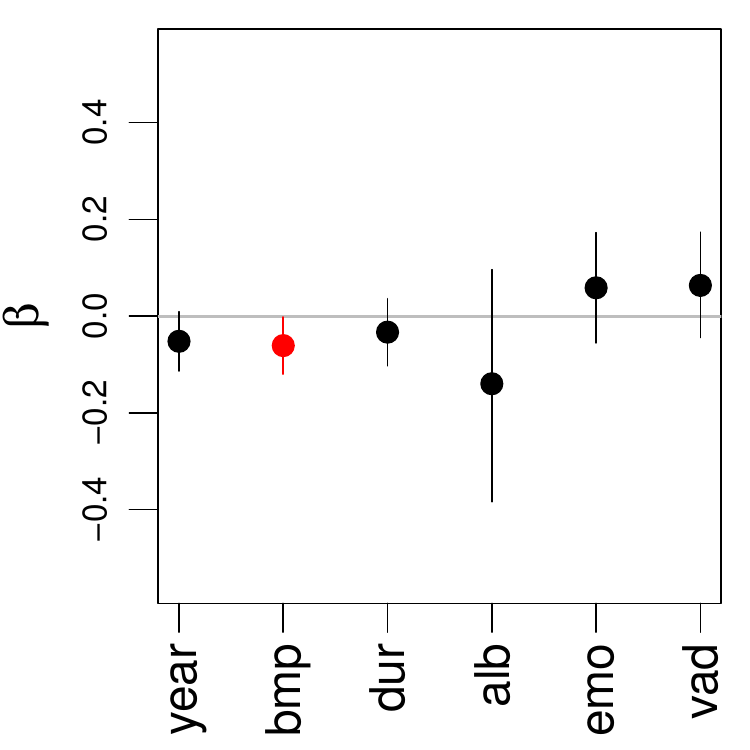}   &
		\includegraphics[scale = 0.28]{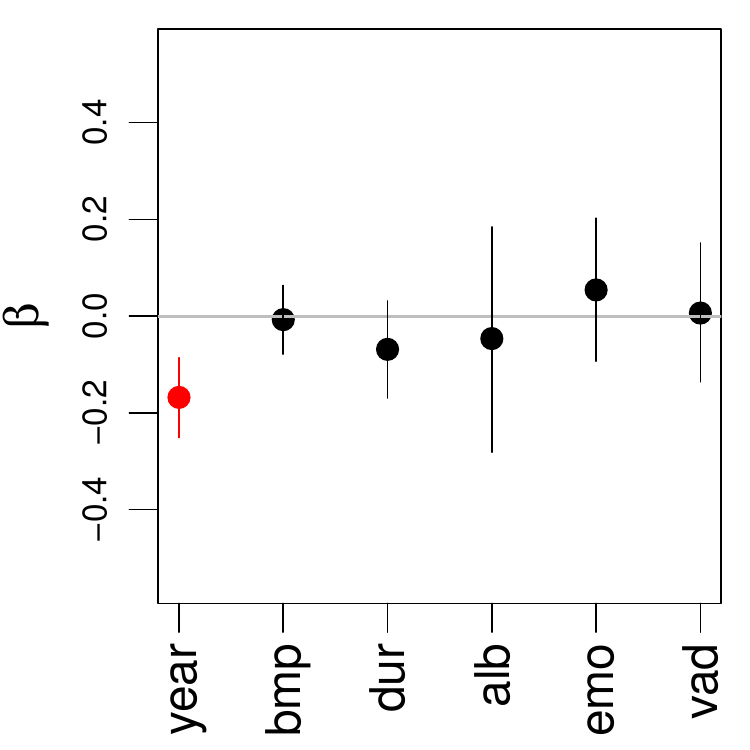}  \\
		\begin{sideways} \hspace{0.9cm} \textbf{Brightness} \end{sideways}             &
		\includegraphics[scale = 0.28]{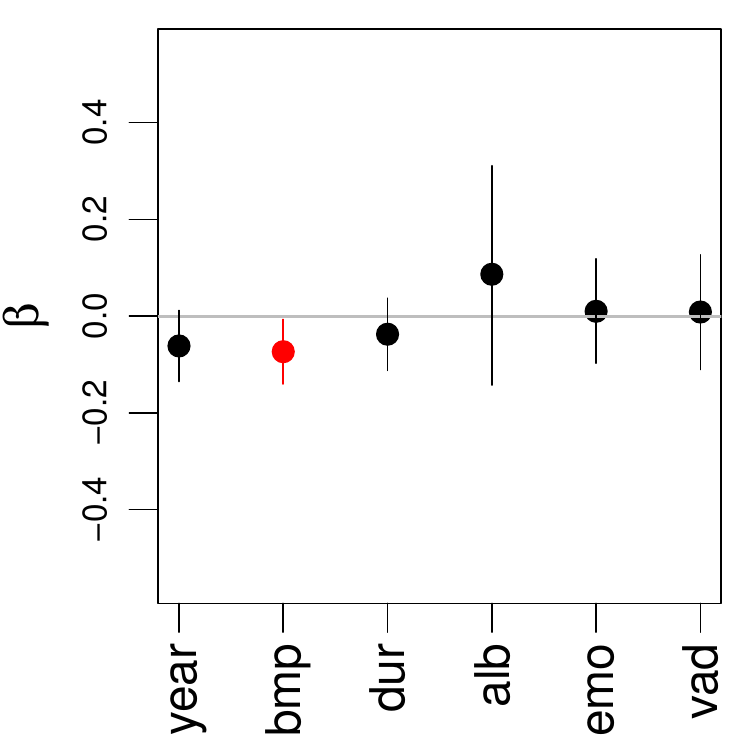}  &
		\includegraphics[scale = 0.28]{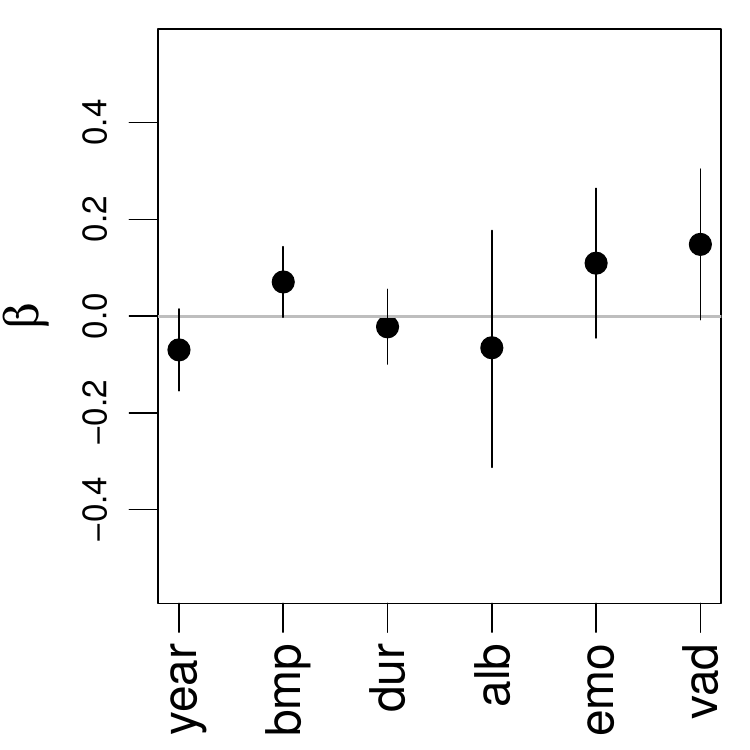}     &
		\includegraphics[scale = 0.28]{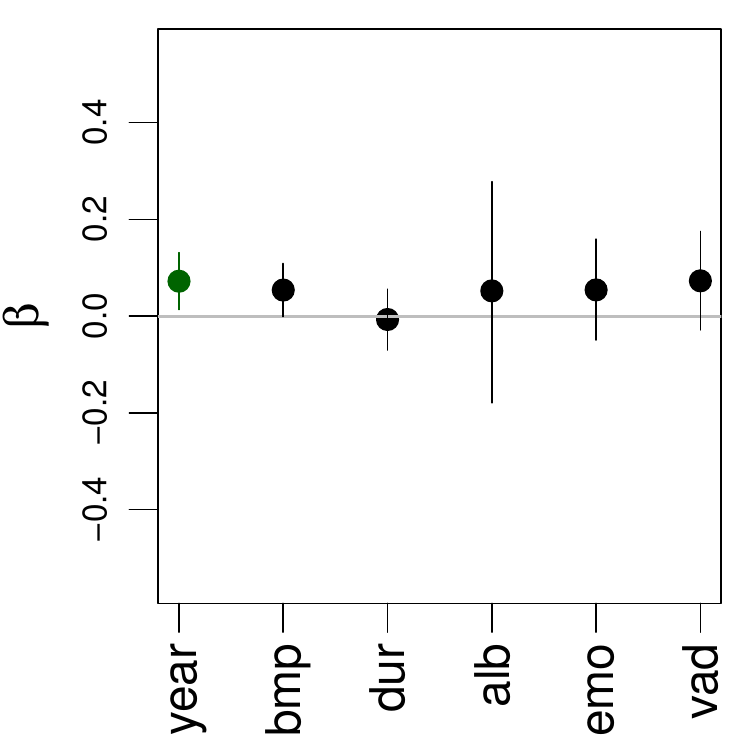}   &
		\includegraphics[scale = 0.28]{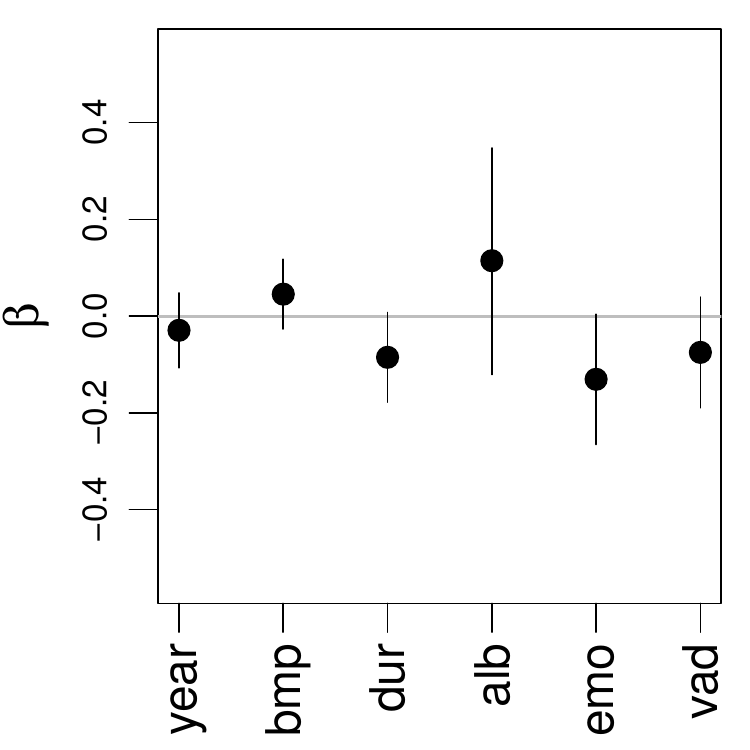}  \\
        \begin{sideways} \hspace{0.9cm} \textbf{Tonality} \end{sideways}               &		\includegraphics[scale = 0.28]{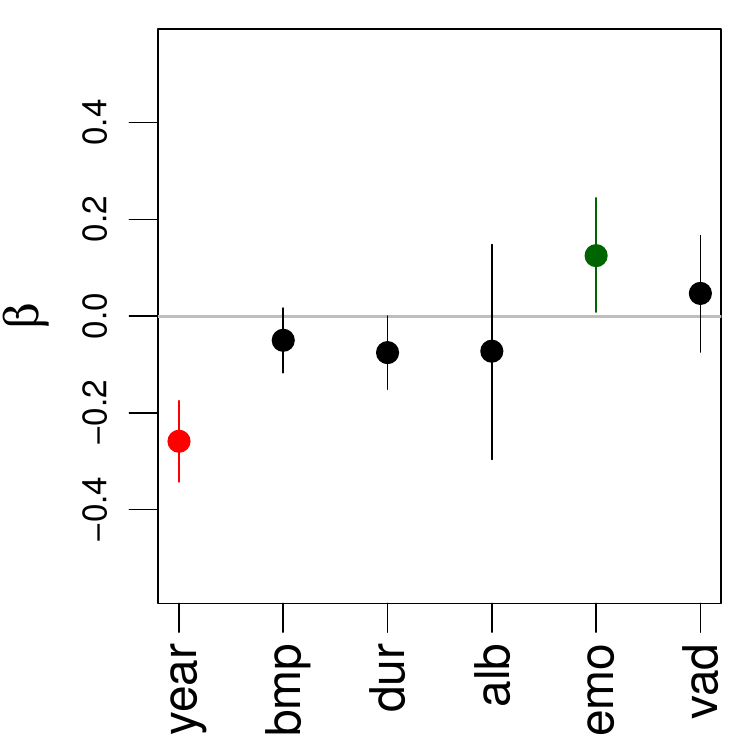}  &
		\includegraphics[scale = 0.28]{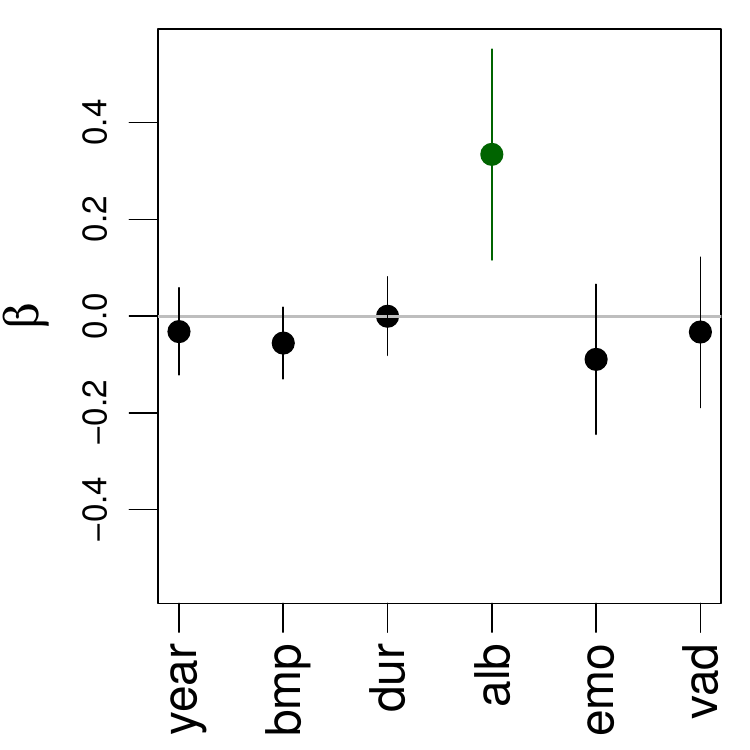}     &
		\includegraphics[scale = 0.28]{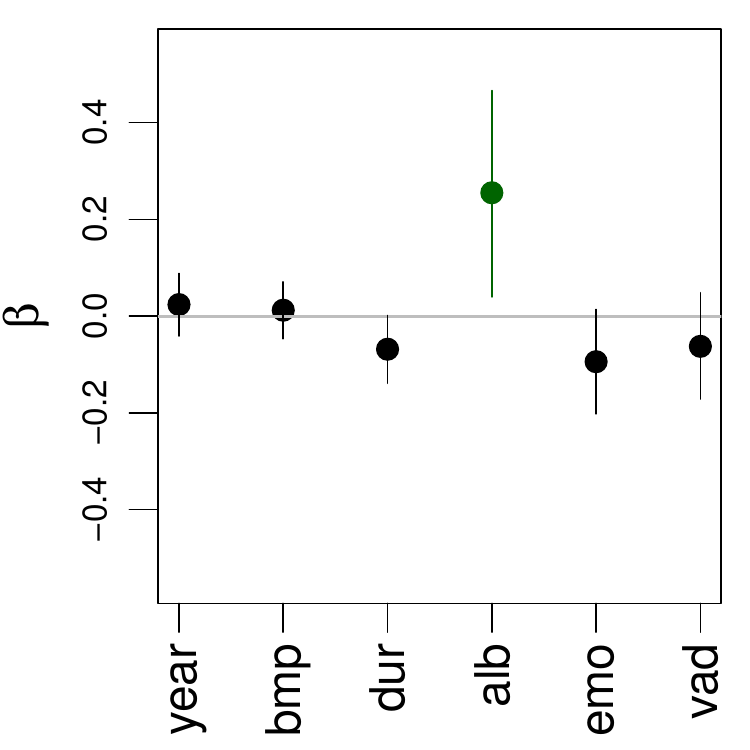}   &
		\includegraphics[scale = 0.28]{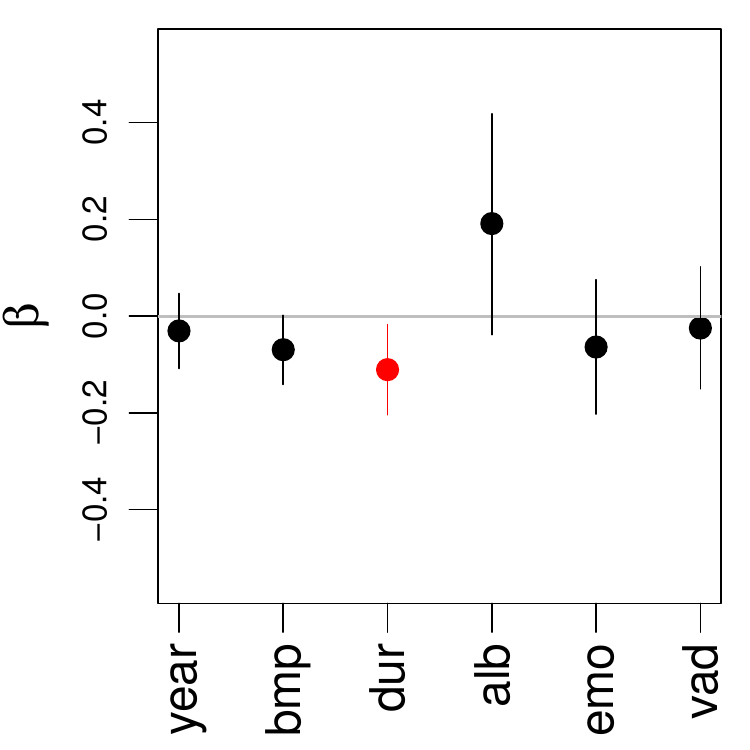}  \\
        \begin{sideways} \hspace{0.9cm} \textbf{Rhythm} \end{sideways}                 &		\includegraphics[scale = 0.28]{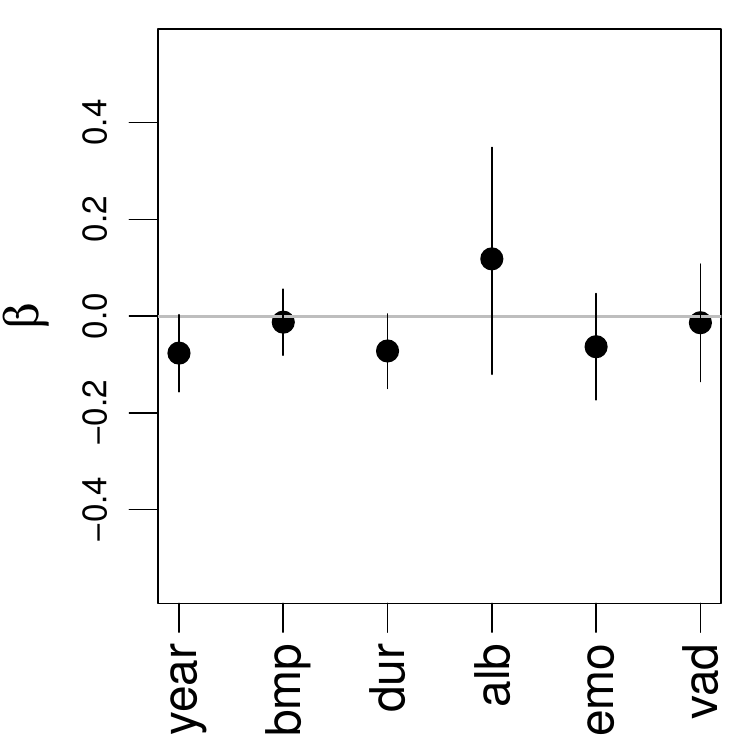}  &
		\includegraphics[scale = 0.28]{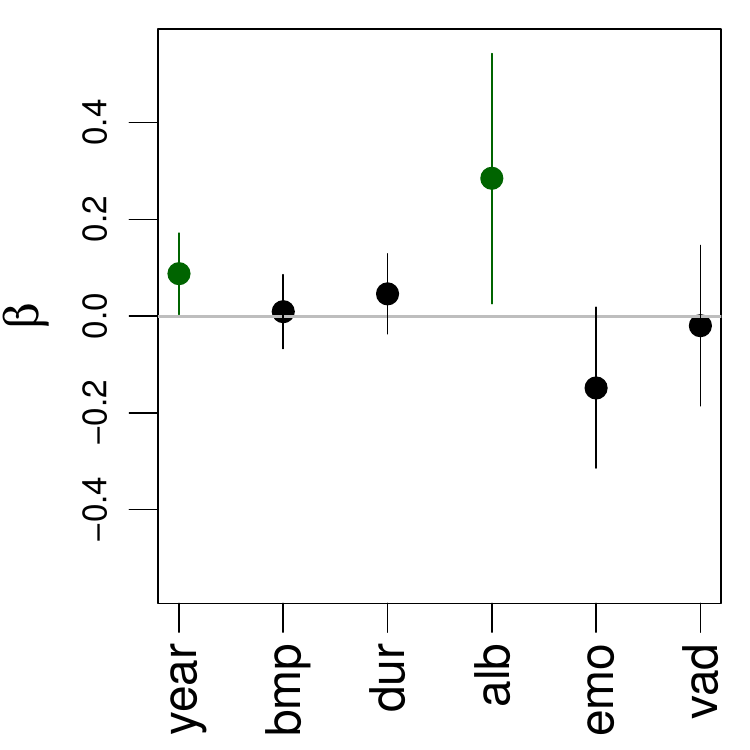}     &
		\includegraphics[scale = 0.28]{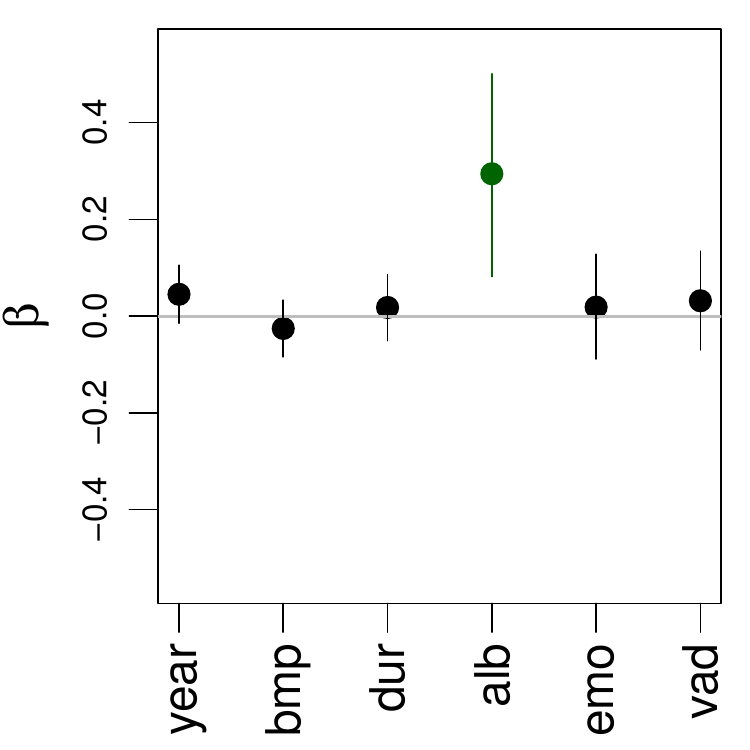}   &
		\includegraphics[scale = 0.28]{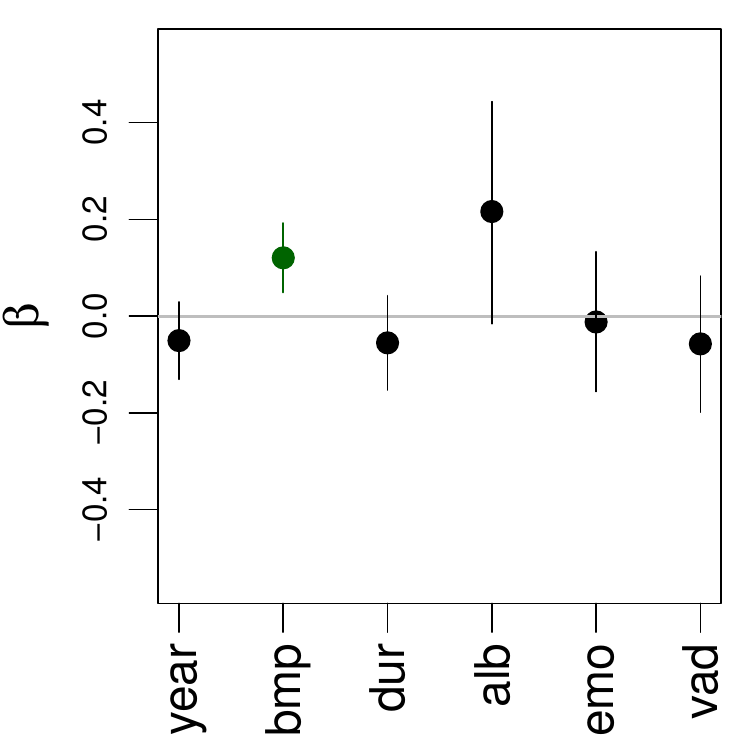}  \\
	\end{tabular}
	\caption{Posterior inference on regression coefficients for the \textsf{SMN-C-SB} model in the Big 4 data. Columns correspond to bands (Metallica, Slayer, Megadeth, Anthrax) and rows to audio layers (Loudness, Brightness, Tonality, Rhythm). Within each panel, points represent posterior means and horizontal bars 90\% credible intervals for the regression coefficients associated with the covariates year (\texttt{year}), BPM (\texttt{bmp}), duration (\texttt{dur}), album (\texttt{alb}), emotion (\texttt{emo}), and VAD (\texttt{vad}). Coefficients whose 90\% credible intervals include zero are shown in black, whereas those with intervals entirely below or above zero are highlighted in red and green, respectively.}
	\label{fig_betas}
\end{figure}

\subsubsection{Regression coefficients}

Across bands and layers, the posterior summaries shown in Figure~\ref{fig_betas} indicate that only a subset of covariates have clearly identifiable effects on song–to–song connectivity, and these effects are generally modest in magnitude. For the distance-type covariates (year, BMP, duration), negative coefficients imply that edges are more likely between songs that are similar on the corresponding attribute, whereas positive coefficients favor pairs that are more dissimilar. In the Loudness layer for Metallica and Anthrax, the negative coefficients on year (and on duration for Metallica) suggest that loudness-based similarity links tend to form between songs released around the same time and of comparable duration, consistent with album- or era-specific production practices. For Metallica, a positive coefficient on the same-album indicator in Loudness and a positive coefficient on emotional similarity in Tonality indicate additional within-album and emotion-based clustering in those layers. By contrast, Slayer and Megadeth show fewer strong covariate effects overall, with only isolated significant coefficients. For example, positive album effects in the Tonality and Rhythm layers for Slayer and Megadeth, and a positive year effect in the Rhythm layer for Slayer, hinting at some cross-era rhythmic ties once other structure is accounted for.

More broadly, the same-album indicator emerges as the most recurrently important covariate, with positive and significant effects in several Tonality and Rhythm layers, pointing to a systematic tendency for within-album songs to be connected beyond what is captured by the latent sociality and block structure. Tempo and duration differences also play a role, although their influence varies by band and layer. For instance, negative BMP effects in Metallica’s Brightness layer and Megadeth’s Loudness layer indicate that songs with similar tempo are more likely to be connected in those feature spaces, while Anthrax’s Tonality layer shows a negative effect of duration, favoring songs of comparable length. By contrast, the textual covariates (emotion and VAD) rarely yield credible intervals that exclude zero, with the exception of Metallica’s Tonality layer where emotional similarity has a positive effect. Overall, these results suggest that, once the rich latent structure is accounted for, covariates mainly provide nuanced refinements to connectivity patterns, for example by highlighting clustering within albums, temporal proximity, and, more sporadically, the influence of tempo, duration, and emotional similarity, rather than dominating the formation of edges in the multilayer similarity networks.

\subsubsection{Variance components}

\begin{figure}[!htb]
    \centering
    \subfigure[Posterior $\omega^2$]  {\includegraphics[scale=0.22]{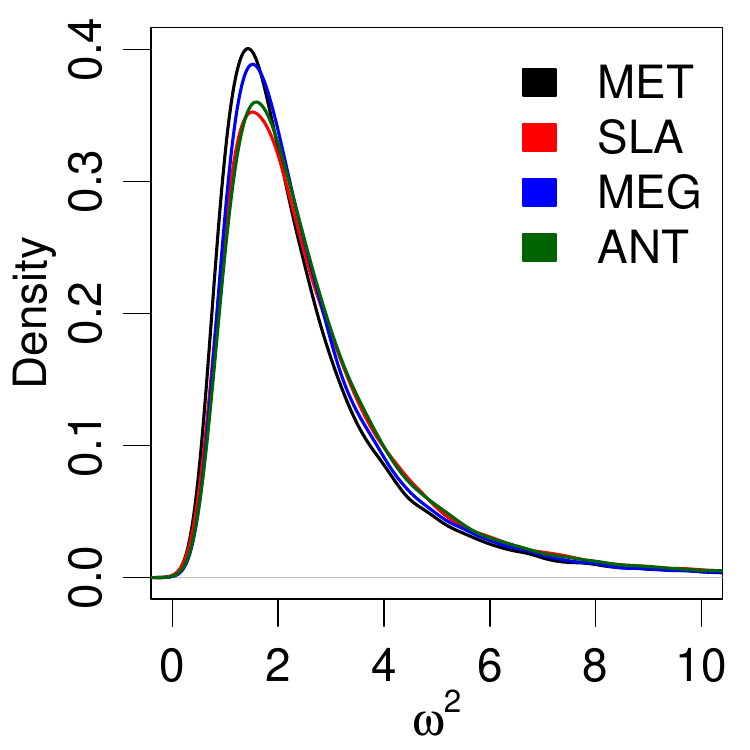}}
    \subfigure[Posterior $\sigma^2$]{\includegraphics[scale=0.22]{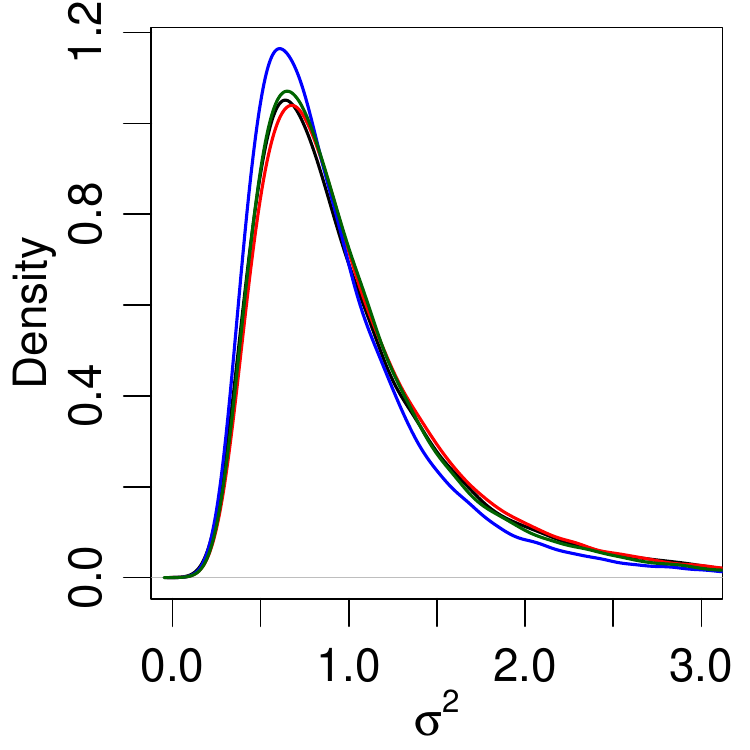}}
    \subfigure[Posterior $\tau^2$]  {\includegraphics[scale=0.22]{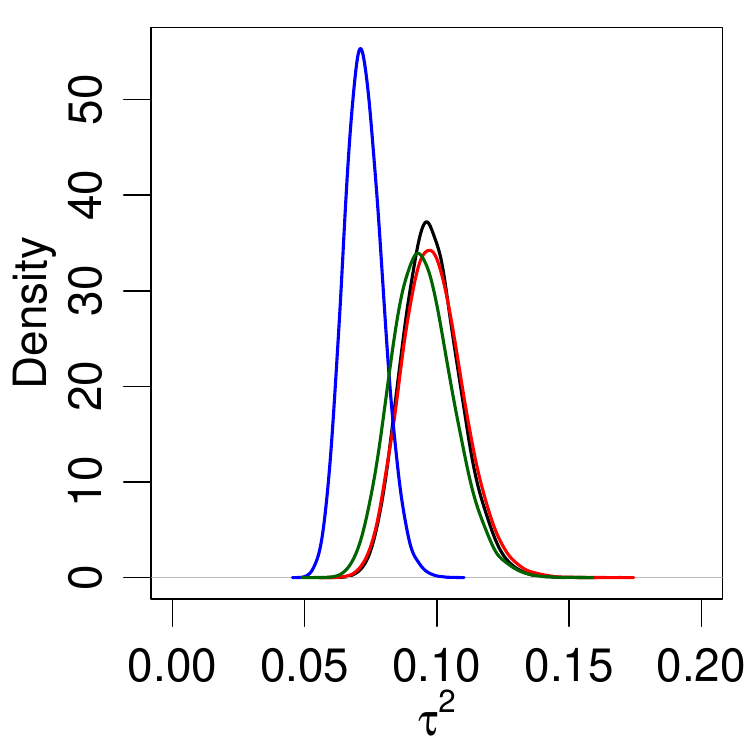}}
    \subfigure[Posterior $\kappa^2$]    {\includegraphics[scale=0.22]{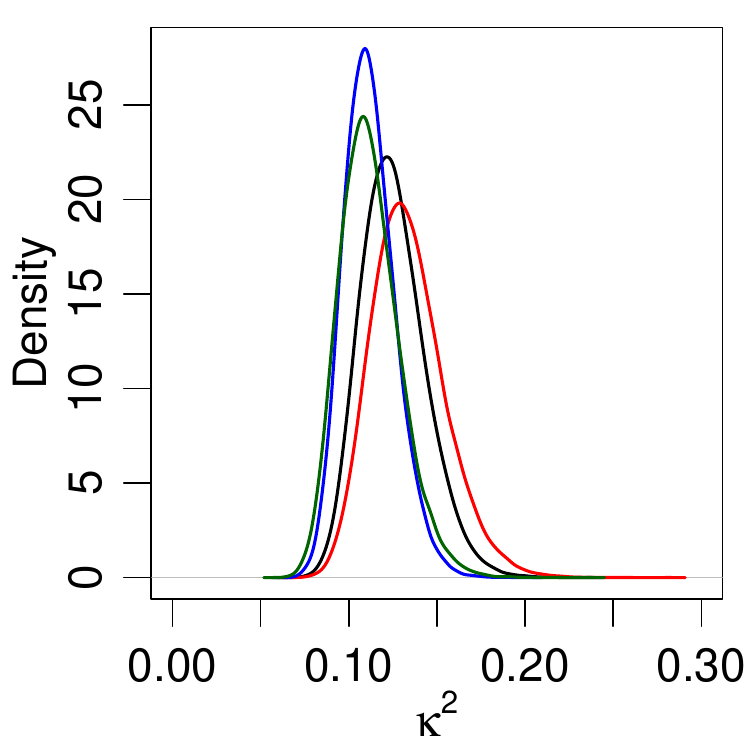}}
    \subfigure[Posterior $\rho^2$]    {\includegraphics[scale=0.22]{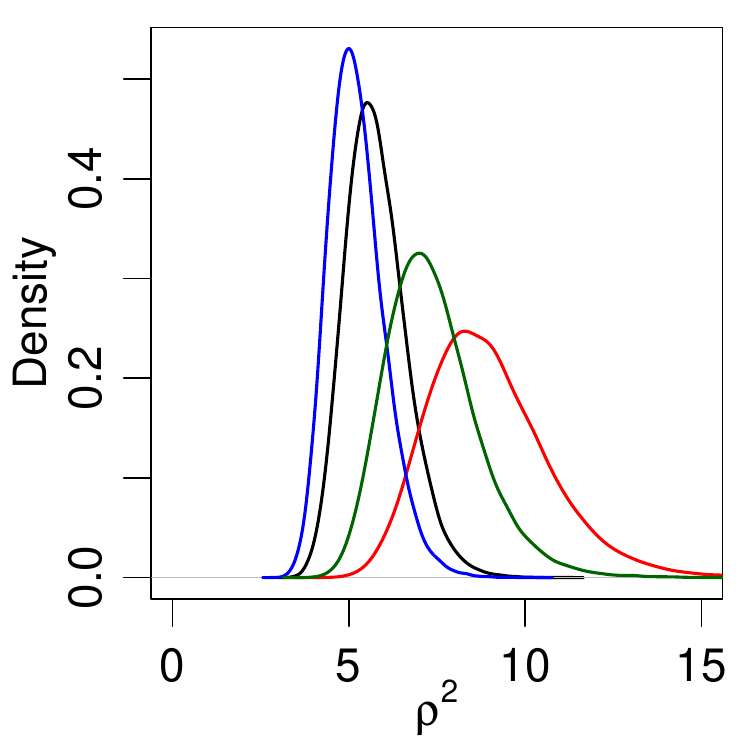}}
    \caption{Posterior inference on the variance components for each layer and band under the \textsf{SMN-C-SB} model. Colors represent bands: Metallica (MET) in black, Slayer (SLA) in red, Megadeth (MEG) in blue, and Anthrax (ANT) in green.}
    \label{fig_var}
\end{figure}

As shown in Figure~\ref{fig_var}, the posterior summaries for the variance components are broadly similar across bands, indicating that the hierarchical priors operate at comparable scales for Metallica, Slayer, Megadeth, and Anthrax. 
The parameter $\omega^2$ controls the variability of the global baseline effect $\zeta$. Posterior means between roughly $2.6$ and $2.9$, suggest a fairly wide, yet still regularized, range of plausible global baselines on the probit scale, with no clear evidence of major between-band differences in this global connectivity level. The variance $\sigma^2$, which governs the dispersion of layer-specific baseline shifts $\mu_k$, has posterior means close to $1$, indicating a moderate degree of heterogeneity in baseline connectivity across layers within each band. Different similarity layers can exhibit distinct baseline densities, but these differences remain tempered by the hierarchical prior.

At the node level, $\tau^2$ controls how much the layer-specific sociability effects $\delta_{i,k}$ vary around the node-specific baseline $\vartheta_i$. Posterior means for $\tau^2$ are modest (compared to other variance components), between roughly $0.07$ and $0.10$, indicating limited but non-negligible layer-to-layer fluctuations in a song's sociability around its baseline $\vartheta_i$. The variance $\kappa^2$ governs the dispersion of the baselines $\vartheta_i$ across songs. Posterior means between roughly $0.11$ and $0.13$, indicate non-negligible differences in overall sociability between songs, of a magnitude comparable to the within-node, across-layer variability captured by $\tau^2$. Together, these values confirm that sociability effects are neither uniform nor highly idiosyncratic, but instead share information across layers and songs. Finally, $\rho^2$ controls the variability of block affinities $\gamma_{a,b,k}$ and hence the strength of community-level effects. Posterior means between roughly $5.2$ and $9.0$ correspond to large dispersion on the probit scale, indicating that block interactions play a major role in explaining connectivity beyond global, layer, and node effects. The somewhat larger values for Slayer and Anthrax suggest particularly pronounced community structure in their multilayer similarity networks.

\subsubsection{Songs with significant mean sociality effects}

To identify the most structurally ``popular'' songs in each band under the \textsf{SMN-C-SB} model, we rank all nodes by the posterior mean of their node-specific baseline sociability parameter $\vartheta_i$ and retain those with the largest and most clearly positive effects. Table~\ref{tab:vartheta_top_songs_extended} reports posterior means and 95\% credible intervals for $\vartheta_i$ for the top songs in each band, with the column \emph{Top} indicating the within-band rank. For Metallica and Megadeth we restrict attention to the ten songs with the highest posterior means, whereas for Slayer and Anthrax we display all songs whose credible intervals provide the strongest evidence of elevated sociability. In this setting, larger values of $\vartheta_i$ correspond to a higher baseline probability that song $i$ forms edges with other songs across layers, even before accounting for layer-specific deviations, covariates, or block effects. Consequently, songs with large positive $\vartheta_i$ are those that tend to be similar to many other tracks in their band's catalogue and can be interpreted as structurally popular, highly connected reference points in the multilayer similarity network.

\begin{table}[!ht]
\centering
\scriptsize
\begin{tabularx}{0.9\textwidth}{c l l c *{3}{>{\centering\arraybackslash}X}}
\toprule
Top & Song & Album & Year & Mean & Lower & Upper \\
\midrule
\multicolumn{7}{c}{\textsf{METALLICA}} \\
\midrule
1  & Turn the Page            & Garage Inc                 & 1998 & 0.823 & 0.479 & 1.167 \\
2  & Battery                  & Master of Puppets         & 1986 & 0.708 & 0.366 & 1.050 \\
3  & Sleepwalk My Life Away   & 72 Seasons                & 2023 & 0.638 & 0.285 & 0.991 \\
4  & That Was Just Your Life  & Death Magnetic            & 2008 & 0.604 & 0.250 & 0.962 \\
5  & Free Speech for the Dumb & Garage Inc                & 1998 & 0.597 & 0.196 & 1.000 \\
6  & Prince Charming          & Reload                    & 1997 & 0.415 & 0.047 & 0.786 \\
7  & Fade to Black            & Ride the Lightning        & 1984 & 0.410 & 0.056 & 0.761 \\
8  & The Small Hours          & Garage Days Re-Revisited  & 1987 & 0.406 & 0.047 & 0.767 \\
9  & Better Than You          & Reload                    & 1997 & 0.392 & 0.035 & 0.752 \\
10 & For Whom the Bell Tolls  & Ride the Lightning        & 1984 & 0.368 & 0.012 & 0.722 \\
\midrule
\multicolumn{7}{c}{\textsf{SLAYER}} \\
\midrule
1 & Read Between the Lies     & South of Heaven           & 1988 & 0.983 & 0.608 & 1.367 \\
2 & Dissident Aggressor       & South of Heaven           & 1988 & 0.694 & 0.330 & 1.060 \\
3 & Hell Awaits               & Hell Awaits               & 1985 & 0.530 & 0.164 & 0.898 \\
4 & World Painted Blood       & World Painted Blood       & 2009 & 0.525 & 0.162 & 0.884 \\
5 & Silent Scream             & South of Heaven           & 1988 & 0.508 & 0.140 & 0.879 \\
6 & Jihad                     & Christ Illusion           & 2006 & 0.425 & 0.053 & 0.793 \\
7 & The Final Command         & Show No Mercy             & 1983 & 0.386 & 0.013 & 0.756 \\
\midrule
\multicolumn{7}{c}{\textsf{MEGADETH}} \\
\midrule
1  & Kingmaker                & Super Collider            & 2013 & 1.020 & 0.709 & 1.329 \\
2  & FFF                      & Cryptic Writings          & 1997 & 0.935 & 0.612 & 1.258 \\
3  & Whose Life Is It Anyways & TH1RT3EN                  & 2011 & 0.832 & 0.507 & 1.146 \\
4  & United Abominations      & United Abominations       & 2007 & 0.594 & 0.277 & 0.913 \\
5  & Don't Turn Your Back     & Super Collider            & 2013 & 0.571 & 0.251 & 0.889 \\
6  & My Last Words            & Peace Sells... but Who's Buying? & 1986 & 0.466 & 0.144 & 0.786 \\
7  & The Threat Is Real       & Dystopia                  & 2016 & 0.447 & 0.133 & 0.764 \\
8  & Conquer or Die           & Dystopia                  & 2016 & 0.400 & 0.031 & 0.765 \\
9  & Silent Scorn             & The World Needs a Hero    & 2001 & 0.385 & 0.015 & 0.754 \\
10 & Trust                    & Cryptic Writings          & 1997 & 0.373 & 0.049 & 0.696 \\
\midrule
\multicolumn{7}{c}{\textsf{ANTHRAX}} \\
\midrule
1 & Contact                   & We've Come for You All    & 2003 & 0.547 & 0.180 & 0.920 \\
2 & Be All, End All           & State of Euphoria         & 1988 & 0.470 & 0.136 & 0.807 \\
3 & Worship Intro             & Worship Music             & 2011 & 0.399 & 0.015 & 0.783 \\
4 & I'm Alive                 & Worship Music             & 2011 & 0.391 & 0.037 & 0.749 \\
5 & Caught in a Mosh          & Among the Living          & 1987 & 0.356 & 0.010 & 0.702 \\
\bottomrule
\end{tabularx}
\caption{Posterior means and 95\% credible intervals for the node-specific baseline sociability parameters $\vartheta_i$ for the songs with the largest and most clearly positive effects in each band under the \textsf{SMN-C-SB} model. The column Top ranks songs within each band. For Metallica and Megadeth we report only the ten songs with the highest posterior means.}
\label{tab:vartheta_top_songs_extended}
\end{table}

The patterns in Table~\ref{tab:vartheta_top_songs_extended} show that structurally popular songs are not confined to a single era or album, but instead occupy musically central positions across each band's discography. For Metallica, the highest-$\vartheta_i$ tracks include covers and deep cuts such as \emph{Turn the Page} and \emph{Free Speech for the Dumb} from \emph{Garage Inc.}, alongside canonical originals like \emph{Battery}, \emph{Fade to Black}, and \emph{For Whom the Bell Tolls}, as well as more recent material from \emph{Death Magnetic} and \emph{72 Seasons}. This blend of classic and newer songs suggests that centrality reflects cross-album stylistic similarity rather than commercial success alone. Slayer's most sociable songs cluster around \emph{South of Heaven} (\emph{Read Between the Lies}, \emph{Dissident Aggressor}, \emph{Silent Scream}), complemented by \emph{Hell Awaits} and later tracks such as \emph{World Painted Blood} and \emph{Jihad}. For Megadeth, top-ranked songs prominently feature late-career albums (\emph{Super Collider}, \emph{TH1RT3EN}, \emph{Dystopia}) together with earlier material like \emph{FFF} and \emph{My Last Words}, indicating that both classic and modern tracks can act as hubs in the similarity space. Anthrax's most central songs include fan favorites such as \emph{Be All, End All} and \emph{Caught in a Mosh}, along with more atmospheric or introductory pieces from \emph{We've Come for You All} and \emph{Worship Music}.

\section{Community detection}\label{sec_community}

For each band and audio layer, we exploit the full posterior output of the \textsf{SMN-C-SB} model to relate the inferred community structure to the discographic organization by album. Specifically, from the MCMC samples we extract the draws of the layer-specific sociability effects $\delta_{i,k}$ and the clustering indicators $\xi_{i,k}$, for nodes $i=1,\ldots,n$ and layers $k=1,\ldots,K$. We summarize sociability by the posterior means, which provide a node- and layer-specific measure of how prone each song is to form connections. To obtain a representative community partition that properly accounts for label switching, we construct the posterior similarity matrix for each layer and apply Dahl's method \citep{dahl2006model} to select the partition that is closest (in squared loss) to the posterior mean co-clustering matrix. This yields, for every band and layer, an estimated partition of songs into latent communities that is robust to label identifiability issues.

\begin{table}[!ht]
\centering
\normalsize
\begin{tabularx}{0.6\textwidth}{l *{4}{>{\centering\arraybackslash}X}}
\toprule
Layer      & Metallica & Slayer & Megadeth & Anthrax \\
\midrule
Loudness   & 0.017     & 0.012  & 0.010    & 0.017   \\
Brightness & 0.023     & 0.061  & 0.007    & 0.006   \\
Tonality   & 0.029     & 0.049  & 0.018    & 0.028   \\
Rhythm     & 0.027     & 0.031  & 0.008    & 0.036   \\
\bottomrule
\end{tabularx}
\caption{Adjusted Rand index (ARI) between the album-based partition and the posterior community partition obtained from the \textsf{SMN-C-SB} model, by band and audio layer.}
\label{tab:ari_album_vs_communities}
\end{table}

\begin{figure}[!htb]
	\centering
	\setlength{\tabcolsep}{0pt}
	\begin{tabular}{ccccc}
		& \textsf{METALLICA} & \textsf{SLAYER} & \textsf{MEGADETH} & \textsf{ANTHRAX} \\
		\begin{sideways} \hspace{0.9cm} \textbf{Loudness} \end{sideways}             &
		\includegraphics[scale = 0.28]{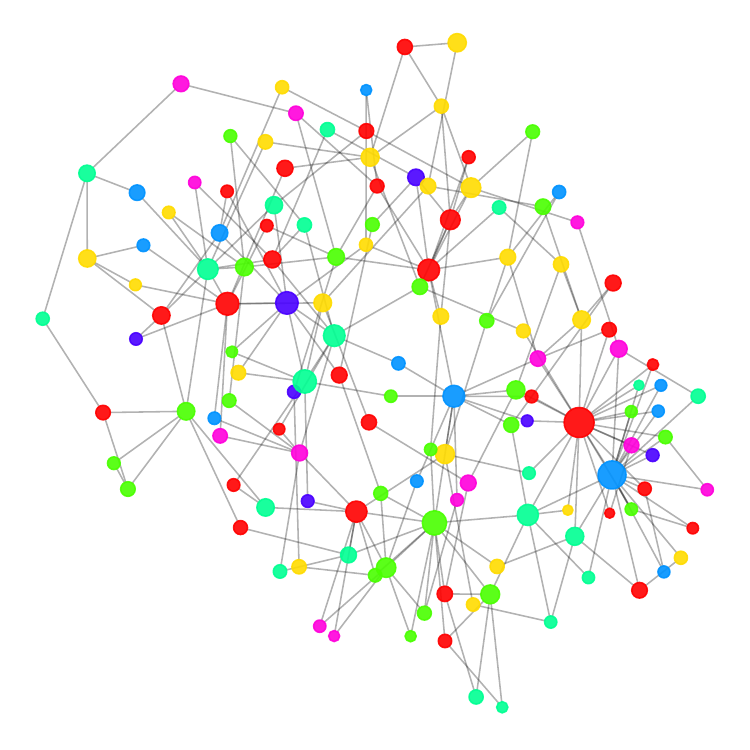} &
		\includegraphics[scale = 0.28]{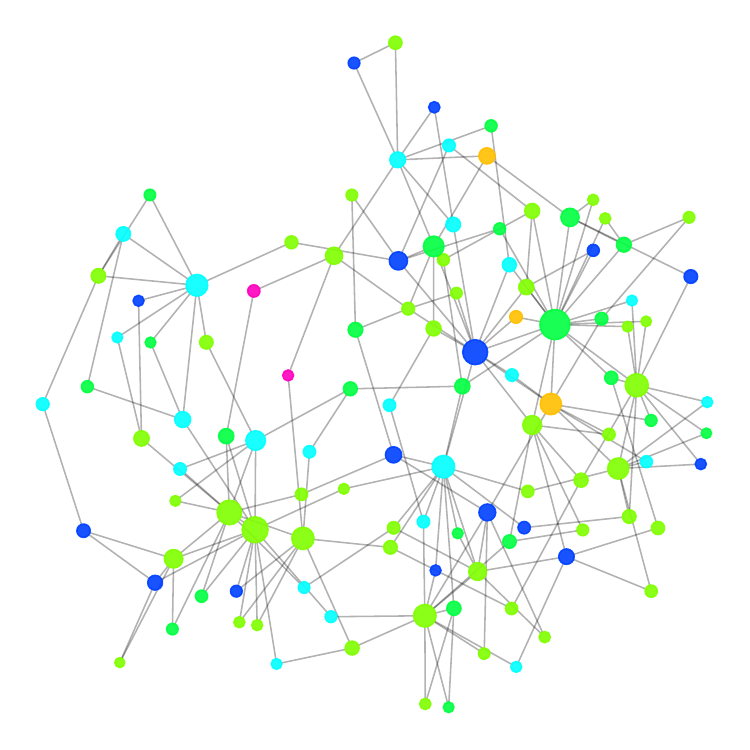}    &
		\includegraphics[scale = 0.28]{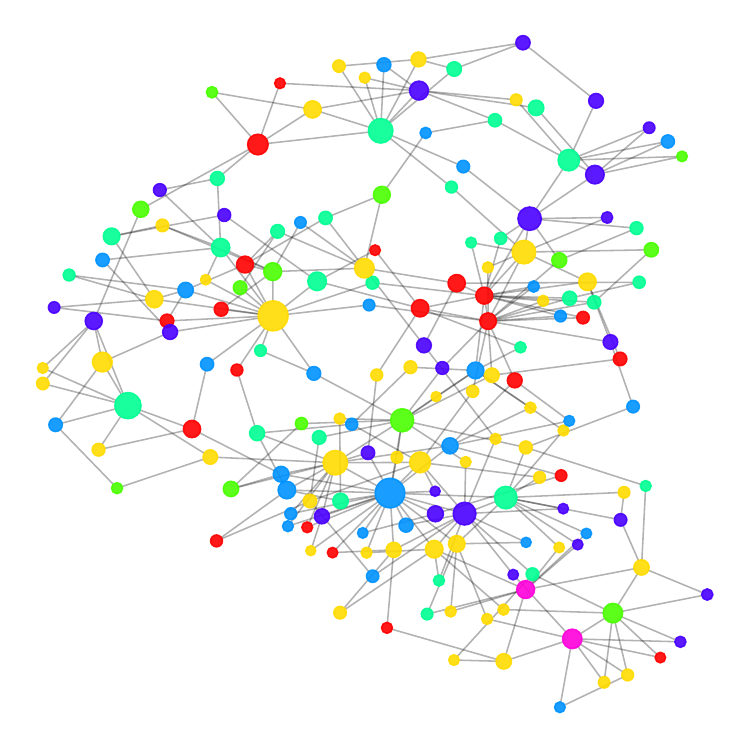}  &
		\includegraphics[scale = 0.28]{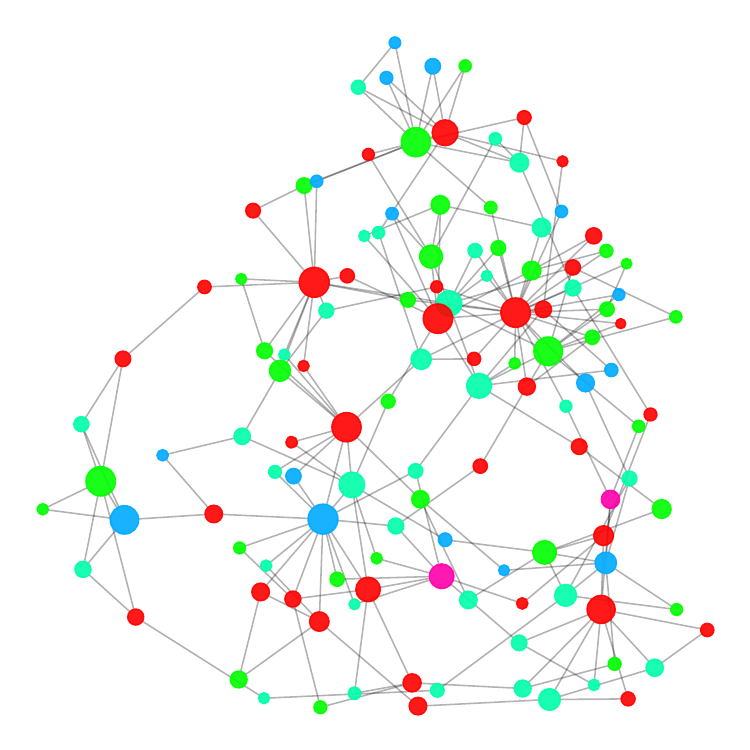}  \\
		\begin{sideways} \hspace{0.9cm} \textbf{Brightness} \end{sideways}             &
		\includegraphics[scale = 0.28]{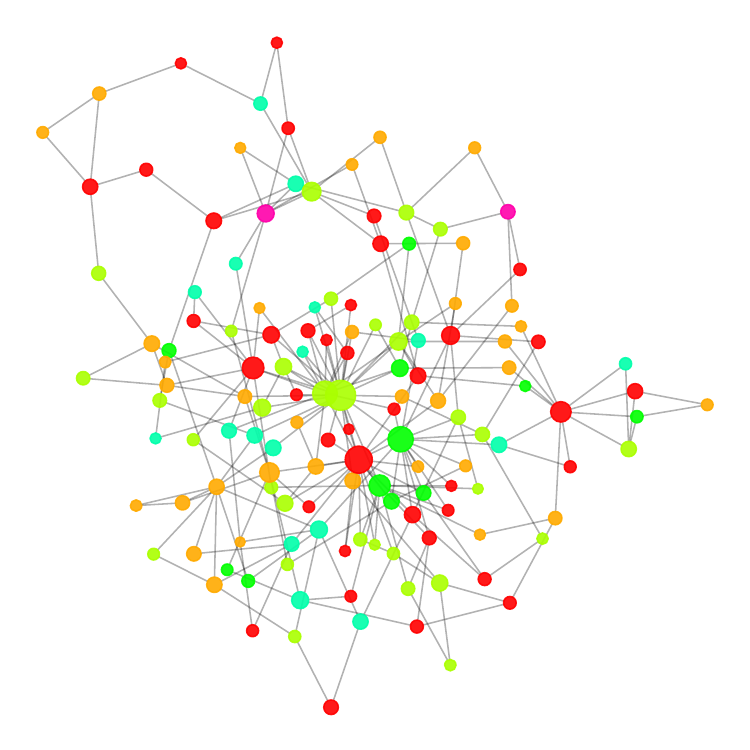} &
		\includegraphics[scale = 0.28]{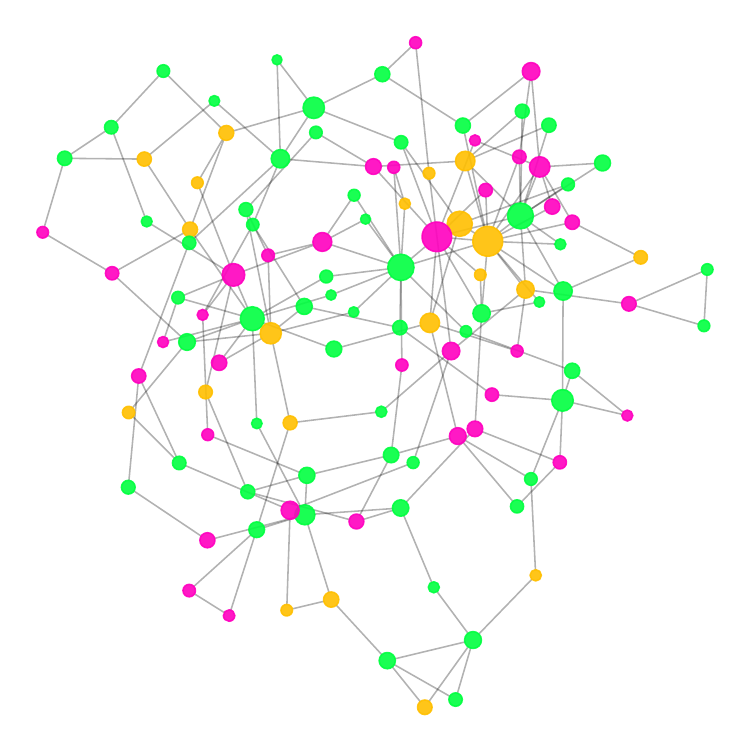}    &
		\includegraphics[scale = 0.28]{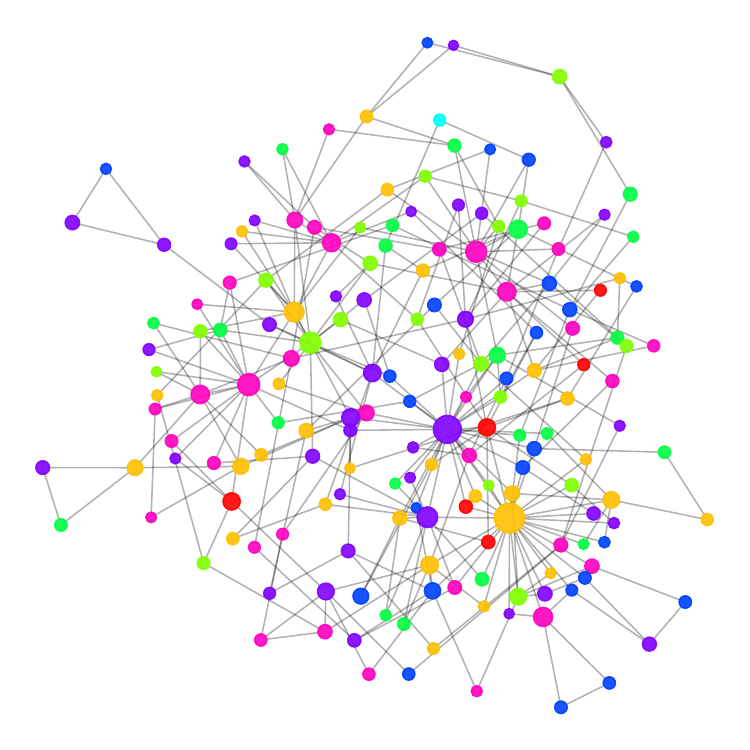}  &
		\includegraphics[scale = 0.29]{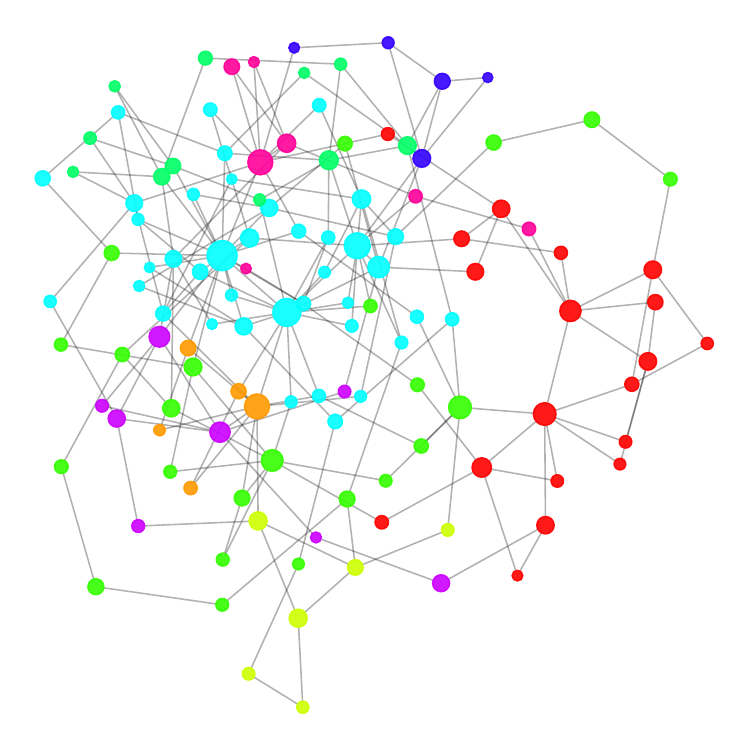}  \\
        \begin{sideways} \hspace{0.9cm} \textbf{Tonality} \end{sideways}             &
		\includegraphics[scale = 0.28]{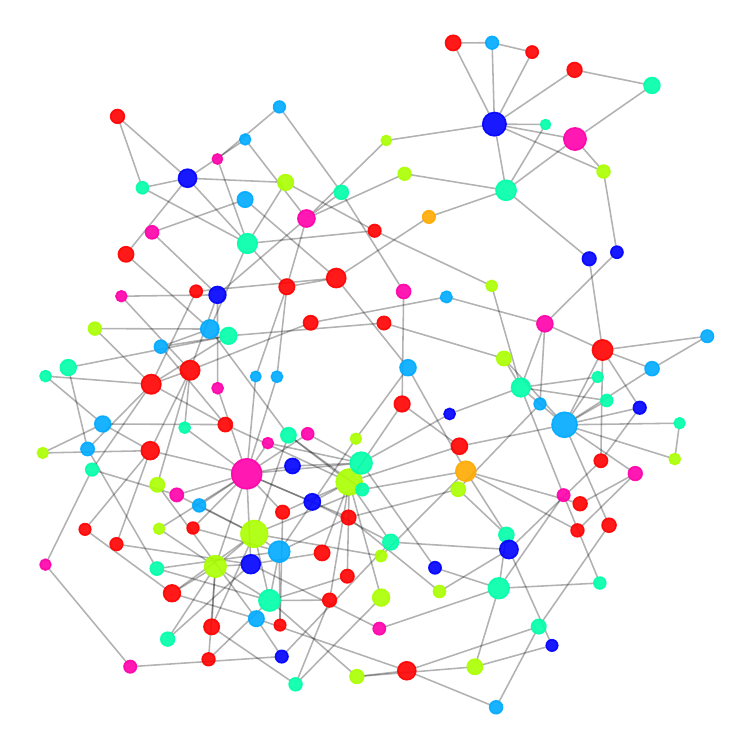} &
		\includegraphics[scale = 0.28]{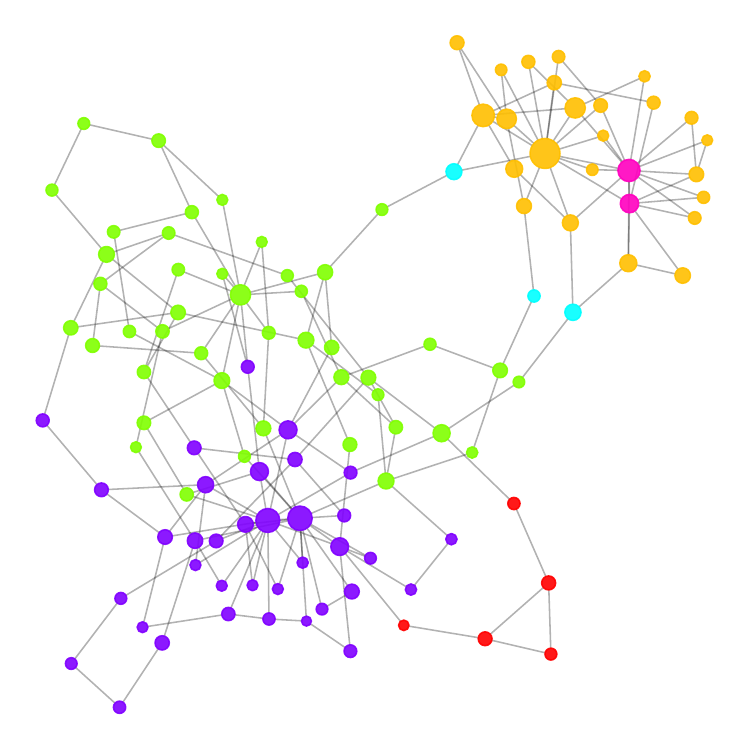}    &
		\includegraphics[scale = 0.28]{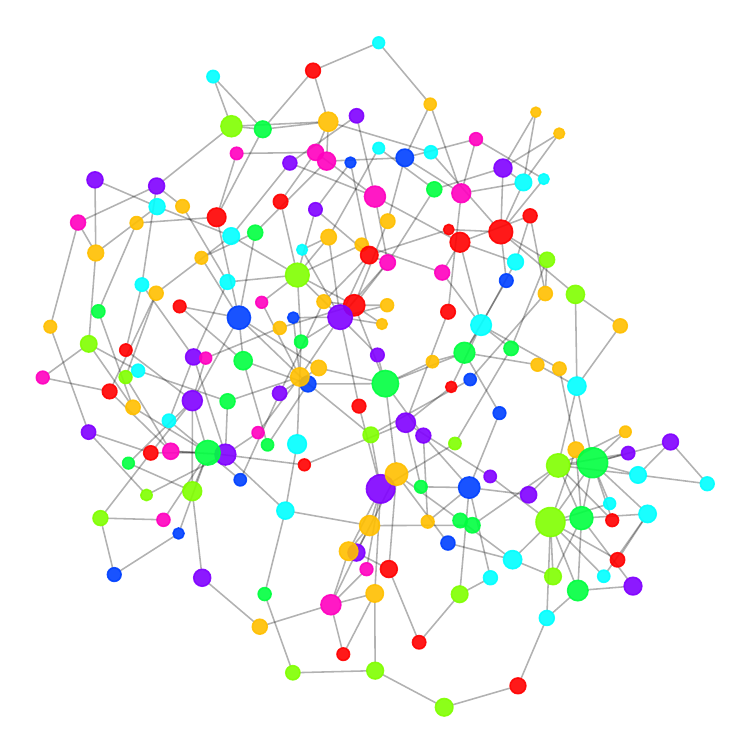}  &
		\includegraphics[scale = 0.29]{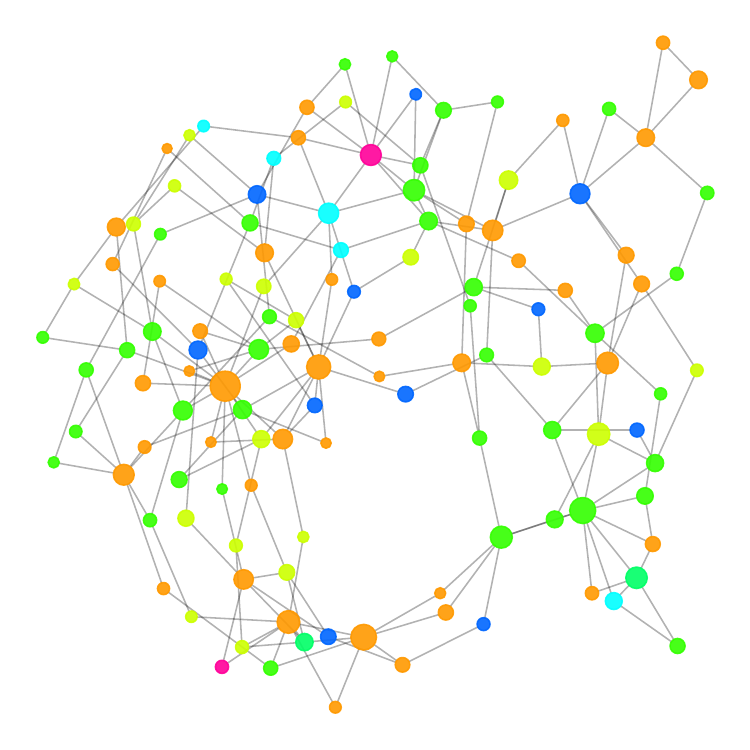}  \\
        \begin{sideways} \hspace{0.9cm} \textbf{Rhythm} \end{sideways}             &
		\includegraphics[scale = 0.28]{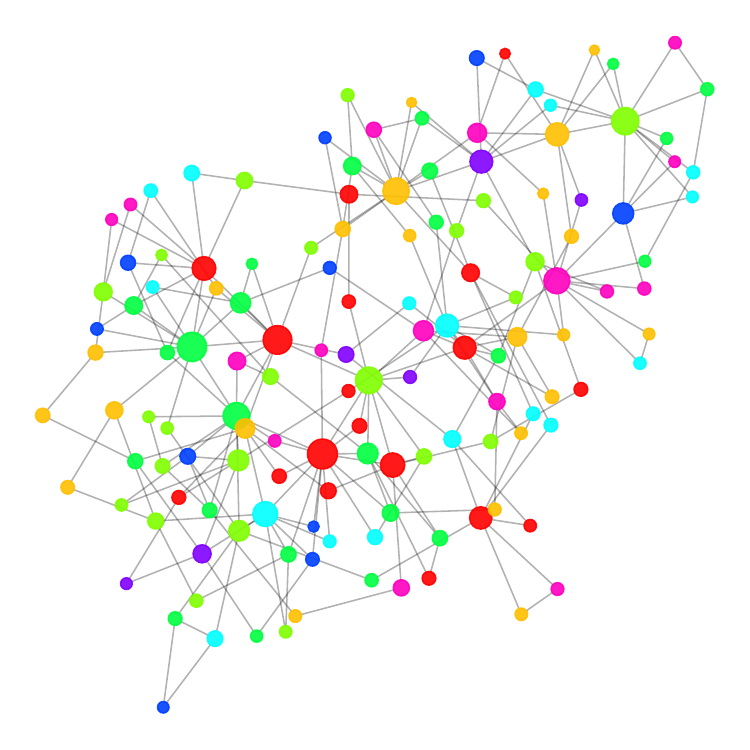} &
		\includegraphics[scale = 0.28]{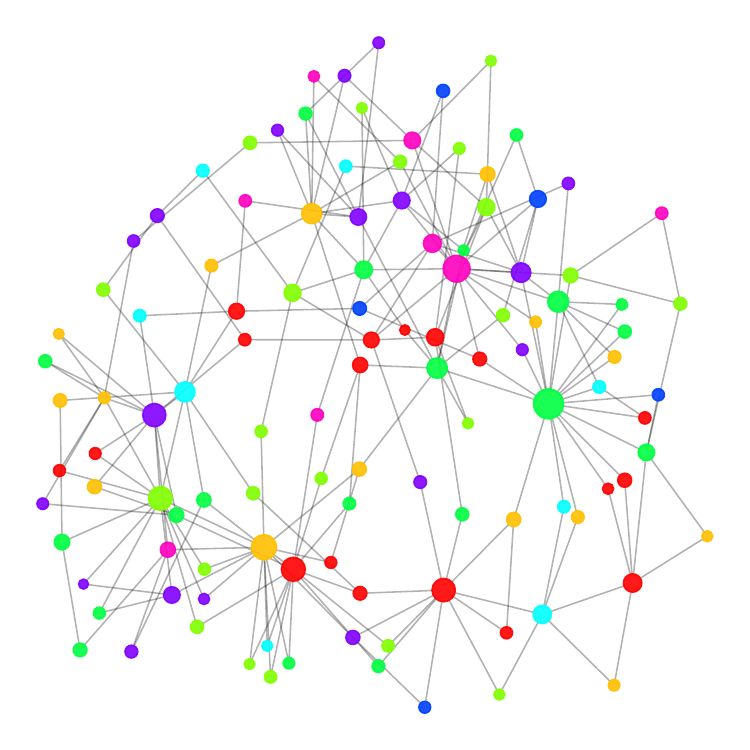}    &
		\includegraphics[scale = 0.28]{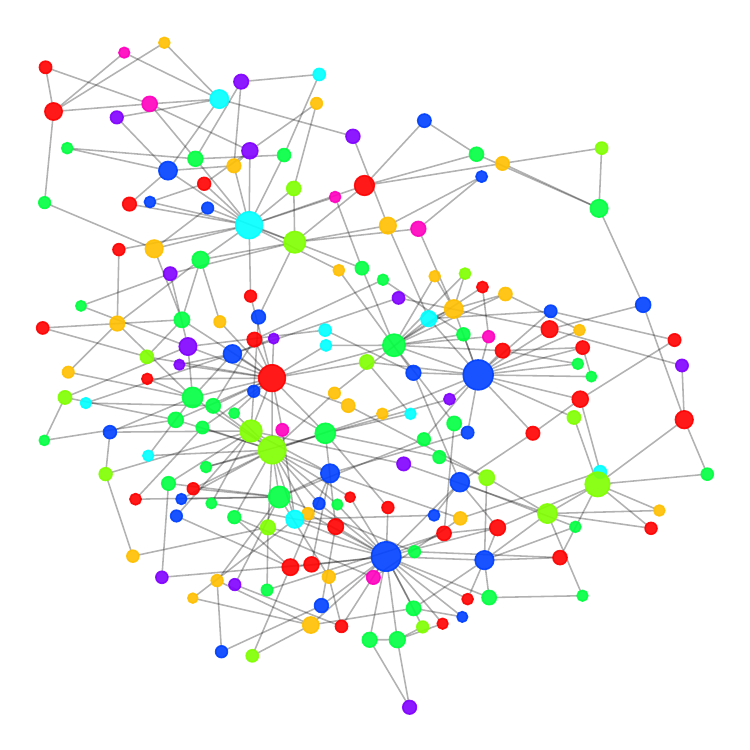}  &
		\includegraphics[scale = 0.29]{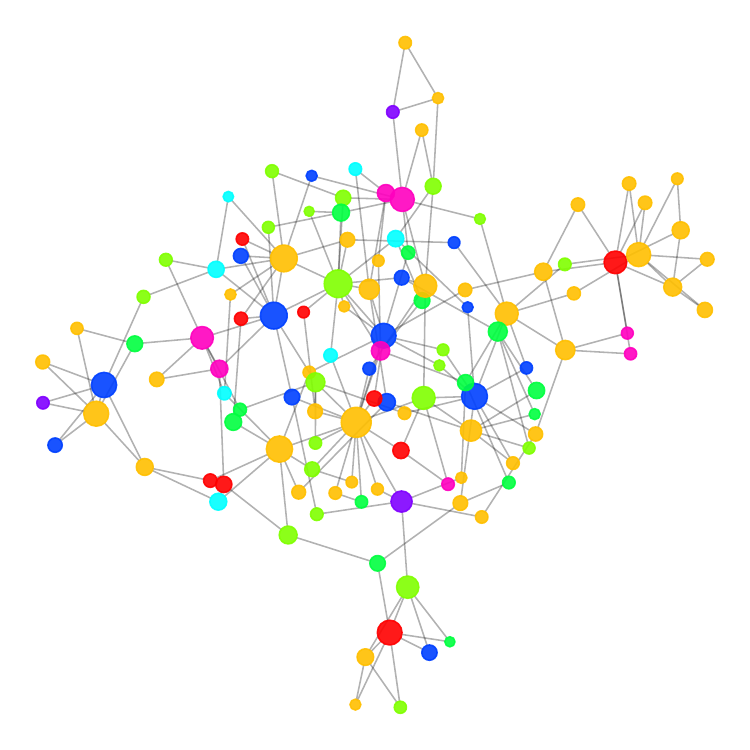}  \\
	\end{tabular}
	\caption{Posterior summaries of song similarity multilayer networks for the Big 4 under the \textsf{SMN-C-SB} model. Columns correspond to bands (Metallica, Slayer, Megadeth, Anthrax) and rows to audio layers (Loudness, Brightness, Tonality, Rhythm). Nodes represent songs, with node size proportional to the posterior mean sociability parameter $\delta_{i,k}$ and node color indicating the estimated community from Dahl’s least–squares partition. Edges correspond to observed similarity links in each layer, so that the plots jointly display how the fitted model concentrates connectivity around highly sociable songs and organizes them into latent communities.}

	\label{fig_clusts}
\end{figure}

We then compare this model-based community structure with the discographic grouping induced by album membership. For each band and layer, we compute the Adjusted Rand Index (ARI; e.g., \citealt{hubert1985comparing}) between the album partition and the Dahl-estimated partition (Table~\ref{tab:ari_album_vs_communities}). Furthermore, we visualize the multilayer networks by plotting each layer with node sizes proportional to the posterior mean of $\delta_{i,k}$ and node colors determined by the estimated community labels, thereby highlighting simultaneously node-level sociability and latent block structure (Figure~\ref{fig_clusts}). The ARI values are uniformly very small across all bands and layers (typically well below $0.1$), indicating only weak agreement between the latent communities recovered by the model and the partition defined by albums. In this context, low ARI implies that the clusters identified by the \textsf{SMN-C-SB} model do not simply reproduce album boundaries. Instead, they capture cross-album groupings driven by deeper similarities in loudness, brightness, tonality, and rhythm. This finding is important because it rules out the trivial explanation that the inferred communities are just re-labeled albums, and confirms that the hierarchical latent structure is uncovering genuinely new patterns of organization in the multilayer song similarity networks.

\section{More datasets}\label{sec_more_dataset}

As an additional assessment, we apply the same modeling and evaluation strategy to a collection of real-world multilayer network datasets that encompass diverse types of actors, sizes, and relational structures (Table~\ref{tab_datasets}). These datasets provide a broad test bed for assessing the robustness and generality of the comparative findings obtained in the thrash metal multilayer network applications. For each dataset, all five model specifications are fitted under the MCMC settings described above, and we evaluate their posterior predictive performance and out-of-sample predictive accuracy using the same set of network summary statistics and scoring rules employed in the metal band analysis. Since these datasets do not include covariate arrays, we specify the baseline linear predictor as $\eta_{i,j,k} = \zeta$ in \textsf{SMN} and $\eta_{i,j,k} = \zeta + \mu_k$ in \textsf{SMN-C}.

\begin{table}[!htb]
	\centering
	\begin{tabular}{clccc}  
		\hline
		Acronym & Reference & Actors & Layers & Edges  \\ 
		\hline
		\textsf{WIRING} & \cite{roethlisberger2003management} &  14 & 4 & 79   \\
		\textsf{TECH}   & \cite{krackhardt-1987}              &  21 & 21& 550  \\
		\textsf{SEVEN}  & \cite{vickers1981representing}      &  29 & 3 & 222  \\
		\textsf{GIRLS}  & \cite{steglich2006applying}         &  50 & 3 & 119  \\           		                
		\textsf{AARHUS} & \cite{magnani2013combinatorial}     &  61 & 5 & 620  \\
 		\textsf{MICRO}  & \cite{banerjee2013diffusion}        &  77 & 6 & 903  \\		
		\hline
	\end{tabular}
    \caption{Multilayer network datasets used in a series of additional experiments comparing all model specifications.}\label{tab_datasets} 
\end{table}

\begin{table}[!htb]
\centering
\scriptsize
\begin{tabularx}{0.9\textwidth}{l *{7}{>{\centering\arraybackslash}X}}
\toprule
Model & Dens. & Trans. & Assor. & M. Deg. & SD Deg. & M. Geo. & Diam. \\
\midrule
\multicolumn{8}{c}{\textsf{WIRING}} \\
\midrule
\textsf{SMN}      & 0.008 & 0.270 & 0.332 & 0.104 & 0.269 & 0.292 & 0.948 \\
\textsf{SMN-C}    & 0.003 & 0.264 & 0.334 & 0.037 & 0.276 & 0.295 & 0.958 \\
\textsf{SMN-C-BG} & 0.004 & 0.107 & \textbf{0.158} & 0.054 & 0.050 & \textbf{0.250} & \textbf{0.854} \\
\textsf{SMN-C-LD} & 0.005 & 0.137 & 0.202 & 0.061 & 0.062 & 0.251 & 0.874 \\
\textsf{SMN-C-SB} & \textbf{0.002} & \textbf{0.062} & \textbf{0.158} & \textbf{0.028} & \textbf{0.043} & 0.277 & 0.941 \\
\midrule
\multicolumn{8}{c}{\textsf{TECH}} \\
\midrule
\textsf{SMN}      & 0.015 & 0.190 & 0.204 & 0.299 & 0.356 & 0.533 & 1.677 \\
\textsf{SMN-C}    & 0.003 & 0.196 & 0.216 & 0.068 & 0.314 & 0.471 & 1.552 \\
\textsf{SMN-C-BG} & 0.003 & 0.193 & 0.218 & 0.068 & 0.279 & \textbf{0.448} & \textbf{1.501} \\
\textsf{SMN-C-LD} & 0.003 & 0.135 & 0.188 & 0.067 & 0.191 & 0.466 & 1.529 \\
\textsf{SMN-C-SB} & \textbf{0.002} & \textbf{0.093} & \textbf{0.185} & \textbf{0.038} & \textbf{0.177} & 0.449 & 1.530 \\
\midrule
\multicolumn{8}{c}{\textsf{SEVEN}} \\
\midrule
\textsf{SMN}      & 0.014 & 0.346 & 0.480 & 0.404 & 0.626 & 0.165 & 1.604 \\
\textsf{SMN-C}    & 0.004 & 0.353 & 0.483 & 0.123 & 0.578 & 0.165 & 1.541 \\
\textsf{SMN-C-BG} & 0.003 & 0.360 & 0.430 & 0.095 & 0.546 & \textbf{0.153} & 1.337 \\
\textsf{SMN-C-LD} & \textbf{0.002} & \textbf{0.029} & \textbf{0.104} & \textbf{0.058} & \textbf{0.163} & 0.318 & \textbf{0.120} \\
\textsf{SMN-C-SB} & \textbf{0.002} & 0.041 & 0.264 & 0.064 & 0.243 & 0.154 & 1.268 \\
\midrule
\multicolumn{8}{c}{\textsf{GIRLS}} \\
\midrule
\textsf{SMN}      & 0.003 & 0.365 & 0.427 & 0.132 & 0.677 & 1.011 & 1.191 \\
\textsf{SMN-C}    & 0.002 & 0.365 & 0.424 & 0.122 & 0.679 & \textbf{0.988} & \textbf{1.101} \\
\textsf{SMN-C-BG} & 0.002 & 0.385 & 0.417 & 0.101 & 0.629 & 1.060 & 1.197 \\
\textsf{SMN-C-LD} & 0.001 & 0.153 & 0.237 & 0.029 & 0.305 & 0.800 & 1.309 \\
\textsf{SMN-C-SB} & \textbf{0.000} & \textbf{0.049} & \textbf{0.159} & \textbf{0.022} & \textbf{0.298} & \textbf{0.731} & 3.218 \\
\midrule
\multicolumn{8}{c}{\textsf{AARHUS}} \\
\midrule
\textsf{SMN}      & 0.005 & 0.289 & 0.155 & 0.277 & 0.528 & 0.987 & 2.556 \\
\textsf{SMN-C}    & 0.002 & 0.292 & 0.169 & 0.105 & 0.539 & 0.791 & 2.075 \\
\textsf{SMN-C-BG} & \textbf{0.001} & 0.153 & 0.139 & 0.052 & 0.217 & 0.486 & 1.302 \\
\textsf{SMN-C-LD} & \textbf{0.001} & 0.137 & 0.141 & 0.048 & 0.171 & 0.347 & 1.089 \\
\textsf{SMN-C-SB} & \textbf{0.001} & \textbf{0.042} & \textbf{0.084} & \textbf{0.033} & \textbf{0.136} & \textbf{0.109} & \textbf{0.711} \\
\midrule
\multicolumn{8}{c}{\textsf{MICRO}} \\
\midrule
\textsf{SMN}      & 0.005 & 0.100 & 0.086 & 0.345 & 0.627 & 0.437 & 1.136 \\
\textsf{SMN-C}    & 0.001 & 0.102 & 0.087 & 0.082 & 0.579 & 0.396 & 1.034 \\
\textsf{SMN-C-BG} & \textbf{0.000} & 0.077 & \textbf{0.069} & \textbf{0.038} & 0.383 & 0.194 & 0.650 \\
\textsf{SMN-C-LD} & 0.001 & 0.071 & 0.075 & 0.044 & \textbf{0.361} & \textbf{0.105} & \textbf{0.427} \\
\textsf{SMN-C-SB} & 0.001 & \textbf{0.022} & 0.085 & 0.077 & 0.499 & 0.311 & 0.921 \\
\bottomrule
\end{tabularx}
\caption{Mean RMSE for posterior predictive network statistics across models and datasets. Columns report density (Dens.), global transitivity (Trans.), degree assortativity (Assor.), mean degree (M. Deg.), standard deviation of degree (SD Deg.), mean geodesic distance (M. Geo.), and diameter (Diam.).}
\label{tab:ppc_rmse_metrics}
\end{table}

\begin{table}[!htb]
\centering
\scriptsize
\begin{tabularx}{0.8\textwidth}{l *{5}{>{\centering\arraybackslash}X}}
\toprule
Model & AUC & BS & LL & DIC & WAIC \\
\midrule
\multicolumn{6}{c}{\textsf{WIRING}} \\
\midrule
\textsf{SMN}      & 0.862 & 0.118 & 0.354 & 301.245 & 300.649 \\
\textsf{SMN-C}    & 0.864 & 0.118 & 0.353 & 301.663 & 303.241 \\
\textsf{SMN-C-BG} & 0.963 & 0.055 & 0.174 & 172.925 & 171.206 \\
\textsf{SMN-C-LD} & 0.959 & 0.063 & 0.198 & 197.550 & 191.251 \\
\textsf{SMN-C-SB} & \textbf{0.973} & \textbf{0.052} & \textbf{0.165} & \textbf{147.272} & \textbf{170.422} \\
\midrule
\multicolumn{6}{c}{\textsf{TECH}} \\
\midrule
\textsf{SMN}      & 0.818 & 0.084 & 0.277 & 2700.144 & 2705.085 \\
\textsf{SMN-C}    & 0.820 & 0.084 & 0.276 & 2680.401 & 2695.170 \\
\textsf{SMN-C-BG} & 0.897 & 0.067 & 0.224 & 2292.874 & 2317.989 \\
\textsf{SMN-C-LD} & \textbf{0.949} & \textbf{0.051} & \textbf{0.168} & \textbf{1797.678} & \textbf{1815.137} \\
\textsf{SMN-C-SB} & 0.937 & 0.055 & 0.176 & 1842.754 & 2020.159 \\
\midrule
\multicolumn{6}{c}{\textsf{SEVEN}} \\
\midrule
\textsf{SMN}      & 0.732 & 0.127 & 0.412 & 1071.139 & 1080.441 \\
\textsf{SMN-C}    & 0.733 & 0.127 & 0.411 & 1066.096 & 1076.376 \\
\textsf{SMN-C-BG} & 0.823 & 0.113 & 0.357 & 1014.390 & 1018.459 \\
\textsf{SMN-C-LD} & \textbf{0.981} & \textbf{0.040} & \textbf{0.127} & \textbf{415.060} & \textbf{415.603} \\
\textsf{SMN-C-SB} & 0.947 & 0.066 & 0.210 & 606.324 & 666.608 \\
\midrule
\multicolumn{6}{c}{\textsf{GIRLS}} \\
\midrule
\textsf{SMN}      & 0.693 & 0.031 & 0.139 & 1106.498 & 1126.149 \\
\textsf{SMN-C}    & 0.695 & 0.031 & 0.139 & 1106.461 & 1126.190 \\
\textsf{SMN-C-BG} & 0.830 & 0.030 & 0.120 & 1060.361 & 1064.565 \\
\textsf{SMN-C-LD} & 0.984 & 0.016 & 0.051 & 499.435 & 528.992 \\
\textsf{SMN-C-SB} & \textbf{0.988} & \textbf{0.015} & \textbf{0.046} & \textbf{397.054} & \textbf{433.085} \\
\midrule
\multicolumn{6}{c}{\textsf{AARHUS}} \\
\midrule
\textsf{SMN}      & 0.814 & 0.054 & 0.194 & 3764.474 & 3764.617 \\
\textsf{SMN-C}    & 0.816 & 0.054 & 0.193 & 3741.639 & 3752.779 \\
\textsf{SMN-C-BG} & 0.967 & 0.030 & 0.102 & 2252.298 & 2243.376 \\
\textsf{SMN-C-LD} & 0.969 & \textbf{0.028} & \textbf{0.097} & 2121.850 & 2155.372 \\
\textsf{SMN-C-SB} & \textbf{0.973} & 0.030 & 0.098 & \textbf{1980.228} & \textbf{2073.874} \\
\midrule
\multicolumn{6}{c}{\textsf{MICRO}} \\
\midrule
\textsf{SMN}      & 0.793 & 0.046 & 0.172 & 6286.238 & 6299.399 \\
\textsf{SMN-C}    & 0.794 & 0.046 & 0.172 & 6277.287 & 6291.145 \\
\textsf{SMN-C-BG} & 0.939 & 0.033 & 0.115 & 4494.034 & 4491.625 \\
\textsf{SMN-C-LD} & \textbf{0.945} & \textbf{0.032} & \textbf{0.111} & \textbf{4304.952} & \textbf{4380.780} \\
\textsf{SMN-C-SB} & 0.888 & 0.039 & 0.140 & 5363.630 & 5506.152 \\
\bottomrule
\end{tabularx}
\caption{Predictive performance and information criteria across models and datasets. 
Columns report mean area under the ROC curve (AUC), Brier score (BS), log-loss (LL), 
deviance information criterion (DIC), and Watanabe--Akaike information criterion (WAIC).}
\label{tab:model_metrics}
\end{table}

Table~\ref{tab:ppc_rmse_metrics} reports the average RMSE between posterior predictive means and observed values for several network statistics, averaged across layers, for all models and datasets. As in the metal band applications, the baseline \textsf{SMN} and the covariate-only extension \textsf{SMN-C} generally exhibit the largest discrepancies, particularly for transitivity, assortativity, and path-based measures. Introducing additional latent structure systematically reduces these RMSEs. For \textsf{WIRING} and \textsf{TECH}, \textsf{SMN-C-SB} attains the smallest errors for most degree-related statistics (density, mean degree, and degree variability), while \textsf{SMN-C-BG} performs slightly better for mean geodesic distance and diameter, indicating that both latent specifications capture higher-order connectivity patterns more accurately than the baselines. In the \textsf{SEVEN} multilayer network, \textsf{SMN-C-LD} clearly dominates for transitivity, assortativity, degree-based statistics, and diameter, with \textsf{SMN-C-BG} showing a modest advantage for mean geodesic distance. For \textsf{GIRLS}, \textsf{SMN-C-SB} yields the smallest RMSE for all local and meso-scale statistics (density, transitivity, assortativity, degree summaries, and mean distance), while \textsf{SMN-C} slightly outperforms the other models in terms of diameter. In the larger \textsf{AARHUS} multilayer network, \textsf{SMN-C-SB} consistently achieves the best fit across all statistics, whereas in \textsf{MICRO} the best performance is shared among \textsf{SMN-C-BG} (for density, assortativity, and mean degree), \textsf{SMN-C-LD} (for degree variability and path-based measures), and \textsf{SMN-C-SB} (for transitivity). Overall, the RMSE results confirm that the latent-structure extensions substantially improve posterior predictive fit relative to \textsf{SMN} and \textsf{SMN-C}, with \textsf{SMN-C-LD} and \textsf{SMN-C-SB} most frequently providing the closest match to the observed multilayer structures.

Table~\ref{tab:model_metrics} summarizes the models' predictive performance in terms of AUC, Brier score, log-loss, DIC, and WAIC for the same set of datasets. The patterns are broadly consistent with those observed in the RMSE analysis. In \textsf{WIRING}, \textsf{SMN-C-SB} attains the highest AUC and the lowest BS, LL, DIC, and WAIC, indicating the best overall discrimination and calibration of edge probabilities among the five specifications. In \textsf{TECH} and \textsf{SEVEN}, \textsf{SMN-C-LD} dominates across all predictive metrics, combining high AUC with low BS and LL, and yielding substantially reduced information criteria relative to \textsf{SMN} and \textsf{SMN-C}. For \textsf{GIRLS}, \textsf{SMN-C-SB} again provides the best predictive performance on all metrics. In \textsf{AARHUS}, \textsf{SMN-C-SB} achieves the largest AUC and the lowest DIC and WAIC, while \textsf{SMN-C-LD} attains slightly smaller BS and LL, suggesting that these two latent extensions offer comparable and clearly superior predictive accuracy compared to the simpler models. Finally, in \textsf{MICRO}, \textsf{SMN-C-LD} attains the best scores across all predictive metrics, with substantial gains over \textsf{SMN} and \textsf{SMN-C}. Taken together, these results show that the conclusions drawn from the thrash metal multilayer networks extend to a broader class of applications: models that enrich the baseline specification with flexible latent structure (\textsf{SMN-C-LD} and \textsf{SMN-C-SB}) consistently deliver superior posterior predictive and out-of-sample performance across heterogeneous multilayer network datasets.

\section{Discussion}\label{sec_discussion}

In this work, we developed and applied a Bayesian framework for the analysis of multilayer networks of musical similarity constructed directly from audio data, integrating multiple acoustic descriptors within a unified hierarchical probabilistic scheme. Through a family of models of increasing complexity, ranging from purely additive formulations (\textsf{SMN}), to covariate-enhanced extensions (\textsf{SMN-C}), continuous latent geometries (\textsf{SMN-C-BG}, \textsf{SMN-C-LD}) and stochastic community structures (\textsf{SMN-C-SB}), we systematically assessed the explanatory and predictive capacity of alternative structural representations of musical similarity.

The empirical results consistently show that models without latent structure are insufficient to adequately represent key properties of the observed networks, such as transitivity, assortativity, and heterogeneity in connectivity patterns. The introduction of latent components leads to substantial improvements in both posterior predictive fit and performance metrics (AUC, Brier score, log-loss, DIC, and WAIC). In particular, the layer-specific stochastic block model \textsf{SMN-C-SB} exhibits the best overall performance across all analyzed datasets, indicating that discrete community organization captures dominant patterns of musical similarity more effectively than the continuous geometric representations considered.

Application to the complete discographies of the \emph{Big 4 of thrash metal} reveals that, after controlling for exogenous covariates and nodal heterogeneity, community structure explains the bulk of the observed connectivity. The inferred latent communities do not trivially coincide with divisions by band, album, or chronological period; rather, they cluster songs according to acoustic affinities that transcend eras and releases, uncovering stylistic links that are not apparent under purely editorial classifications. Furthermore, the detection of structurally central nodes suggests the existence of songs that act as stylistic convergence points across multiple communities.

From a methodological perspective, latent geometry models provide useful continuous representations for visualization and spatial interpretation of musical similarity, albeit at increased computational cost and with inherent identifiability limitations. In contrast, the stochastic block formulation offers a favorable balance between predictive performance, interpretability, and inferential stability, positioning it as the most robust and practical alternative for the analysis of moderate-to-large multilayer acoustic networks.

Among the main limitations of this study are the static treatment of networks, without explicit modeling of the temporal dynamics of musical styles; the binarization of originally continuous similarity measures; and the restriction to a specific set of acoustic descriptors. Future work may extend the proposed framework toward dynamic multilayer models, weighted network representations, and expanded feature spaces, including semantic, textual, or deep embeddings derived from modern machine learning approaches. Additionally, the development of more efficient computational strategies will be crucial to scaling the methodology to substantially larger and more complex musical catalogs.

Overall, this study demonstrates that Bayesian multilayer modeling constitutes a rigorous methodological tool for translating large volumes of high dimensional musical data into interpretable network structures, facilitating the quantitative analysis of stylistic similarity, community organization, and musical interconnectivity within extensive and heterogeneous collections.

\section*{Statements and declarations}

The authors declare that they have no known competing financial interests or personal relationships that could have appeared to influence the work reported in this article.

All R and C++ code required to reproduce our results is publicly available at \url{https://github.com/jstats1702/the-big-4}. The repository includes a detailed README with step by step instructions, and the scripts are well documented. All datasets used in the applications and cross validation exercises are also included in the repository.

During the preparation of this work the authors used ChatGPT-5-Thinking in order to improve language and readability. After using this tool, the authors reviewed and edited the content as needed and take full responsibility for the content of the publication.

\bibliography{references.bib}
\bibliographystyle{apalike}

\appendix

\section{Album covarage}\label{sec_albums}

Below we report the album coverage for each band (years in parentheses), as listed on their official websites:
\url{https://www.metallica.com/}, \url{https://www.slayer.net}, \url{https://www.megadeth.com}, and \url{https://www.anthrax.com}.

\textsf{METALLICA:} \textit{Kill ’Em All} (1983), \textit{Ride the Lightning} (1984), \textit{Master of Puppets} (1986), \textit{Garage Days Re-Revisited} (1987), \textit{…And Justice for All} (1988), \textit{Metallica} (1991), \textit{Load} (1996), \textit{Reload} (1997), \textit{Garage Inc.} (1998), \textit{St. Anger} (2003), \textit{Death Magnetic} (2008), \textit{Hardwired… to Self-Destruct} (2016), \textit{72 Seasons} (2023).

\textsf{SLAYER:} \textit{Show No Mercy} (1983), \textit{Hell Awaits} (1985), \textit{Reign in Blood} (1986), \textit{South of Heaven} (1988), \textit{Seasons in the Abyss} (1990), \textit{Divine Intervention} (1994), \textit{Diabolus in Musica} (1998), \textit{God Hates Us All} (2001), \textit{Christ Illusion} (2006), \textit{World Painted Blood} (2009), \textit{Repentless} (2015).

\textsf{MEGADETH:} \textit{Killing Is My Business… and Business Is Good!} (1985), \textit{Peace Sells… but Who’s Buying?} (1986), \textit{So Far, So Good… So What!} (1988), \textit{Rust in Peace} (1990), \textit{Countdown to Extinction} (1992), \textit{Youthanasia} (1994), \textit{Cryptic Writings} (1997), \textit{Risk} (1999), \textit{The World Needs a Hero} (2001), \textit{The System Has Failed} (2004), \textit{United Abominations} (2007), \textit{Endgame} (2009), \textit{TH1RT3EN} (2011), \textit{Super Collider} (2013), \textit{Dystopia} (2016), \textit{The Sick, the Dying… and the Dead!} (2022).

\textsf{ANTHRAX:} \textit{Fistful of Metal} (1984), \textit{Spreading the Disease} (1985), \textit{Among the Living} (1987), \textit{State of Euphoria} (1988), \textit{Persistence of Time} (1990), \textit{Sound of White Noise} (1993), \textit{Stomp 442} (1995), \textit{Volume 8: The Threat Is Real} (1998), \textit{We’ve Come for You All} (2003), \textit{Worship Music} (2011), \textit{For All Kings} (2016).

\section{Notation}

The cardinality of a set \(A\) is denoted by \(|A|\). If \(P\) is a logical proposition, its indicator is \(\mathbb{I}\{P\}\in\{0,1\}\), with \(I\{P\}=1\) when \(P\) is true and \(I\{P\}=0\) otherwise. The Gamma function is defined by \(\Gamma(x)=\int_{0}^{\infty} u^{x-1} e^{-u}\, \mathrm{d}u\). Vectors and matrices whose entries are subscripted variables are written in boldface. For example, \(\boldsymbol{x}=(x_1,\ldots,x_n)^\top\) denotes an \(n\times 1\) column vector with elements \(x_1,\ldots,x_n\). We use \(\boldsymbol{0}\) and \(\boldsymbol{1}\) for column vectors of zeros and ones, and \(\mathbf{I}\) for the identity matrix (a subscript indicates dimension, e.g., \(\mathbf{I}_n\) is the \(n\times n\) identity). The transpose of a vector \(\boldsymbol{x}\) is \(\boldsymbol{x}^\top\) (and analogously for matrices). For a square matrix \(\mathbf{X}\), \(\operatorname{tr}(\mathbf{X})\) denotes its trace and \(\mathbf{X}^{-1}\) its inverse. The Euclidean norm of \(\boldsymbol{x}\) is \(\|\boldsymbol{x}\|=\sqrt{\boldsymbol{x}^\top \boldsymbol{x}}\).

Now, we present the form of some standard probability distributions:
\begin{itemize}

    \item A random variable $X$ has a Normal distribution with parameters $\mu\in\mathbb{R}$ and $\sigma^2>0$, denoted $X\mid\mu,\sigma^2\sim\textsf{N}(\mu,\sigma^2)$, if its density is
    $$
    p(x\mid\mu,\sigma^2)=(2\pi\sigma^2)^{-1/2}\,\exp\!\left\{-\frac{(x-\mu)^2}{2\sigma^2}\right\},\qquad x\in\mathbb{R}.
    $$
    
    \item A $d\times 1$ random vector $\boldsymbol{X}=(X_1,\ldots,X_d)$ has a multivariate Normal distribution with parameters $\boldsymbol{\mu}$ and $\mathbf{\Sigma}$, denoted by $\boldsymbol{X}\mid\boldsymbol{\mu},\mathbf{\Sigma} \sim \textsf{N}_d(\boldsymbol{\mu},\mathbf{\Sigma})$, if its density is
    $$
    p(\mathbf{x}\mid\boldsymbol{\mu},\mathbf{\Sigma}) = (2\pi)^{-d/2}\,|\mathbf{\Sigma}|^{-1/2}\,\exp\!\left\{-\tfrac12 (\mathbf{x} - \boldsymbol{\mu})^{\top}\mathbf{\Sigma}^{-1}(\mathbf{x} - \boldsymbol{\mu}) \right\}.
    $$
    
    \item 
    A random variable $X$ has a Gamma distribution with shape $\alpha>0$ and rate $\beta>0$, denoted $X\mid \alpha,\beta\sim\textsf{G}(\alpha,\beta)$, if its density is
    $$
    p(x\mid \alpha,\beta)=\frac{\beta^{\alpha}}{\Gamma(\alpha)}\,x^{\alpha-1}\,e^{-\beta x},\qquad x>0.
    $$

    \item
    A random variable $X$ has an Inverse--Gamma distribution with shape $\alpha>0$ and scale $\beta>0$, denoted $X\mid \alpha,\beta\sim\textsf{IG}(\alpha,\beta)$, if its density is
    $$
    p(x\mid \alpha,\beta)=\frac{\beta^{\alpha}}{\Gamma(\alpha)}\,x^{-(\alpha+1)}\,e^{-\beta/x},\qquad x>0.
    $$

    \item
    A $K\times 1$ random vector $\boldsymbol{X} = (X_1,\ldots, X_K)$ has a Dirichlet distribution with parameter vector $\boldsymbol{\alpha} = (\alpha_1,\ldots , \alpha_K)$, each $\alpha_k > 0$, denoted $\boldsymbol{X}\mid\boldsymbol{\alpha}\sim\textsf{Dir}(\boldsymbol{\alpha})$, if its density is
    $$
    p(\mathbf{x}\mid\boldsymbol{\alpha}) =
    \begin{cases}
      \displaystyle \frac{\Gamma\!\left(\sum_{k=1}^K\alpha_k\right)}{\prod_{k=1}^K\Gamma(\alpha_k)}\prod_{k=1}^K x_k^{\alpha_k-1}, & \text{if }\sum_{k=1}^K x_k = 1\text{ and }x_k\ge 0,\\[8pt]
      0, & \text{otherwise.}
    \end{cases}
    $$

    \item
    A random variable $X$ has a Categorical distribution with parameter vector $\boldsymbol{\pi} = (\pi_1,\ldots , \pi_K)$, where $\sum_{k=1}^K \pi_k =1$ and $\pi_k\geq 0$, denoted $X\mid\boldsymbol{\pi}\sim\textsf{Cat}(\boldsymbol{\pi})$, if its probability mass function is
    $$
    p(x \mid\boldsymbol{\pi}) =
    \begin{cases}
      \displaystyle \prod_{k=1}^K \pi_k^{1_{\{x = k\}}}, & x\in\{1,\ldots,K\},\\[6pt]
      0, & \text{otherwise.}
    \end{cases}
    $$

\end{itemize}

\end{document}